\documentclass[preprint,11pt]{elsarticle} % Single-column format without the forced double-spacing

\usepackage[margin=1.0in]{geometry} % Explicitly sets tight, professional margins
\usepackage{lineno} % Line numbers in the margin for reviewers
\usepackage{graphicx}
\usepackage{subcaption}
\usepackage{float} % explicit: provides the [H] float specifier used below

\usepackage{amsmath}
\usepackage{amssymb}
\usepackage{amsthm}
\newtheorem{theorem}{Theorem}

\usepackage{booktabs}
\usepackage{multirow}
\usepackage{rotating}
\usepackage{microtype}
\usepackage{hyperref}
\usepackage{color}

\usepackage{algorithm}
\usepackage{algpseudocode}
\usepackage{tikz} % Required for flow chart
\usetikzlibrary{shapes,arrows, arrows.meta, calc} % Tikz libraries required for the flow chart in the template

\tikzstyle{mybox} = [draw=black, fill=gray!5, thick,
    rectangle, inner sep=5pt, inner ysep=10pt]
\tikzstyle{fancytitle} =[fill=black!60, text=white]
\usetikzlibrary{decorations.pathreplacing,arrows.meta}

\definecolor{purple_heart}{RGB}{100,50,210}
\colorlet{lpurple_heart}{purple_heart!65}
\colorlet{llpurple_heart}{purple_heart!45}
\colorlet{lllpurple_heart}{purple_heart!20}
\definecolor{sand}{RGB}{255,220,83}
\definecolor{lightgreen}{RGB}{0,204,102}
\definecolor{grass}{RGB}{178,255,153}
\definecolor{truegreen}{RGB}{0, 155, 0}
\definecolor{flashpurple}{RGB}{255, 255, 0}
\definecolor{graincolor}{RGB}{255, 209, 78}
\definecolor{orangered}{RGB}{255, 100, 0}
\definecolor{darkblue}{RGB}{5, 0, 186}
\definecolor{gold}{RGB}{255, 231, 0}
\definecolor{silver}{RGB}{207, 207, 207}

\newcommand{\mycirc}[1][black]{\footnotesize\textcolor{#1}{\ensuremath\bullet}}
\newcommand{\mysquare}[1][black]{\footnotesize\textcolor{#1}{\ensuremath\blacksquare}}
\newcommand{\myrect}[1][black]{\footnotesize\textcolor{#1}{\rule{1.2em}{0.4em}}}
\newcommand{\myemptysquare}[2][black]{%
  \tikz[scale=#2]{%
    \fill[#1] (0,0) rectangle (0.12,0.12);
    \fill[white] (0.03,0.03) rectangle (0.09,0.09);
  }%
}
\newcommand{\myemptydisk}[2][black]{%
  \tikz[scale=#2]{%
    \fill[#1] (0.06,0.06) circle (0.06);
    \fill[white] (0.06,0.06) circle (0.03);
  }%
}

\NewDocumentCommand{\mycross}{O{black} O{1.5}}{%
  \tikz[scale=#2, baseline=-0.5ex]{%
    \draw[#1, very thick] (-0.06, 0) -- (0.06, 0); % Horizontal line
    \draw[#1, very thick] (0, -0.06) -- (0, 0.06); % Vertical line
  }%
}
\journal{Computer Methods in Applied Mechanics and Engineering}

\begin{document}

\begin{frontmatter}

\title{Topology-Preserving Mesh Adaptation for Sharp-Interface Multiphase PFEM}

\author[immc]{Félix Ruyffelaere\corref{cor1}}
\ead{felix.ruyffelaere@uclouvain.be}
\author[inria]{Michel Henry}
\author[immc]{Jonathan Lambrechts}
\author[immc]{Jean-François Remacle}

\address[immc]{UCLouvain - iMMC, Avenue Georges Lemaître 4, 1348 Louvain-la-Neuve, Belgium}
\address[inria]{Univ. Grenoble Alpes, Inria, 38000 Grenoble, France}
\cortext[cor1]{Corresponding author}

\begin{abstract}
This paper presents a robust, fully Lagrangian framework based on the Particle Finite Element Method (PFEM) capable of simulating multiphase flows with an arbitrary number of immiscible phases. Interface-capturing methods can suffer from numerical diffusion, while interface-tracking methods can allow the underlying mesh resolution to prematurely dictate topological changes. To address these limitations, we introduce a dynamic mesh adaptation strategy that naturally preserves sharp geometric interfaces without relying on classical constrained triangulation. A protecting, node-empty disk is assigned to each segment of the discretized interface, ensuring that the edge is part of the Delaunay triangulation. Our approach decouples the interface physics from the grid size, allowing the integration of sub-grid physical models to properly govern topological changes independently of the user-defined mesh size. The capabilities and accuracy of the framework are validated against standard multiphase benchmarks, closely matching references while maintaining a remarkably low overall node count. We demonstrate the scalability and geometric versatility of the method, in particular with a challenging 16-phase Rayleigh-Taylor simulation.
\end{abstract}


\end{frontmatter}

%% -------------------------------------------------------------------------
\section{Introduction}
\label{sec:intro}

Immiscible multiphase flows are ubiquitous in both natural phenomena and industrial engineering applications. From environmental processes such as wave breaking and atmospheric droplet dynamics to advanced industrial applications, such as chemical bubble column reactors, microfluidics, and metallurgical casting, the interaction between distinct fluid phases governs the macroscale behavior of the system. Accurate numerical simulation of these processes is of paramount importance for the optimization of industrial designs, risk assessment, and predictive modeling of complex fluid-structure systems. Consequently, developing robust computational frameworks capable of tracking and resolving these multi-fluid interactions remains a critical focus of contemporary research in computational fluid dynamics.
\bigskip

Despite decades of methodological advancements, the high-fidelity simulation of immiscible multiphase flows presents severe mathematical and computational challenges. Chief among these is the distinct multi-scale disparity characterizing the flow; macroscale convective transport often relies heavily on localized, microscopic interfacial phenomena. Furthermore, fluid interfaces often experience extreme deformations and complex topological bifurcations such as merging, pinching, and droplet atomization \cite{fragmentation_cohesion}. Capturing these transitions accurately is highly difficult because the underlying physics at the interface is not fully encapsulated by the standard bulk Navier-Stokes equations alone \cite{ns_limits}. Interfacial mechanics requires the precise evaluation of surface tension forces, which are highly sensitive to local geometric curvature. At the macroscopic scale, these forces manifest as sharp jump discontinuities in the pressure and velocity gradient fields. Crucially, the exact physical mechanisms governing topological transitions remain an open area of research, meaning numerical models must be capable of handling changing boundaries without introducing non-physical numerical artifacts.
\medskip

Historically, Eulerian formulations have served as the dominant paradigm for fluid simulations. In these frameworks, the governing equations are solved on a fixed spatial grid and the moving interfaces are tracked either explicitly or implicitly. Prominent among these are the Volume-of-Fluid (VOF) \cite{vof}, Level-Set \cite{level_set}, and Diffuse Interface (or Phase-Field) methods. While highly successful in specific regimes, each of these methods exhibits inherent limitations when applied to sharp, immiscible boundaries. Diffuse interface models regularize the boundary over a finite thickness, which inherently prevents the strict enforcement of sharp pressure jumps and introduces non-physical numerical diffusion across the phases \cite{phase_field}. Conversely, while the Level-Set method preserves sharp boundaries and provides smooth geometrical curvatures, it is fundamentally prone to numerical mass loss, although new re-initialization techniques try to correct this flaw \cite{level_set_mass}. The VOF method enforces strict mass conservation but requires complex geometrical reconstruction algorithms to compute accurate interface normals and curvatures.
\bigskip

Driven by exponential advancements in high-performance computing, researchers have increasingly explored Lagrangian descriptions of fluid motion. While classical Finite Element Method (FEM) frameworks have been extended to accommodate Lagrangian descriptions, maintaining mesh integrity during extreme nodal advection continues to pose a significant challenge. To circumvent these mesh-distortion bottlenecks, alternative mesh-free paradigms have been developed, most notably Smoothed Particle Hydrodynamics (SPH) \cite{sph} and the Moving Particle Semi-implicit (MPS) \cite{mps} method. However, a primary drawback of these formulations is their fundamental reliance on empirical smoothing kernels or localized interpolation stencils, which inherently introduce numerical diffusion and can compromise spatial accuracy. To mitigate grid distortion without abandoning the structural advantages of a mesh, the Arbitrary Lagrangian-Eulerian (ALE) formulation \cite{ale}, originally conceived for solid mechanics, decouples the grid motion from the underlying material trajectories, frequently restricting purely Lagrangian displacement to regions in the immediate vicinity of phase interfaces. Nevertheless, in its classical implementation, the ALE method cannot natively accommodate topological transitions or extreme fluid shear. Under such scenarios, frequent global remeshing becomes unavoidable, which inevitably introduces undesirable projection and discretization errors. Although alternative approaches have been proposed to address these interface-tracking limitations \cite{xmesh}, they frequently introduce distinct and complex algorithmic challenges.
\medskip

Operating under the premise that severe deformations and topological transitions are inherently unavoidable in multiphase flows, the Particle Finite Element Method (PFEM), a comprehensive review of which is provided in \cite{pfem_review}, maintains a fully Lagrangian framework across the entire computational domain. By executing a global remeshing step immediately following the nodal advection phase, the PFEM inherently preserves element quality and prevents mesh entanglement. Crucially, the purely Lagrangian viewpoint eliminates the convective acceleration term from the material derivative in the Navier-Stokes equations (\autoref{sec:governing_equations}). In its standard form, the PFEM consists of three successive operations: solving the governing equations on the current mesh, advecting the mesh nodes according to the computed velocity field, and reconstructing a new triangulation of the resulting point cloud. Additional mesh adaptation procedures are generally incorporated to maintain an appropriate point distribution throughout the computation \cite{thomas_refine, pfem_liege}. To date, PFEM has proven exceptionally robust in resolving violent free-surface flows, including breaking waves \cite{thomas_pfemdem} and complex fluid-structure interactions \cite{pfem_fsi, pfem_artery}. This success is largely based on geometric boundary reconstruction techniques, such as the $\alpha$-shape method, which recovers free surfaces from the unordered point cloud after remeshing.
\bigskip

Extending this framework to immiscible multiphase flows is considerably more challenging. Unlike free surfaces, internal interfaces cannot simply be reconstructed by filtering the triangulation: they must remain explicit and conforming throughout successive remeshing operations so that discontinuous material properties remain confined to their respective phases. Furthermore, unconstrained Delaunay triangulations may alter the interface connectivity through edge flips, causing topological changes that are dictated by the mesh rather than by the underlying physics. Substantial efforts by Idelsohn et al.~\cite{pfem_nfluids1, pfem_nfluids2, pfem_nfluids3} have been made to address these issues within the PFEM framework, where interfaces are represented by tracked interface nodes. To reduce unintended connectivity changes, localized mesh refinement is introduced along the interfaces. However, no mechanism controls the triangulation when elements become entirely composed of interface nodes, making topological transitions strongly dependent on the local mesh resolution.
\bigskip

In this paper, we propose a mesh-adaptive framework for immiscible multiphase PFEM simulations that provides strict geometric and topological guarantees during remeshing. Existing methods implicitly couple topology changes to the mesh size. In contrast, our method completely decouples geometry preservation from the mesh resolution, allowing physical models, not the discretization, to decide when topological transitions occur. As a result, such transitions can be properly handled by dedicated physical or empirical models, such as \cite{filaments_of_death}. We further incorporate established bulk mesh adaptation algorithms \cite{thomas_refine} to improve element quality throughout the fluid domain. By embedding geometric constraints directly into the remeshing stage, the method preserves sharp interfaces without diffuse regularization or majority-node approximations. The remainder of this paper presents the proposed algorithm and evaluates its mass conservation, computational efficiency, and robustness on classical multiphase benchmarks.

\section{Method}

\begin{figure}[h!]
    \centering
    \scalebox{0.9}{
        \begin{tikzpicture}
                \node at (0, 0) {\includegraphics[height=4cm]{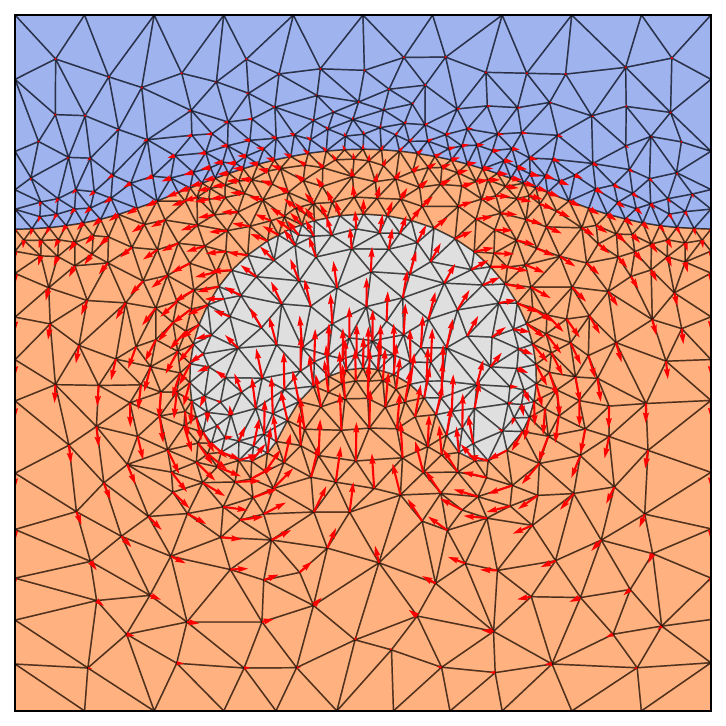}};
                \node[text width=4cm, align=center] at (0, -2.75) {\footnotesize \textbf{(0)}: Solving the governing equations (FEM). The velocity field is obtained.};
                \draw[->, black, very thick] (0.4, 2.1) -- (0.4, 2.6) -- (3.8, 2.6) -- (3.8, 2.1);
                \node at (2.05, 2.85) {$t_i = t_{i+1}$}; 
                \node at (4.2, 0) {\includegraphics[height=4cm]{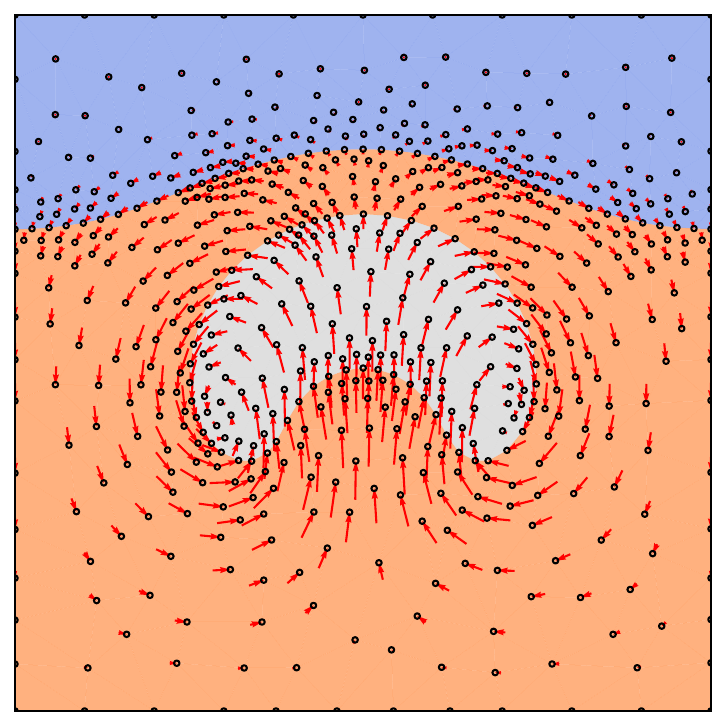}};
                \node[text width=4cm, align=center] at (4.1, -2.75) {\footnotesize \textbf{(1)}: Advection of the material points and indicator field.};
                \draw[->, black, very thick] (4.6, 2.1) -- (4.6, 2.6) -- (8, 2.6) -- (8, 2.1);
                \node at (8.4, 0) {\includegraphics[height=4cm]{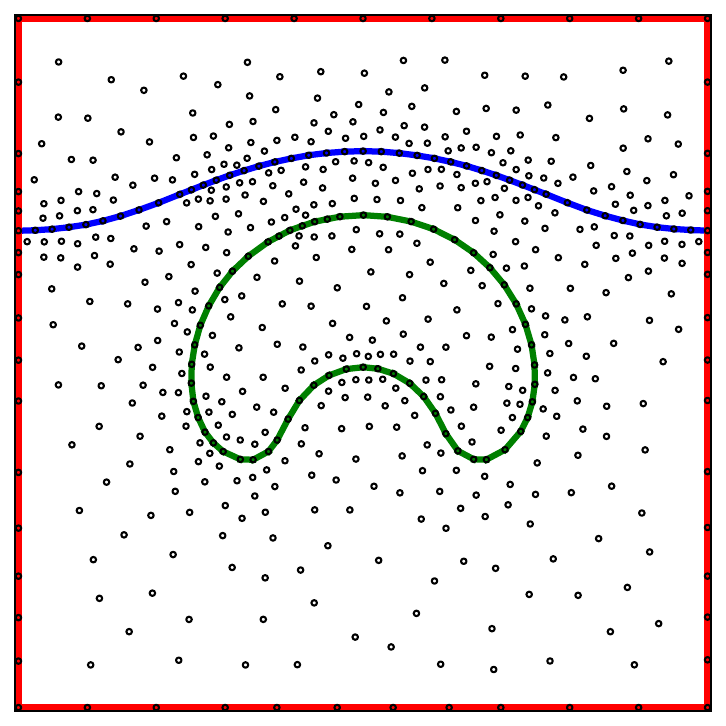}};
                \node[text width=4cm, align=center] at (8.4, -2.75) {\footnotesize \textbf{(2)}: Boundaries \mysquare[red] and interfaces \mysquare[blue] \mysquare[truegreen] are extracted and adapted.};
                \draw[->, black, very thick] (8.8, 2.1) -- (8.8, 2.6) -- (12.2, 2.6) -- (12.2, 2.1);
                \node at (12.6, 0) {\includegraphics[height=4cm]{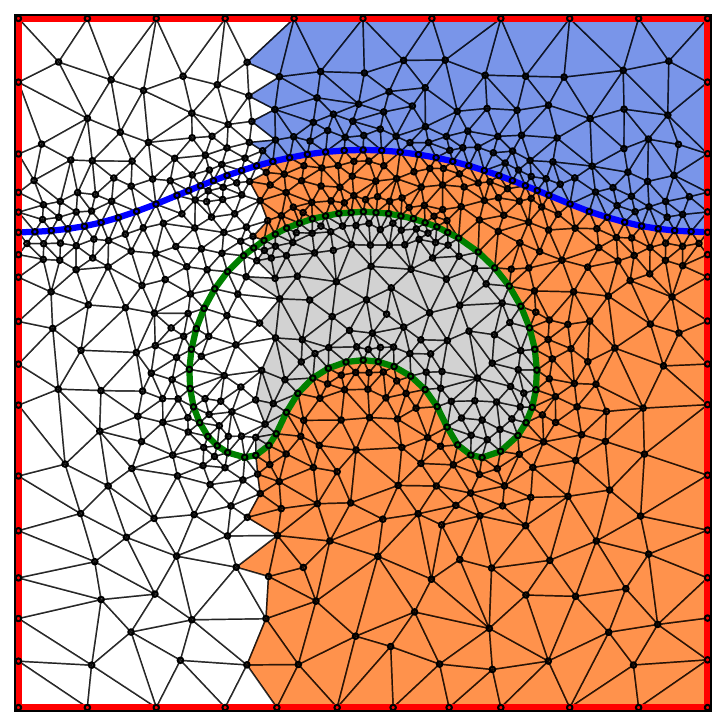}};
                \node[text width=4cm, align=center] at (12.6, -3) {\footnotesize \textbf{(3)}: Delaunay connectivity and mesh adaptation of the bulk. The indicator field is reconstructed.};
                \draw[->, black, very thick] (12.2, -4.2) -- (12.2, -4.35) -- (0.4, -4.35) -- (0.4, -3.6);
        \end{tikzpicture}
    }
    \caption{Workflow of the Particle Finite Element Method for multiphase flows. The fluid is treated in a fully Lagrangian manner, where the material points here coincide with the mesh nodes. The points are advected with the solved velocity field. A remeshing procedure, at the interfaces and in the fluid's bulk, ensures good mesh quality at all times.\label{fig:pfem_schematic}}
\end{figure}

This section presents the proposed remeshing framework for immiscible multiphase flows within the Particle Finite Element Method. An overview of the complete algorithm is provided in \autoref{fig:pfem_schematic}. Starting from a conforming mesh, the governing equations, presented in \autoref{sec:governing_equations}, are first solved to obtain the fluid velocity. The mesh nodes are then advected according to this velocity, after which a new computational mesh is reconstructed.
\bigskip

The remeshing procedure constitutes the main contribution of this work and is divided into two stages. First, \autoref{sec:interfaces} introduces the interface reconstruction algorithm. By explicitly reconstructing the interfaces and adapting the local point distribution in their vicinity, the proposed approach theoretically guarantees that interface geometry and topology are preserved during the subsequent triangulation, a result developed in \autoref{sec:protecting_balls} for completeness. Next, \autoref{sec:mesh_adaptation} describes the adaptation of the fluid bulk using established mesh refinement algorithms and their integration into the proposed framework. Finally, \autoref{sec:coloring} presents the transfer of physical quantities from the previous mesh to the reconstructed one, completing the remeshing procedure before the next time step. Importantly, while the methodology and numerical validation in this study are formulated within a two-dimensional framework, the underlying geometric philosophy and theoretical guarantees remain fully valid in three dimensions.
\bigskip

% ------ MODIFIED VERSION -----------

\subsection{Governing equations}
\label{sec:governing_equations}
% describe navier-stokes equations 
% show how having an explicit interface makes things easy 
% describe the advection of the particles

To dictate the motion of the particles, in the first step of our method, the incompressible Navier-Stokes equations are considered in a domain with multiple fluid phases. Let $\Omega \subset \mathbb{R}^{d}$, with $d=2$ in this work, be a domain containing $n$ distinct and immiscible phases. Let $t$ denote time and $\mathbf{x}$ the Cartesian coordinates. The evolution of the velocity field $\mathbf{u}(\mathbf{x}, t)$ and the pressure field $p(\mathbf{x}, t)$ is governed by:

\begin{equation}
    \begin{aligned}
        \rho \frac{\mathrm{d} \mathbf{u}}{\mathrm{d} t} &= \nabla \cdot \boldsymbol{\sigma} + \rho \mathbf{g} - \gamma \kappa \mathbf{n} \delta_\Gamma\\ 
        \nabla \cdot \mathbf{u} &= 0
    \end{aligned}
    \label{eq:ns}
\end{equation}
where $\rho$ is the density, $\boldsymbol{\sigma}$ is the Cauchy stress tensor, $\mathbf{g}$ is the gravitational acceleration vector. The last term of \autoref{eq:ns}, which we will denote $\mathbf{f}_\gamma$, accounts for the forces acting on the $(d-1)$-dimensional interfaces under surface tension $\gamma$, with $\kappa$ the curvature, $\mathbf{n}$ the unit normal and $\delta_\Gamma$ the Dirac distribution supported on the interface. The sign of this term follows from the orientation conventions adopted below, for which the capillary force is directed along $-\mathbf{n}$ for a convex inclusion of phase $1$. This work considers an isotropic Newtonian fluid such that:

\begin{equation}
    \boldsymbol{\sigma} = -p\mathbf{I} + 2 \mu \, \mathbf{D}(\mathbf{u}), \qquad \mathbf{D}(\mathbf{u}) = \frac{1}{2} \left( \nabla \mathbf{u} + (\nabla \mathbf{u})^T \right)
    \label{eq:cauchy}
\end{equation}

where $\mu$ is the dynamic viscosity, $\mathbf{D}(\mathbf{u})$ is the strain-rate tensor and $\mathbf{I}$ is the identity tensor. Along the domain boundaries, partitioned as $\partial \Omega = \partial \Omega_D \cup \partial \Omega_N$ with disjoint interiors, the following conditions are prescribed:

\begin{equation}
    \begin{aligned} 
        \mathbf{u} &= \hat{\mathbf{u}} \quad \text{on } \partial \Omega_D \\ 
        \boldsymbol{\sigma} \cdot \mathbf{n}_{\partial \Omega} &= \hat{\boldsymbol{\sigma}}_n \quad \text{on } \partial \Omega_N 
    \end{aligned}
    \label{eq:bc}
\end{equation}

where $\mathbf{n}_{\partial \Omega}$ denotes the outward unit normal to $\partial \Omega$, not to be confused with the interface normal $\mathbf{n}$ introduced below.
\bigskip

Let us now consider the open and pairwise disjoint subdomains $\{\Omega_k\}$ such that $\overline{\Omega} = \bigcup_k \overline{\Omega_k}$, where each subdomain corresponds to a distinct phase $k$. We define the interfaces between the phases by $\Gamma(t) = \bigcup_{k \neq l} \left( \partial \Omega_k \cap \partial \Omega_l \right)$. The physical properties within the domain are defined piecewise for each phase $k$:

\begin{equation}
    \rho(\mathbf{x}, t) = \rho_k \quad \text{and} \quad \mu(\mathbf{x}, t) = \mu_k \quad \text{if } \mathbf{x} \in \Omega_k(t)
    \label{eq:properties}
\end{equation}

Since the phases are immiscible and transported by the flow, and no mass transfer occurs across $\Gamma$, these properties are constant along the trajectories. At the interface $\Gamma(t)$, defining the jump of a quantity $\phi$ across the interface as $[\phi] = \phi_1 - \phi_2$, the coupling conditions are given by

\begin{equation}
    \begin{aligned} 
        {[}\mathbf{u}] &= \mathbf{0} \quad \text{on } \Gamma \\
    [\boldsymbol{\sigma}] \cdot \mathbf{n} &= -\gamma \kappa \mathbf{n} \quad \text{on } \Gamma
    \end{aligned}
    \label{eq:interface_conditions}
\end{equation}

where $\gamma$ is the surface tension coefficient between the two phases around $\Gamma$, assumed constant along each interface, so that Marangoni effects are not considered, and $\kappa = \nabla \cdot \mathbf{n}$ is the local interface curvature. The unit normal $\mathbf{n}$ is oriented from phase $1$ towards phase $2$, consistently with the definition of the jump, so that a drop of phase $1$ satisfies $\kappa > 0$ and $[p] = \gamma \kappa > 0$.
\bigskip

% --------------- SURFACE TENSION

In this manuscript, the equations of \autoref{eq:ns} are solved at each time step using the FEM with a stabilized $P_1-P_1$ formulation. We stress that this equal-order pair is not a constraint imposed by the method. Nothing in the remeshing framework presented hereafter precludes the use of inf-sup stable pairs: the meshes produced by our procedure are standard conforming simplicial meshes and support any element type. The temporal discretization is performed using an implicit Euler scheme. 
\bigskip

A benefit of using a conforming mesh at the interface is that piecewise-linear basis functions are able to represent the kinks of $p$ and $\mathbf{u}$ that arise due to a density contrast between multiple phases. However, the pressure space is unable to capture the pressure discontinuity

\begin{equation}
    [p] = \gamma \kappa + 2 \, \left[ \mu \, \frac{\partial u_n}{\partial n}\right],
    \label{eq:pressure_jump}
\end{equation}

where $u_n = \mathbf{u} \cdot \mathbf{n}$ and $\partial \, \cdot / \partial n = \mathbf{n} \cdot \nabla ( \cdot )$ denotes the derivative along the interface normal. To account for the surface tension force $\mathbf{f}_\gamma$ and to model the subsequent jump of $p$, we use here a method introduced by the authors of \cite{xmesh}, inspired from the Ghost-Fluid-Method \cite{surf_tension_review, ghost_fluid_method}. The key idea is to modify the discrete pressure gradient by adding to it an auxiliary field $p^*$, that is constructed explicitly to capture the jump given by \autoref{eq:pressure_jump}.
\bigskip

In a variational setting, the problem becomes: Let $\mathcal{V}_{\hat{\mathbf{u}}} = \{ \mathbf{v} \in [H^1(\Omega)]^d \ : \ \mathbf{v} = \hat{\mathbf{u}} \text{ on } \partial \Omega_D \}$, $\mathcal{V}_0$ the corresponding space with $\hat{\mathbf{u}} = \mathbf{0}$, and $\mathcal{Q} = H^1(\Omega)$. Find $(\mathbf{u}, p) \in \mathcal{V}_{\hat{\mathbf{u}}} \times \mathcal{Q}$ such that, for all $(\mathbf{v}, q) \in \mathcal{V}_0 \times \mathcal{Q}$,

\begin{equation}
    \begin{aligned}
        \int_{\Omega} \rho \frac{\mathrm{d} \mathbf{u}}{\mathrm{d} t} \cdot \mathbf{v} \, \mathrm{d}\Omega
        + \int_{\Omega} \left( 2 \mu \, \mathbf{D}(\mathbf{u}) : \mathbf{D}(\mathbf{v}) + \nabla(p+p^*) \cdot \mathbf{v} \right) \mathrm{d}\Omega
        &= \int_{\Omega} \rho \, \mathbf{g} \cdot \mathbf{v} \, \mathrm{d}\Omega \\
        &\quad + \int_{\partial \Omega_N} \left( \hat{\boldsymbol{\sigma}}_n + (p + p^*) \, \mathbf{n}_{\partial \Omega} \right) \cdot \mathbf{v} \, \mathrm{d}S, \\
        \int_{\Omega} q \, \nabla \cdot \mathbf{u} \, \mathrm{d}\Omega &= 0.
    \end{aligned}
    \label{eq:weak_form}
\end{equation}

% ------------------------------

The pressure term is deliberately retained in its non-integrated form, which enables the pressure jump to be introduced through $p^*$. The Neumann contribution now involves $\hat{\boldsymbol{\sigma}}_n + (p + p^*)\,\mathbf{n}_{\partial\Omega}$, rather than $\hat{\boldsymbol{\sigma}}_n$ alone. In the present work, the second contribution in \autoref{eq:pressure_jump} is neglected: we assume that capillary effects dominate the pressure jump, which is the regime targeted by the test cases proposed below whenever surface tension is taken into account. Under this hypothesis, the pressure jump reduces to $[p] = \gamma\kappa$, which is known explicitly once the curvature is available, and which we impose directly through $p^*$. With $p^*$ constructed as detailed below, the effect of surface tension is taken into account in \autoref{eq:weak_form} by the modified pressure gradient operator. Finally, \autoref{eq:weak_form} is presented in its Galerkin form. The stabilization terms required by the equal-order $P_1 - P_1$ finite-element pair are introduced at the discrete level and are detailed in \cite{thomas_pfemdem}.
\bigskip

In the discrete setting, $p_h$ is a continuous $P_1$ field, while $p_h^*$ is double-valued at the interface and $\nabla p_h^*$ is evaluated element by element. The field $p_h^*$ is defined to be zero across $\Omega_2$. In $\Omega_1$, it vanishes everywhere except on elements bordering the interface, where it is defined as a piecewise linear function equal to $\gamma\kappa$ at the interface nodes. At each interface node $\mathbf{x}_i$, the curvature is obtained from the circle passing through the three consecutive interface nodes $\mathbf{x}_{i-1}$, $\mathbf{x}_i$ and $\mathbf{x}_{i+1}$: it is defined as the inverse of the corresponding circumradius $R_i$, with its sign chosen consistently with the convention adopted in \autoref{eq:interface_conditions}. This three-point construction provides a purely local estimate based solely on the explicit polyline representation of $\Gamma$, which is maintained by the remeshing procedure throughout the simulation. Since $p_h$ is continuous across $\Gamma$, the total pressure $p_h + p_h^*$ carries, by construction, the prescribed jump at the interface.
\bigskip

The explicit construction of $p^*$ avoids duplicating the degrees of freedom associated with the pressure at the interface. This treatment is specific to the present choice of surface tension model, and alternative formulations could be employed within the same framework; it has so far been developed for two-phase flows, the presence of triple points introducing additional complexity in the construction of $p^*$. 
\bigskip

The non-linear convective component of the inertial term in the Navier-Stokes equations is inherently absorbed by the Lagrangian framework. The elimination of this advective term linearizes the momentum equations, significantly simplifying the resulting FEM problem, but requires augmenting the framework with a kinematic constraint to track node positions over time. The trajectory of each node is governed by the ordinary differential equation:

\begin{equation}
    \mathbf{x}'(t) = \mathbf{u}\left( \mathbf{x}(t), t \right).
    \label{eq:trajectory}
\end{equation}

Conventionally, this relation is discretized with an explicit-Euler scheme. It should be emphasized that this explicit treatment is unrelated to the stability of the flow solver: the Navier-Stokes equations are integrated implicitly, and only the kinematic update \autoref{eq:trajectory} remains explicit. The accuracy of this single explicit step is nevertheless critical in a multiphase context. The interfaces are material lines carried by the nodes themselves, so any error committed while integrating \autoref{eq:trajectory} is directly a geometric error on $\Gamma$, and therefore an error on the volume occupied by each phase. This error vanishes as $\Delta t$ decreases, but a first-order scheme requires prohibitively small time steps for it to become negligible, as demonstrated in \autoref{sec:vortex}. While higher-order multistage schemes (e.g., Runge-Kutta methods) could mitigate this issue through predictor-corrector sub-steps, their implementation is severely constrained by the evolving mesh topology: evaluating intermediate stages at a frozen topology risks generating inverted elements or invalid geometries. Independently of accuracy, the explicit advection is also subject to a CFL condition, bounded by the local mesh size and maximum nodal velocity, which safeguards the topological integrity of the elements during the step. 
\bigskip

To circumvent these limitations without introducing computationally expensive sub-iterations, we utilize the temporal history of the velocity field. The velocity field computed at the previous time step is interpolated onto the newly generated mesh prior to the solution phase, yielding an advected velocity field $\tilde{\mathbf{u}}^{n-1}$ mapped onto the current nodal configuration. This historical data enables a second-order accurate position update via an explicit second-order Adams-Bashforth (AB2) scheme:

\begin{equation}
    \mathbf{x}^{n+1} = \mathbf{x}^n + \Delta t \left( \frac{3}{2}\mathbf{u}^n - \frac{1}{2}\tilde{\mathbf{u}}^{n-1} \right),
    \label{eq:ab2}
\end{equation}

where $\mathbf{u}^n$ denotes the velocity obtained from the solve performed on the configuration $\mathbf{x}^n$, and where a constant time step is assumed. As demonstrated later, this multi-step formulation substantially mitigates mass conservation errors, while preserving the algorithmic simplicity of the overarching framework, as it requires no auxiliary linear system solves or inner iteration loops. The construction of $\tilde{\mathbf{u}}^{n-1}$ is markedly cheaper than a full field projection: material points retained from one mesh to the next, which represent the vast majority of the nodes, simply keep their velocity, and an interpolation is required only for the nodes created by the refinement stage. Each of them is located in the previous advected mesh, where the linear velocity field of the containing element is evaluated.

\subsection{Protecting-ball criterion}
\label{sec:protecting_balls}

Following the advection step, the fixed mesh topology causes elements to distort, severely degrading mesh quality and necessitating a remeshing procedure. While classical PFEM relies on a standard Delaunay triangulation of the point cloud combined with an $\alpha$-shape filter (a geometry-based criterion) to reconstruct external boundaries (e.g., free surfaces), our multiphase framework introduces a higher level of complexity. We must explicitly preserve the discrete interfaces, defined by a specific subset of edges of the advected mesh, deep within the fluid bulk. Preventing these interfacial edges from flipping is essential to rigorously maintain the physical topology of each phase. 
\bigskip

Before detailing the algorithmic steps of the remeshing procedure required to achieve this conforming mesh, this subsection introduces the theoretical foundations upon which our approach is constructed. The central mathematical criterion guiding this method is presented in \autoref{the:protecting_balls}.
\medskip

\begin{theorem}[Protecting-ball criterion]
\label{the:protecting_balls}
Let $V\subset\mathbb{R}^2$ be a set of vertices and let $\Gamma=\{e_i\}$ be a collection of non-intersecting interface segments joining pairs of vertices of $V$.
Assume that every interface segment $e_i=(v_a,v_b)$ admits an open disk $B_i$ such that
\begin{enumerate}
    \item $v_a,v_b\in\partial B_i$;
    \item $B_i\cap V=\emptyset$.
\end{enumerate}
Then every interface segment $e_i$ belongs to the Delaunay triangulation of $V$. Consequently, the interface $\Gamma$ is exactly recovered by the unconstrained Delaunay triangulation.
\end{theorem}
\medskip

This criterion is derived from the fundamental duality between the Voronoï diagram and the Delaunay triangulation, a well-established property in computational geometry (e.g., \cite{Gabriel_condition, edelsbrunner}). For simplicity, this duality is illustrated in \autoref{fig:duality} for a two-dimensional space; however, the underlying principles generalize to any arbitrary dimension.
\medskip

\begin{minipage}{0.4\textwidth}
    \noindent
    \begin{tikzpicture}[scale=8]
    % Paramètres du cercle
    \def\r{2} % rayon
    \coordinate (a) at (0,0); 
    \coordinate (b) at (0.3,0.4); 
    \coordinate (c) at (0.1,0.7); 
    \coordinate (d) at (-0.15,0.4); 

    \coordinate(c1) at (3/40, 7/15);
    \coordinate(c2) at (3/40, 41/160);

    \fill (a) circle (0.01) node[below left] {$v_k$};
    \fill (b) circle (0.01) node[above right] {$v_i$};
    \fill (c) circle (0.01) node[above left] {$v_l$};
    \fill (d) circle (0.01) node[above left] {$v_j$};

    \fill (c1) [orange] circle (0.01);
    \fill (c2) [orange] circle (0.01);

    \draw[thick] (c) -- (d);
    \draw[thick] (d) -- (b);
    \draw[thick] (b) -- (c);
    \draw[thick] (d) -- (a);
    \draw[thick] (a) -- (b);

    \draw[color=orange] (c1) -- ($(c1) + 2.2*0.5*(c) + 2.2*0.5*(b) - 2.2*(c1)$) node[pos=1, sloped, anchor=west] {$\cdots$};
    \draw[color=orange] (c1) -- ($(c1) + 2.2*0.5*(c) + 2.2*0.5*(d) - 2.2*(c1)$) node[pos=1, sloped, anchor=east] {$\dots$};
    \draw[color=orange] (c1) -- (c2);
    \draw[color=orange] (c2) -- ($(c2) + 2*0.5*(d) + 2*0.5*(a) - 2*(c2)$) node[pos=1, sloped, anchor=east] {$\cdots$};
    \draw[color=orange] (c2) -- ($(c2) + 3*0.5*(a) + 3*0.5*(b) - 3*(c2)$) node[pos=1, sloped, anchor=west] {$\cdots$};

    \coordinate (ball) at (3/40, 0.32);
    \draw (ball) [blue] circle (0.24);
    \fill (ball) [blue] circle (0.01) node[left] {$\mathcal{B}(c, r)$};

    \node[orange] at (0.28, 0) {\small $\mathrm{Vor}(V)$};
    \node[black] at (0.23, 0.67) {\small $\mathcal{D}(V)$};

\end{tikzpicture}

    \hspace{-0.2cm} \captionof{figure}{Duality between the Voronoï diagram $\mathrm{Vor}(V)$ and the Delaunay triangulation $\mathcal{D}(V)$.}
    \label{fig:duality}
\end{minipage}
\hfill
\begin{minipage}{0.55\textwidth}
    Let us define the Voronoï diagram as a collection of cells, where the cell associated with a node $\mathbf{v}_i \in V$ is defined as
    \begin{equation}
        \mathrm{Vor}(\mathbf{v}_i) = \{\mathbf{p} \in \mathbb{R}^2 \ \text{s.t.} \ |\mathbf{p} - \mathbf{v}_i| \le |\mathbf{p} - \mathbf{v}_j|, \ \forall \mathbf{v}_j \in V\},
    \end{equation}
    where $|\cdot|$ denotes the Euclidean norm. The cell boundaries therefore represent the geometric locus of points equidistant between adjacent nodes, and the vertices of the diagram correspond to the circumcenters of the Delaunay triangles. For an arbitrary edge $e = (\mathbf{v}_i, \mathbf{v}_j)$, if there exists an open disk $B(c, r)$ having $\mathbf{v}_i$ and $\mathbf{v}_j$ on its boundary and containing no node of $V$, then these two nodes must share a boundary in the Voronoï diagram. This follows from the fact that the center $c$ has $\mathbf{v}_i$ and $\mathbf{v}_j$ as its unique closest nodes. By duality, $\mathbf{v}_i$ and $\mathbf{v}_j$ are therefore connected in the corresponding Delaunay triangulation, guaranteeing the existence of the edge $e$.
\end{minipage} \\

Consequently, to ensure the presence of the interface edges $\{e\}$ in the remeshed domain, one only needs to define $V$ so that every edge $e_i$ can be associated with an empty disk $B_i$, which does not contain other nodes than those that form the edge. This is precisely \autoref{the:protecting_balls}.

\subsection{Interface recovery and adaptation}
\label{sec:interfaces}

At this stage, the interfaces can be extracted from the previous advected mesh as a discrete set of edges by comparing the phases of adjacent elements. This subsection, which corresponds to the third step of our framework, presents an approach designed to robustly preserve these interfacial edges during the subsequent remeshing stage. Guided by the condition established in \autoref{the:protecting_balls}, we show how to modify the advected point cloud to guarantee the existence of at least one protecting-ball for every discrete segment composing $\Gamma$, ensuring that they emerge naturally during the Delaunay triangulation of the modified point cloud. Our algorithm thus eliminates the reliance on constrained triangulations, which suffer from high computational overhead and poor generalization to three dimensions, while inherently preserving the topological features of the multiphase system.
\bigskip

In practice, modifying the point cloud to achieve this relies on a sequential combination of interface resolution control and local node filtering. First, because the interface representation is piecewise linear, a user-defined baseline resolution, $s_{\min}$, is enforced by splitting excessively long edges. While this step helps maintain a critical resolution along the interfaces, it does not by itself guarantee the preservation of these edges in the subsequent triangulation.
\bigskip 

The core of our approach to addressing this is illustrated in \autoref{fig:interfaces}. Although the existence of any protecting-ball $B_i$ (i.e., an empty open disk) is mathematically sufficient to preserve an edge $e_i$, we specifically enforce a strict diametral empty-disk condition, requiring the disk to be centered at the edge's midpoint with a radius equal to half its length. To satisfy this geometric condition without altering the interface topology, we employ two distinct operations. First, edge splitting is performed directly on the interface, mirroring the initial resolution adaptation. By dividing the edge into smaller segments, this operation effectively reduces the size of the required diametral disk. Second, surrounding nodes that encroach upon this disk are filtered out. However, a strict constraint applies: removing a node that belongs to the interface itself would irreversibly alter the topology of $\Gamma$ and is therefore strictly prohibited. Conversely, removing a volume node located within the fluid bulk is entirely permissible, as these nodes do not carry critical topological information.
\bigskip

\begin{figure}[h!]
    \centering
    % First subfigure
    \begin{subfigure}[t]{0.45\textwidth}
        \centering
        \begin{tikzpicture}[scale=0.8]
    \def\offsetx{8}
    \def\offsety{6}

    %%%%%%%%%% FIRST BOX -> INITIAL CONFIG %%%%%%%%%%%
    \begin{scope}[xshift=0, yshift=5cm]
        \draw[black, fill=white] (-0.2, 0) -- (7.5, 0) -- (7.5, 5) -- (-0.2, 5) -- cycle;

        \coordinate (n1) at (1,1.3); 
        \coordinate (n2) at (2,2.1); 
        \coordinate (n3) at (3,2.3); 
        \coordinate (n4) at (5,1.8); 
        \coordinate (n5) at (6,1.3);

        \coordinate (n6) at (0.8,3); 
        \coordinate (n7) at (2,3); 
        \coordinate (n8) at (4,2.7);
        \coordinate (n9) at (5,2.7);
        \coordinate (n10) at (7,3.2);

        \coordinate (v1) at (1.5, 0.7); 
        \coordinate (v2) at (2, 1.5);
        \coordinate (v3) at (3, 0.9);
        \coordinate (v4) at (4, 0.7); 
        \coordinate (v5) at (3.5, 1.7);
        \coordinate (v6) at (5.5, 0.5);
        \coordinate (v7) at (4.75, 1.2);
        \coordinate (v8) at (2, 2.6);
        \coordinate (v9) at (6.1, 2.3);
        \coordinate (v10) at (0.7, 2.4);
        \coordinate (v11) at (0.8, 4.2);
        \coordinate (v12) at (2, 3.8);
        \coordinate (v13) at (3.3, 4);
        \coordinate (v14) at (4.7, 3.4);
        \coordinate (v15) at (6, 3.4);
        \coordinate (v16) at (5.5, 4.3);

        \draw[very thick]
        \foreach \i [count=\j from 2] in {1,2,3,4} {
        (n\i) -- (n\j)
        };

        \draw[very thick]
        \foreach \i [count=\j from 6] in {7,8,9, 10} {
        (n\i) -- (n\j)
        };

        % Gamma 1 
        \draw[thick,truegreen] let \p1 = (n6), \p2 = (n7), \p3 = ($ (n6)!0.5!(n7) $), \n1 = {veclen(\x2-\x1,\y2-\y1)/2} in (\p3) circle (\n1);
        \draw[very thick,orange] let \p1 = (n7), \p2 = (n8), \p3 = ($ (n7)!0.5!(n8) $), \n1 = {veclen(\x2-\x1,\y2-\y1)/2} in (\p3) circle (\n1);
        \draw[thick,truegreen] let \p1 = (n8), \p2 = (n9), \p3 = ($ (n8)!0.5!(n9) $), \n1 = {veclen(\x2-\x1,\y2-\y1)/2} in (\p3) circle (\n1);
        \draw[thick,truegreen] let \p1 = (n9), \p2 = (n10), \p3 = ($ (n9)!0.5!(n10) $), \n1 = {veclen(\x2-\x1,\y2-\y1)/2} in (\p3) circle (\n1);

        % Gamma 2
        \draw[thick,truegreen] let \p1 = (n1), \p2 = (n2), \p3 = ($ (n1)!0.5!(n2) $), \n1 = {veclen(\x2-\x1,\y2-\y1)/2} in (\p3) circle (\n1);
        \draw[thick,truegreen] let \p1 = (n2), \p2 = (n3), \p3 = ($ (n2)!0.5!(n3) $), \n1 = {veclen(\x2-\x1,\y2-\y1)/2} in (\p3) circle (\n1);
        \draw[very thick,orange] let \p1 = (n3), \p2 = (n4), \p3 = ($ (n3)!0.5!(n4) $), \n1 = {veclen(\x2-\x1,\y2-\y1)/2} in (\p3) circle (\n1);
        \draw[thick,truegreen] let \p1 = (n4), \p2 = (n5), \p3 = ($ (n4)!0.5!(n5) $), \n1 = {veclen(\x2-\x1,\y2-\y1)/2} in (\p3) circle (\n1);

        \fill[black]
        \foreach \i in {1,...,10}
            {(n\i) circle (0.08)};

        \fill[black!40]
        \foreach \i in {1,...,16}
            {(v\i) circle (0.06)};

        \node[black] at (0.35, 3) {$\Gamma_1$};
        \node[black] at (0.5, 1.3) {$\Gamma_2$};

    \end{scope}

\end{tikzpicture}
        \caption{Detection of non-empty diametral disks.}
        \label{fig:interfaces1}
    \end{subfigure}
    % Third subfigure
    \begin{subfigure}[t]{0.45\textwidth}
        \centering
        \begin{tikzpicture}[scale=0.8]
    \def\offsetx{8}
    \def\offsety{6}

    \begin{scope}[xshift=9cm, yshift=5cm]
        \draw[black, fill=white] (-0.2, 0) -- (7.5, 0) -- (7.5, 5) -- (-0.2, 5) -- cycle;

        \coordinate (n1) at (1,1.3); 
        \coordinate (n2) at (2,2.1); 
        \coordinate (n3) at (3,2.3); 
        \coordinate (n4) at (5,1.8); 
        \coordinate (n5) at (6,1.3);
        \coordinate (add1) at ($ (n3)!0.5!(n4) $);
        \coordinate (n6) at (0.8,3); 
        \coordinate (n7) at (2,3); 
        \coordinate (n8) at (4,2.7);
        \coordinate (n9) at (5,2.7);
        \coordinate (n10) at (7,3.2); 
        \coordinate (add2) at ($ (n7)!0.5!(n8) $);

        \coordinate (v1) at (1.5, 0.7); 
        \coordinate (v2) at (2, 1.5);
        \coordinate (v3) at (3, 0.9);
        \coordinate (v4) at (4, 0.7); 
        \coordinate (v5) at (3.5, 1.7);
        \coordinate (v6) at (5.5, 0.5);
        \coordinate (v7) at (4.75, 1.2);
        \coordinate (v8) at (2, 2.6);
        \coordinate (v9) at (6.1, 2.3);
        \coordinate (v10) at (0.7, 2.4);
        \coordinate (v11) at (0.8, 4.2);
        \coordinate (v12) at (2, 3.8);
        \coordinate (v13) at (3.3, 4);
        \coordinate (v14) at (4.7, 3.4);
        \coordinate (v15) at (6, 3.4);
        \coordinate (v16) at (5.5, 4.3);

        \draw[very thick]
            \foreach \i [count=\j from 2] in {1,2,3,4} {
            (n\i) -- (n\j)
        };

        \draw[very thick]
            \foreach \i [count=\j from 6] in {7,8,9, 10} {
            (n\i) -- (n\j)
        };

        % Gamma 1 
        \draw[thick,truegreen] let \p1 = (n6), \p2 = (n7), \p3 = ($ (n6)!0.5!(n7) $), \n1 = {veclen(\x2-\x1,\y2-\y1)/2} in (\p3) circle (\n1);
        \draw[thick,truegreen] let \p1 = (n7), \p2 = (add2), \p3 = ($ (n7)!0.5!(add2) $), \n1 = {veclen(\x2-\x1,\y2-\y1)/2} in (\p3) circle (\n1);
        \draw[thick,truegreen] let \p1 = (add2), \p2 = (n8), \p3 = ($ (add2)!0.5!(n8) $), \n1 = {veclen(\x2-\x1,\y2-\y1)/2} in (\p3) circle (\n1);
        \draw[thick,truegreen] let \p1 = (n8), \p2 = (n9), \p3 = ($ (n8)!0.5!(n9) $), \n1 = {veclen(\x2-\x1,\y2-\y1)/2} in (\p3) circle (\n1);
        \draw[thick,truegreen] let \p1 = (n9), \p2 = (n10), \p3 = ($ (n9)!0.5!(n10) $), \n1 = {veclen(\x2-\x1,\y2-\y1)/2} in (\p3) circle (\n1);

        % Gamma 2
        \draw[thick,truegreen] let \p1 = (n1), \p2 = (n2), \p3 = ($ (n1)!0.5!(n2) $), \n1 = {veclen(\x2-\x1,\y2-\y1)/2} in (\p3) circle (\n1);
        \draw[thick,truegreen] let \p1 = (n2), \p2 = (n3), \p3 = ($ (n2)!0.5!(n3) $), \n1 = {veclen(\x2-\x1,\y2-\y1)/2} in (\p3) circle (\n1);
        \draw[thick,truegreen] let \p1 = (n3), \p2 = (add1), \p3 = ($ (n3)!0.5!(add1) $), \n1 = {veclen(\x2-\x1,\y2-\y1)/2} in (\p3) circle (\n1);
        \draw[thick,truegreen] let \p1 = (add1), \p2 = (n4), \p3 = ($ (add1)!0.5!(n4) $), \n1 = {veclen(\x2-\x1,\y2-\y1)/2} in (\p3) circle (\n1);
        \draw[thick,truegreen] let \p1 = (n4), \p2 = (n5), \p3 = ($ (n4)!0.5!(n5) $), \n1 = {veclen(\x2-\x1,\y2-\y1)/2} in (\p3) circle (\n1);

        \fill[black]
        \foreach \i in {1,...,10}
            {(n\i) circle (0.08)};
        
        \fill[orange] (add1) circle (0.1);
        \fill[orange] (add2) circle (0.1);

        \fill[black!40]
        \foreach \i in {1,...,16}
            {(v\i) circle (0.06)};

        \node[black] at (0.35, 3) {$\Gamma_1$};
        \node[black] at (0.5, 1.3) {$\Gamma_2$};

    \end{scope}

\end{tikzpicture}
        \caption{Edge splitting.}
        \label{fig:interfaces2}
    \end{subfigure}
    \vspace{0.3cm}

    % Second subfigure
    \begin{subfigure}[t]{0.45\textwidth}
        \centering
        \begin{tikzpicture}[scale=0.8]
    \def\offsetx{8}
    \def\offsety{6}

    %%%%%%%%%% THIRD BOX -> VOLUME CONFIG %%%%%%%%%%%
    \begin{scope}[xshift=0cm, yshift=-1cm]
        
        \draw[black, fill=white] (-0.2, 0) -- (7.5, 0) -- (7.5, 5) -- (-0.2, 5) -- cycle;

        \coordinate (n1) at (1,1.3); 
        \coordinate (n2) at (2,2.1); 
        \coordinate (n3) at (3,2.3); 
        \coordinate (n4) at (5,1.8); 
        \coordinate (n5) at (6,1.3);
        \coordinate (add1) at ($ (n3)!0.5!(n4) $);
        \coordinate (n6) at (0.8,3); 
        \coordinate (n7) at (2,3); 
        \coordinate (n8) at (4,2.7);
        \coordinate (n9) at (5,2.7);
        \coordinate (n10) at (7,3.2); 
        \coordinate (add2) at ($ (n7)!0.5!(n8) $);

        \coordinate (v1) at (1.5, 0.7); 
        \coordinate (v2) at (2, 1.5);
        \coordinate (v3) at (3, 0.9);
        \coordinate (v4) at (4, 0.7); 
        \coordinate (v5) at (3.5, 1.7);
        \coordinate (v6) at (5.5, 0.5);
        \coordinate (v7) at (4.75, 1.2);
        \coordinate (v8) at (2, 2.6);
        \coordinate (v9) at (6.1, 2.3);
        \coordinate (v10) at (0.7, 2.4);
        \coordinate (v11) at (0.8, 4.2);
        \coordinate (v12) at (2, 3.8);
        \coordinate (v13) at (3.3, 4);
        \coordinate (v14) at (4.7, 3.4);
        \coordinate (v15) at (6, 3.4);
        \coordinate (v16) at (5.5, 4.3);

        \draw[very thick]
            \foreach \i [count=\j from 2] in {1,2,3,4} {
            (n\i) -- (n\j)
        };

        \draw[very thick]
            \foreach \i [count=\j from 6] in {7,8,9, 10} {
            (n\i) -- (n\j)
        };

        % Gamma 1 
        \draw[black] let \p1 = (n6), \p2 = (n7), \p3 = ($ (n6)!0.5!(n7) $), \n1 = {veclen(\x2-\x1,\y2-\y1)/2} in (\p3) circle (\n1);
        \draw[black] let \p1 = (n7), \p2 = (add2), \p3 = ($ (n7)!0.5!(add2) $), \n1 = {veclen(\x2-\x1,\y2-\y1)/2} in (\p3) circle (\n1);
        \draw[black] let \p1 = (add2), \p2 = (n8), \p3 = ($ (add2)!0.5!(n8) $), \n1 = {veclen(\x2-\x1,\y2-\y1)/2} in (\p3) circle (\n1);
        \draw[black] let \p1 = (n8), \p2 = (n9), \p3 = ($ (n8)!0.5!(n9) $), \n1 = {veclen(\x2-\x1,\y2-\y1)/2} in (\p3) circle (\n1);
        \draw[black] let \p1 = (n9), \p2 = (n10), \p3 = ($ (n9)!0.5!(n10) $), \n1 = {veclen(\x2-\x1,\y2-\y1)/2} in (\p3) circle (\n1);

        % Gamma 2
        \draw[black] let \p1 = (n1), \p2 = (n2), \p3 = ($ (n1)!0.5!(n2) $), \n1 = {veclen(\x2-\x1,\y2-\y1)/2} in (\p3) circle (\n1);
        \draw[black] let \p1 = (n2), \p2 = (n3), \p3 = ($ (n2)!0.5!(n3) $), \n1 = {veclen(\x2-\x1,\y2-\y1)/2} in (\p3) circle (\n1);
        \draw[black] let \p1 = (n3), \p2 = (add1), \p3 = ($ (n3)!0.5!(add1) $), \n1 = {veclen(\x2-\x1,\y2-\y1)/2} in (\p3) circle (\n1);
        \draw[black] let \p1 = (add1), \p2 = (n4), \p3 = ($ (add1)!0.5!(n4) $), \n1 = {veclen(\x2-\x1,\y2-\y1)/2} in (\p3) circle (\n1);
        \draw[black] let \p1 = (n4), \p2 = (n5), \p3 = ($ (n4)!0.5!(n5) $), \n1 = {veclen(\x2-\x1,\y2-\y1)/2} in (\p3) circle (\n1);

        \fill[black]
        \foreach \i in {1,...,10}
            {(n\i) circle (0.06)};
        
        \fill[black] (add1) circle (0.06);
        \fill[black] (add2) circle (0.06);
        
        \fill[truegreen] (v1) circle (0.08);
        \fill[red] (v2) circle (0.08);
        \fill[truegreen] (v3) circle (0.08);
        \fill[truegreen] (v4) circle (0.08);
        \fill[red] (v5) circle (0.08);
        \fill[truegreen] (v6) circle (0.08);
        \fill[truegreen] (v7) circle (0.08);
        \fill[truegreen] (v8) circle (0.08);
        \fill[red] (v9) circle (0.08);
        \fill[truegreen] (v10) circle (0.08);
        \fill[truegreen] (v11) circle (0.08);
        \fill[truegreen] (v12) circle (0.08);
        \fill[truegreen] (v13) circle (0.08);
        \fill[truegreen] (v14) circle (0.08);
        \fill[red] (v15) circle (0.08);
        \fill[truegreen] (v16) circle (0.08);

        \node[black] at (0.35, 3) {$\Gamma_1$};
        \node[black] at (0.5, 1.3) {$\Gamma_2$};

    \end{scope}

\end{tikzpicture}
        \caption{Bulk nodes filtering.}
        \label{fig:interfaces3}
    \end{subfigure}
    % Third subfigure
    \begin{subfigure}[t]{0.45\textwidth}
        \centering
        \begin{tikzpicture}[scale=0.8]
    \def\offsetx{8}
    \def\offsety{6}

    %%%%%%%%%% FOURTH BOX -> FINAL CONFIG (mesh) %%%%%%%%%%%
    \begin{scope}[xshift=9cm, yshift=-1cm]
        \draw[black, fill=white] (-0.2, 0) -- (7.5, 0) -- (7.5, 5) -- (-0.2, 5) -- cycle;

        \coordinate (n1) at (1,1.3); 
        \coordinate (n2) at (2,2.1); 
        \coordinate (n3) at (3,2.3); 
        \coordinate (n4) at (5,1.8); 
        \coordinate (n5) at (6,1.3);
        \coordinate (add1) at ($ (n3)!0.5!(n4) $);
        \coordinate (n6) at (0.8,3); 
        \coordinate (n7) at (2,3); 
        \coordinate (n8) at (4,2.7);
        \coordinate (n9) at (5,2.7);
        \coordinate (n10) at (7,3.2); 
        \coordinate (add2) at ($ (n7)!0.5!(n8) $);

        \coordinate (v1) at (1.5, 0.7); 
        \coordinate (v2) at (2, 1.5);
        \coordinate (v3) at (3, 0.9);
        \coordinate (v4) at (4, 0.7); 
        \coordinate (v5) at (3.5, 1.7);
        \coordinate (v6) at (5.5, 0.5);
        \coordinate (v7) at (4.75, 1.2);
        \coordinate (v8) at (2, 2.6);
        \coordinate (v9) at (6.1, 2.3);
        \coordinate (v10) at (0.7, 2.4);
        \coordinate (v11) at (0.8, 4.2);
        \coordinate (v12) at (2, 3.8);
        \coordinate (v13) at (3.3, 4);
        \coordinate (v14) at (4.7, 3.4);
        \coordinate (v15) at (6, 3.4);
        \coordinate (v16) at (5.5, 4.3);

        \draw[very thick, truegreen]
            \foreach \i [count=\j from 2] in {1,2,3,4} {
            (n\i) -- (n\j)
        };

        \draw[very thick, truegreen]
            \foreach \i [count=\j from 6] in {7,8,9, 10} {
            (n\i) -- (n\j)
        };

        \draw [black!60] (n6) -- (v11);
        \draw [black!60] (v11) -- (v12);
        \draw [black!60] (n6) -- (v12);
        \draw [black!60] (n7) -- (v12);
        \draw [black!60] (add2) -- (v12);
        \draw [black!60] (add2) -- (v13);
        \draw [black!60] (v12) -- (v13);
        \draw [black!60] (n8) -- (v13);
        \draw [black!60] (v13) -- (v14);
        \draw [black!60] (n8) -- (v14);
        \draw [black!60] (n9) -- (v14);
        \draw [black!60] (v14) -- (v16);
        \draw [black!60] (v14) -- (n10);
        \draw [black!60] (v16) -- (n10);

        \draw [black!60] (v1) -- (n1);
        \draw [black!60] (v1) -- (n2);
        \draw [black!60] (v3) -- (n2);
        \draw [black!60] (v1) -- (v3);
        \draw [black!60] (v3) -- (n3);
        \draw [black!60] (v3) -- (v4);
        \draw [black!60] (v3) -- (add1);
        \draw [black!60] (v4) -- (v7);
        \draw [black!60] (v4) -- (add1);
        \draw [black!60] (n4) -- (v7);
        \draw [black!60] (v7) -- (add1);
        \draw [black!60] (v7) -- (n5);
        \draw [black!60] (v7) -- (v6);
        \draw [black!60] (v6) -- (n5);

        \draw [black!60] (n6) -- (v10);
        \draw [black!60] (n1) -- (v10);
        \draw [black!60] (v10) -- (v8);
        \draw [black!60] (v8) -- (n7);
        \draw [black!60] (v8) -- (n2);
        \draw [black!60] (v10) -- (n2);
        \draw [black!60] (n6) -- (v8);
        \draw [black!60] (v8) -- (n3);
        \draw [black!60] (v8) -- (add2);
        \draw [black!60] (add2) -- (n3);
        \draw [black!60] (n3) -- (n8);
        \draw [black!60] (n8) -- (add1);
        \draw [black!60] (n9) -- (add1);
        \draw [black!60] (n9) -- (n4);
        \draw [black!60] (n9) -- (n5);
        \draw [black!60] (n5) -- (n10);

        \fill[black]
        \foreach \i in {1,...,10}
            {(n\i) circle (0.06)};
        
        \fill[black] (add1) circle (0.06);
        \fill[black] (add2) circle (0.06);
        
        \fill[black] (v1) circle (0.06);
        \fill[black] (v3) circle (0.06);
        \fill[black] (v4) circle (0.06);
        \fill[black] (v6) circle (0.06);
        \fill[black] (v7) circle (0.06);
        \fill[black] (v8) circle (0.06);
        \fill[black] (v10) circle (0.06);
        \fill[black] (v11) circle (0.06);
        \fill[black] (v12) circle (0.06);
        \fill[black] (v13) circle (0.06);
        \fill[black] (v14) circle (0.06);
        \fill[black] (v16) circle (0.06);

        \node[truegreen] at (0.35, 3) {$\Gamma_1$};
        \node[truegreen] at (0.5, 1.3) {$\Gamma_2$};

    \end{scope}

\end{tikzpicture}
        \caption{Classical Delaunay triangulation.}
        \label{fig:interfaces4}
    \end{subfigure}

    \caption{Illustration of the methodology used to ensure the preservation of advected interfacial edges in the new Delaunay triangulation. By enforcing an empty-disk criterion, each edge is protected through two specific operations: local edge splitting (b) and volume node filtering (c).}
    \label{fig:interfaces}
\end{figure}

\medskip
\begin{algorithm}[h!]
\caption{Interface adaptation}
\label{alg:interface}
\begin{algorithmic}[1]
    \State $V_\Gamma \gets \text{interfacial nodes}$
        \Comment{Extracted from the previously advected mesh}
    \State $E_\Gamma \gets \text{interfacial edges}$
    \State $V_B \gets \text{bulk nodes}$

    \Statex \textbf{Phase 1: Edge splitting}
    \State $\textsc{SplitEdges}(\{e_i \in E_\Gamma : |e_i| > s_{\min}\})$
        \Comment{Baseline interface resolution}
    \While{$\exists e_i \in E_\Gamma$ such that
    $B(c_i,r_i) \cap V_\Gamma \neq \emptyset$ and $|e_i| > w_{\min}$}
        \State $\mathcal{E}_{\mathrm{enc}} \gets
            \textsc{DetectEncroachments}(E_\Gamma,V_\Gamma)$
            \Comment{In parallel}
        \State $\textsc{SplitEdges}(\mathcal{E}_{\mathrm{enc}})$
            \Comment{In parallel; update $V_\Gamma$ and $E_\Gamma$}
    \EndWhile

    \Statex \textbf{Phase 2: Bulk node deletion}
    \State $B_\Gamma \gets
        \text{diametral balls associated with the interfacial edges}$
    \ForAll{$v_b \in V_B$} \Comment{In parallel}
        \If{$\exists B \in B_\Gamma : v_b \in B$}
            \State $\textsc{MarkForDeletion}(v_b)$
        \EndIf
    \EndFor
\end{algorithmic}
\end{algorithm}
\clearpage

\begin{figure}[H]
    \centering
    % First subfigure
    \begin{subfigure}[t]{0.35\textwidth}
        \centering
        \includegraphics[width=\textwidth]{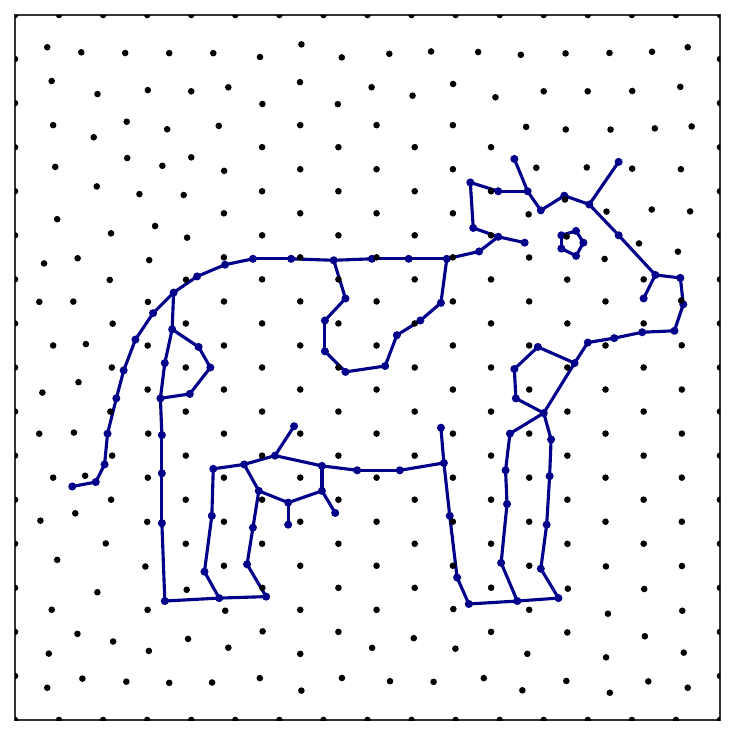}
        \caption{Target discretized interface, aimed to be recovered among the triangulation of the point cloud.}
        \label{fig:cow1}
    \end{subfigure}
    \hspace{0.5cm}
    % Third subfigure
    \begin{subfigure}[t]{0.35\textwidth}
        \centering
        \includegraphics[width=\textwidth]{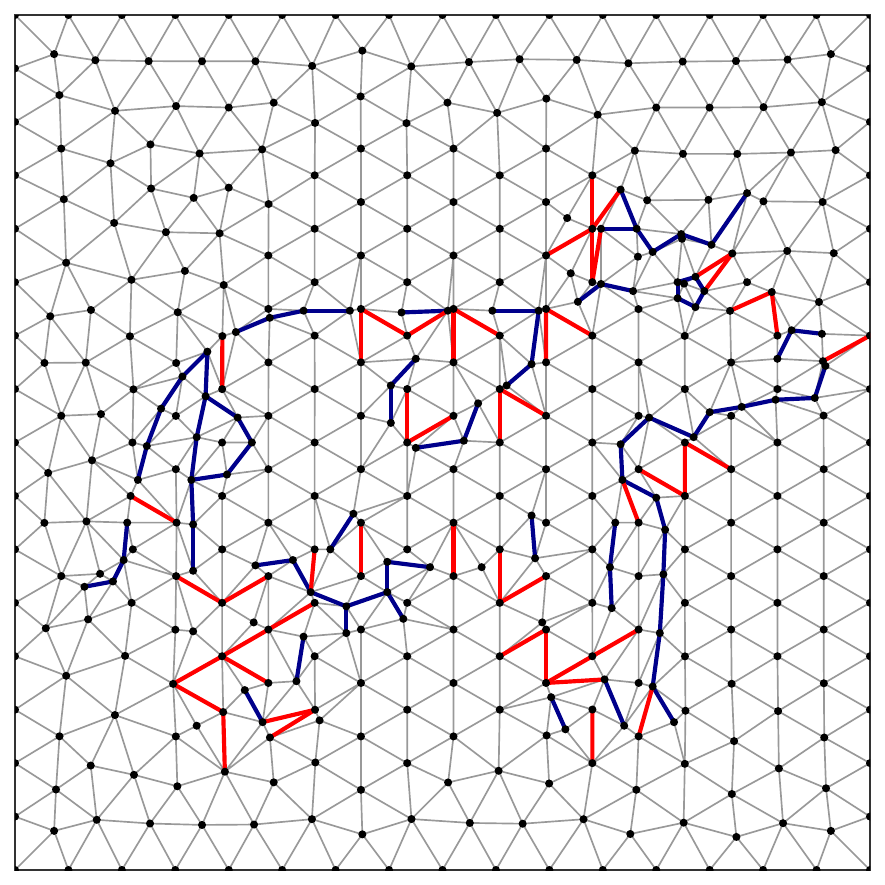}
    \caption{Delaunay triangulation without adaptation. Edges conforming to the target {\myrect[darkblue]} have no guarantee to appear, resulting in edges crossing the target {\myrect[red]}.}
        \label{fig:cow3}
    \end{subfigure}
    % Second subfigure
    \begin{subfigure}[t]{0.35\textwidth}
        \centering
        \includegraphics[width=\textwidth]{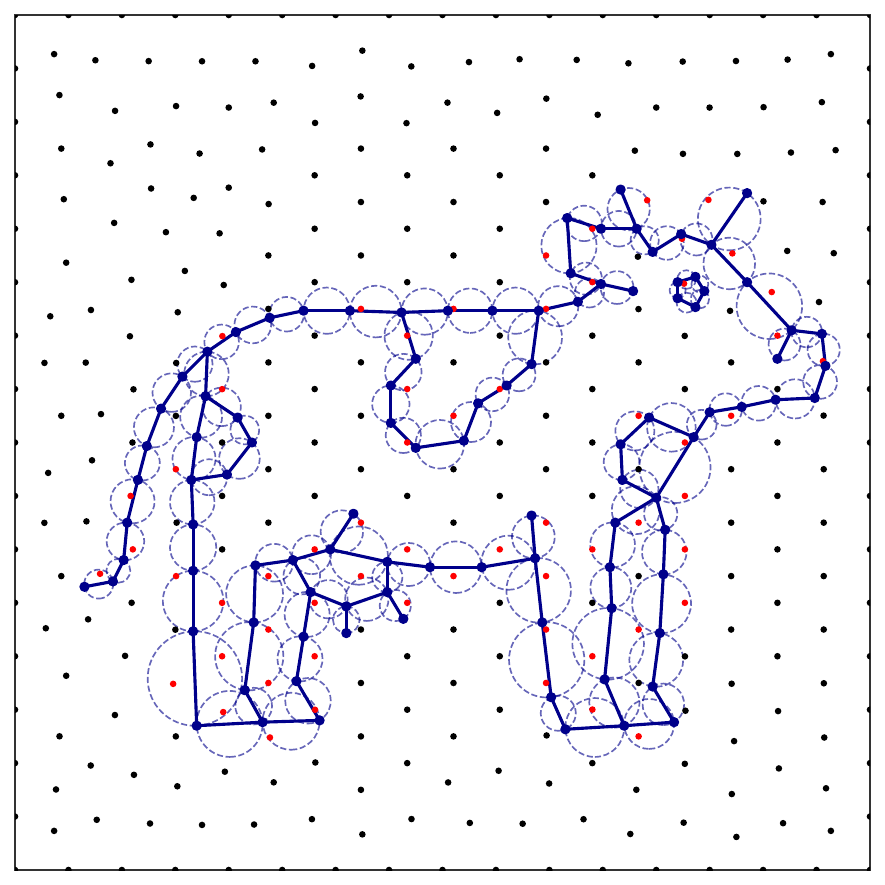}
        \caption{Bulk nodes filtering {\mycirc[red]} to restore the empty ball condition.}
        \label{fig:cow2}
    \end{subfigure}
    \hspace{0.5cm}
    % Third subfigure
    \begin{subfigure}[t]{0.35\textwidth}
        \centering
        \includegraphics[width=\textwidth]{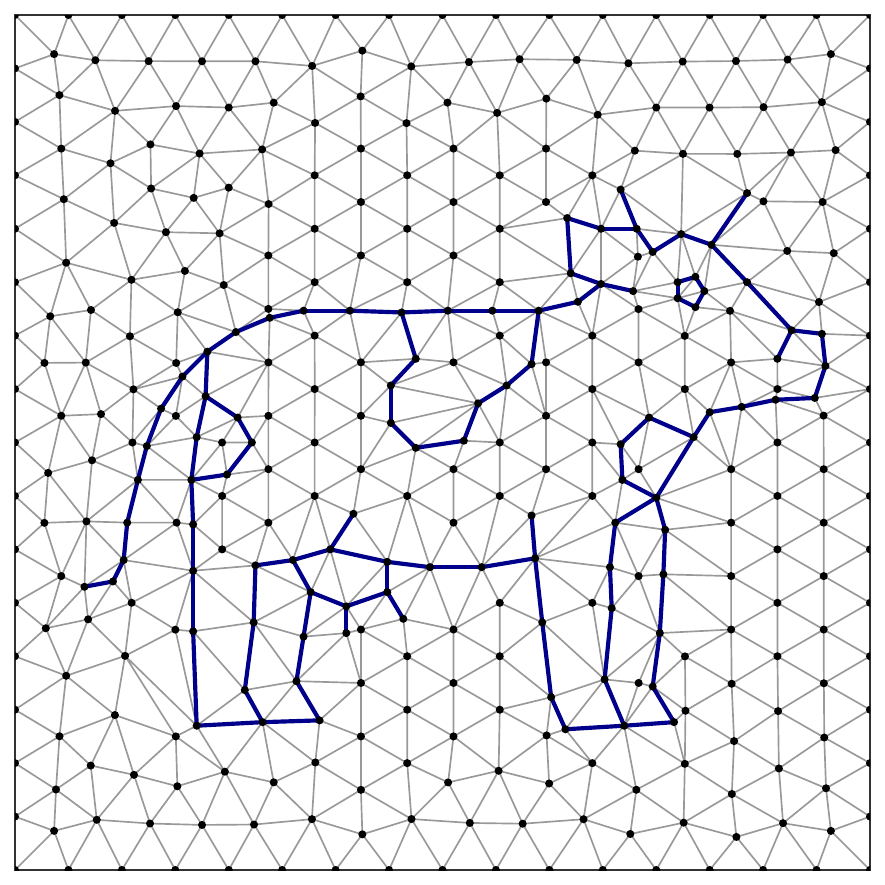}
        \caption{Delaunay triangulation after adaptation. The target edges are fully recovered within the mesh connectivity.}
        \label{fig:cow4}
    \end{subfigure}
    \caption{Schematic of the protecting-ball method for edge recovery. The algorithm aims to recover a set of edges among a point cloud (a). With a naive Delaunay triangulation, connectivity breaks the interface (b). A filtering procedure (c) allows us to guarantee the existence of the target edges (d).}
    \label{fig:cow}
\end{figure}

\autoref{alg:interface} describes the implementation of the strategy illustrated in \autoref{fig:interfaces}. The first phase focuses on protecting interface edges from mutually encroaching upon one another while temporarily disregarding volume nodes (i.e., nodes that do not belong to any interface). A spatial search is performed to identify edges at risk of flipping, specifically, those that lack an associated diametral empty disk, during the subsequent triangulation. For these interface edges, only splitting is permissible; the interfaces are recursively refined until no interface node encroaches upon the diametral empty disk of any non-incident edge. This operation is illustrated in steps (a) and (b). Following this, the algorithm addresses the volume nodes, filtering them out if they encroach upon a protected empty disk, as depicted in step (c). Finally, a standard Delaunay triangulation is applied to the adapted point cloud, successfully recovering the conforming phase interfaces $\Gamma$. This edge recovery algorithm is further demonstrated in \autoref{fig:cow}, where a complex cow geometry is seamlessly incorporated into the mesh.
\bigskip

From an analytical standpoint, this algorithm preserves topology regardless of how close two interfaces may be. In practice, however, the procedure operates within predefined refinement tolerances, and edge splitting terminates at a critical threshold. Generally, a target size $s_{\min}$ is specified for the interface edges to ensure a baseline resolution. When two distinct interfaces approach one another, their constituent edges are further refined for protection until they reach a minimum allowable size, $w_{\min}$. This minimum size $w_{\min}$ is completely independent of the baseline mesh resolution and can be chosen to be arbitrarily small. Conceptually, setting $w_{\min}$ to zero would result in a complete prohibition of topological changes. Below this size threshold, the empty-disk criterion is no longer strictly enforced during remeshing, meaning that topological changes may occur natively within the Delaunay triangulation. Consequently, as discussed above, intentional topological transitions should be algorithmically detected and executed before this critical size is reached, thereby accurately capturing the underlying physical phenomena before reaching the resolution limit of the mesh.
\bigskip

While this procedure preserves the interface topology without altering its discrete representation, continuously adding nodes is computationally unsustainable. Segments of the interface that were heavily refined below $s_{\min}$ for proximity protection no longer require such high resolution once the interacting interfaces separate. Ideally, these segments should be coarsened back toward the baseline interface size $s_{\min}$ to maintain consistency and computational efficiency. However, removing an interface node alters the discrete geometry; theoretically, this operation is only perfectly conservative if three consecutive nodes are strictly collinear. In practice, over-refined interface regions are coarsened if they are deemed sufficiently flat according to a user-defined geometric tolerance. This flatness metric is evaluated by analyzing the triangle formed by three consecutive interface nodes, specifically computing the ratio between the longest side and the sum of the two shorter sides. Nodes located at the junction of three or more phases (triple points) are strictly protected from deletion. For regular interface nodes, deletion introduces minor geometric alterations, and consequently, small phase-volume or mass conservation errors, which are rigorously bounded by both the flatness tolerance and the baseline constraint $s_{\min}$. 

\subsection{Volume mesh adaptation}
\label{sec:mesh_adaptation}

\begin{figure}[h!]
    \centering
    \includegraphics[width=0.7\linewidth]{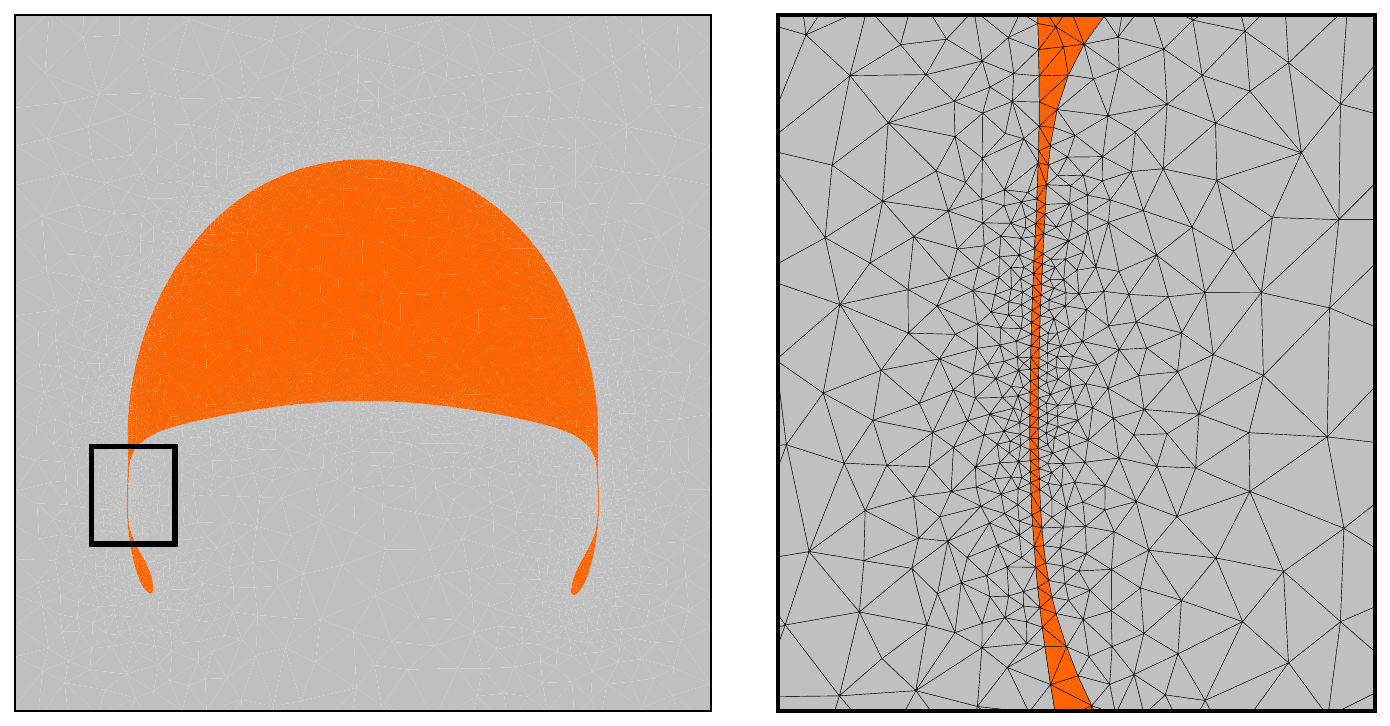}
    \caption{Local mesh adaptation and refinement around a fluid filament developing within a rising bubble. The smooth size gradient is driven by a user-defined size field that matches the localized resolution of the interfaces.}
    \label{fig:mesh_adaptation}
\end{figure}

During the filtering step of our edge recovery method, volume nodes in the vicinity of the interfaces may have been filtered out from the point cloud. Furthermore, due to the subdivision of certain critical edges, specific regions of the domain may become excessively refined. While performing a standard Delaunay triangulation on this modified point cloud now guarantees a conforming mesh along the interfaces, the highly disparate node density inevitably leads to poorly shaped elements near the phase boundaries. To address this issue, an additional mesh adaptation step is performed to enhance element quality within the fluid bulk. The purpose of this section is to present an algorithm that regularizes the mesh to ensure a smooth transition in element-size around the interface, guided by a user-defined sizing field. A typical outcome of this procedure is illustrated in \autoref{fig:mesh_adaptation}, where the element-size varies continuously near the interfaces, seamlessly matching the localized interface discretization to achieve a smooth size gradient. 
\bigskip

Our approach is based on the work of Chew \cite{chew}, which introduces a refinement procedure that ensures the final mesh guarantees a minimum bound on element angles while conforming to a user-defined element-size field. More recently, this algorithm has been adapted within the PFEM framework to optimize the mesh in the vicinity of a free surface for single-phase flows \cite{thomas_refine}. The general methodology of this procedure is summarized in \autoref{alg:chew_refinement}. The core concept involves sequentially refining the mesh by inserting vertices to split elements whose circumradii violate the prescribed size field, up to tolerance $\alpha$. Each new node is placed precisely at the circumcenter of the targeted element. This choice inherently optimizes the local aspect ratio when the Delaunay property, which requires an empty circumcircle for each element, is restored via local edge flips following the insertion. To orchestrate this process, a priority queue sorts elements based on the ratio between their target sizes and circumradius; although not strictly mandatory for convergence, this sorting strategy has been shown to yield superior final mesh distributions. To maintain the local protection of interfaces, a circumcenter cannot be inserted if its location falls within the protected empty disk of an interface edge. Consequently, near the phase boundaries, our adapted mesh relaxes the strict theoretical bounds on minimum angles originally guaranteed by Chew's algorithm. In practice, however, the mesh does not suffer from noticeable degradation. Finally, an exception is made for circumcenters that fall outside the computational domain. In such cases, special care is taken to add points on the boundary segment which intersects the path between the element intended for refinement and its circumcenter. Further implementation details and proofs of final quality are available in \cite{chew}.
\bigskip

To govern the refinement process, a spatial sizing field must be defined over the computational domain. As established in \autoref{sec:interfaces}, the target size at nodes belonging to interface edges is directly dictated either by $s_{\min}$, if no special protection is required, or by the minimum length of the interfacial edges incident to that node, bounded from below by $w_{\min}$. For the volume nodes, we define a sizing field that evolves as a function of the Euclidean distance to the nearest interface. Let $d_i$ be the distance from a volume node to the closest interface, and $s_i$ be the local interface sizing value at that point. The target element-size $s(d_i)$ within the bulk fluid is computed as:
\begin{equation}
    s(d_i) = \begin{cases}
                s_i & \text{if } d_i \le d_{\min}, \\
                s_{\max} & \text{if } d_i \ge d_{\max}, \\
                s_i \left( \frac{d_{\max} - d_i}{d_{\max} - d_{\min}} \right) + s_{\max} \left( \frac{d_i - d_{\min}}{d_{\max} - d_{\min}} \right) & \text{otherwise.}
            \end{cases}
    \label{eq:sizefield}
\end{equation}
Under this formulation, the mesh maintains a highly refined resolution, matched to the local interface size, within a band of width $d_{\min}$ surrounding the interfaces. It then transitions linearly up to a distance of $d_{\max}$, where it reaches the coarsest bulk fluid resolution, $s_{\max}$. This sizing field can be defined arbitrarily; users can easily couple it with other physical fields (e.g., velocity gradients or error estimators) to dynamically refine localized regions of interest. Similarly, the parameters defining the size field in \autoref{eq:sizefield} can be adjusted by the user to achieve the desired mesh resolution, balance accuracy against computational cost, or target the characteristic length scale at which topological transitions are expected to occur from a physical perspective.

\begin{algorithm}[h!]
\caption{Mesh refinement}
\label{alg:chew_refinement}
\begin{algorithmic}[1]
    \State $Q \gets \textsc{InitializePriorityQueue}
        (\{K \mid \alpha R_K > s(c_K)\})$
        \Comment{By size deviation against target}
    \While{$Q$ is not empty}
        \State $K \gets Q\texttt{.pop}()$
        \If{$\alpha R_K \le s(c_K)$ \textbf{or} $c_K \in \Omega_{\mathrm{protected}}$}
            \State \textbf{continue}
        \EndIf

        \If{$c_K \notin \Omega$}
            \Comment{Circumcenter outside the domain}
            \State $e \gets \textsc{FindIntersectedBoundaryEdge}(c_K)$
            \State $\textsc{SplitEdge}(e)$
            \State $\textsc{FilterCloseNodes}()$
        \Else
            \Comment{Circumcenter inside the domain}
            \State $\textsc{InsertVertex}(c_K)$
        \EndIf

        \State $\textsc{RestoreDelaunay}()$
        \State $Q\texttt{.push}\bigl(
            \textsc{GetNewAndModifiedElements}()\bigr)$
    \EndWhile
\end{algorithmic}
\end{algorithm}

While adapting the mesh according to the aforementioned strategy directly enhances element quality, it inevitably comes at the cost of significantly increasing the total number of nodes. Similarly to \autoref{sec:interfaces}, a region that undergoes localized refinement at a given iteration may not require maximum resolution throughout the entire simulation if the phase interfaces move away from it. Consequently, a mesh coarsening mechanism is essential. In our framework, coarsening is achieved via vertex removal. The set of bulk edges (excluding those belonging to $\Gamma$) is scanned, and their lengths are evaluated against the local size field. If an edge is found to be smaller than the target size by a predefined multiplicative threshold $\beta$, the vertex that least satisfies the sizing criterion is selected for deletion, with an exception made for interface nodes. From an implementation standpoint, the refinement tolerance $\alpha$ and the coarsening tolerance $\beta$ must be chosen to allow for sufficient relaxation; an inadequate dead-band between these two parameters can lead to numerical chattering, where nodes are successively inserted and deleted by the competing algorithms.

\subsection{Mesh coloring and velocity interpolation}
\label{sec:coloring}

\begin{figure}[h!]
    \centering
    \resizebox{\textwidth}{!}{%
        \begin{tikzpicture}[>=Latex]
            % Fixed bounding box so the sub-captions cannot skew the centering
            \useasboundingbox (-3.3,-2.8) rectangle (16.3,2.3);

            % Images
            \node (img1) at (0,0)    {\includegraphics[height=5cm]{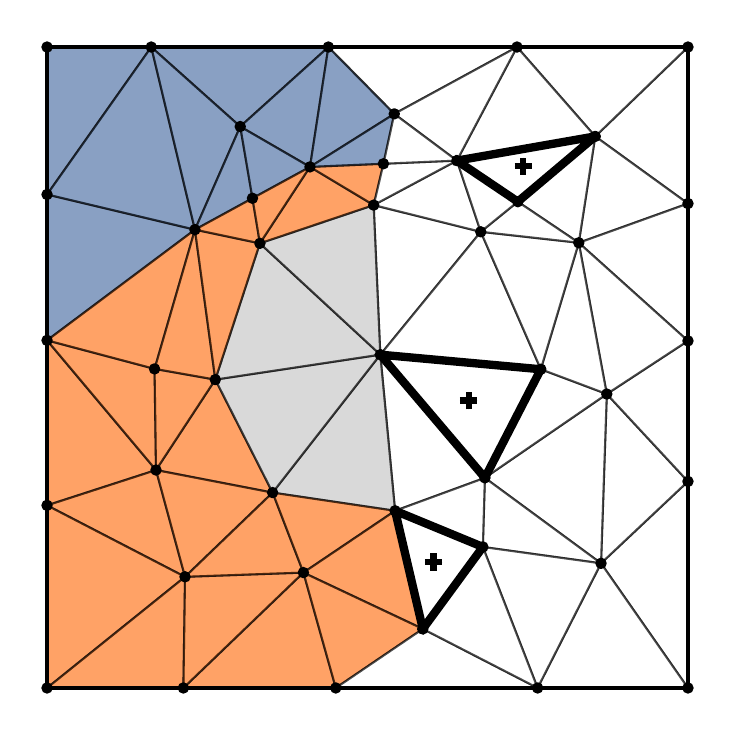}};
            \node (img2) at (6.5,0)  {\includegraphics[height=5cm]{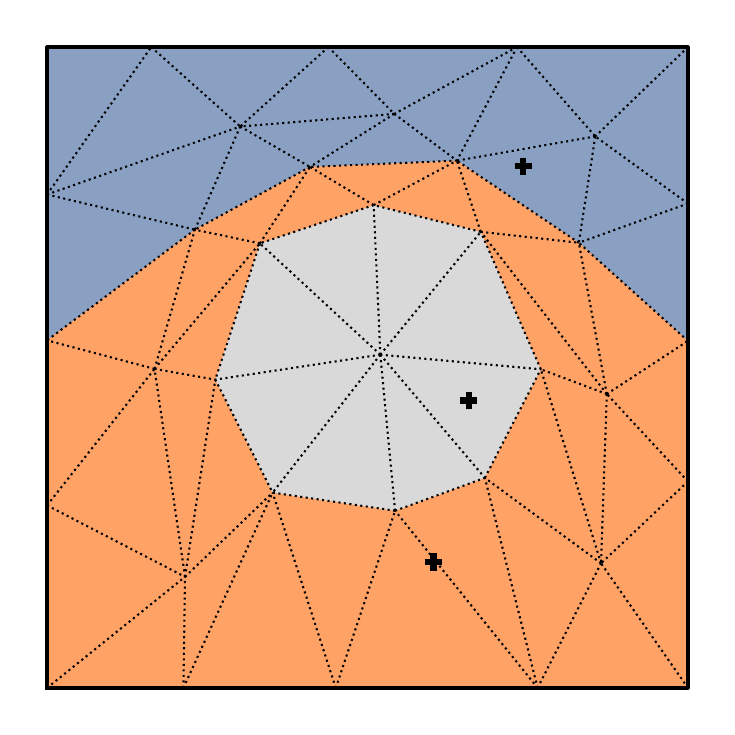}};
            \node (img3) at (13,0)   {\includegraphics[height=5cm]{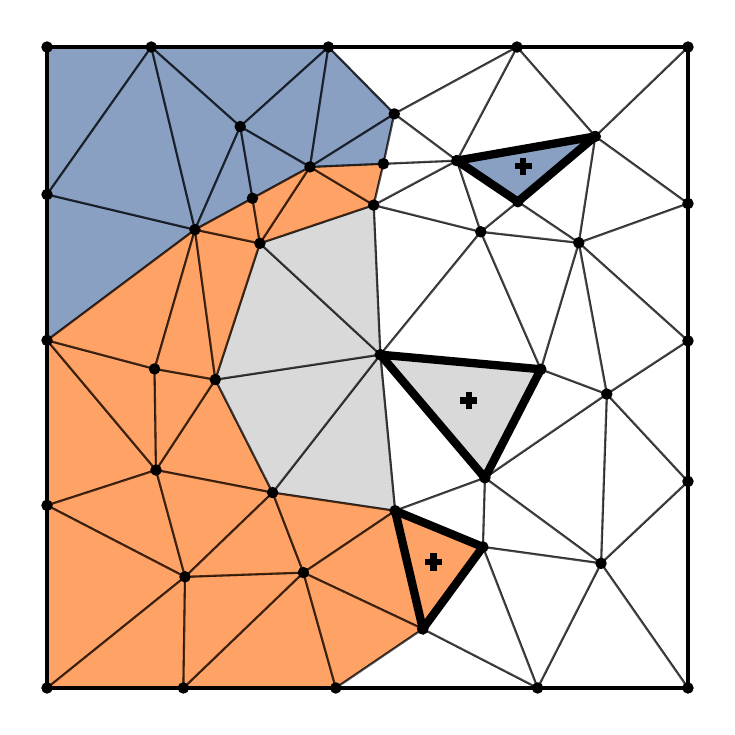}};

            % Curly brace next to the first image
            \draw[
                line width=1.2pt,
                decorate,
                decoration={brace,mirror,amplitude=6pt}
            ] (2.5,-2.4) -- (2.5,2.4);

            % Arrow from brace to second image
            \draw[-{Latex[length=3mm]}, thick]
                (2.8,0) -- (4.2,0)
                node[midway, above=6pt, xshift=-3pt,
                    draw,
                    fill=white,
                    rounded corners,
                    inner sep=2pt] {\scriptsize Phase?};

            \draw[-{Latex[length=3mm, fill=blue!70]}, thick, draw=black, double=blue!70, double distance=1.5pt]
                (7.9,1.35) -- (13.5,1.35)
                node[midway, above=6pt, xshift=-28pt,
                    draw=blue,
                    text=blue,
                    fill=white,
                    rounded corners,
                    inner sep=2pt] {Phase 1};

            \draw[-{Latex[length=3mm, fill=gray]}, thick, draw=black, double=gray, double distance=1.5pt]
                (7.4,-0.25) -- (13.3,-0.25)
                node[midway, above=6pt, xshift=-17pt,
                    draw=gray,
                    text=gray,
                    fill=white,
                    rounded corners,
                    inner sep=2pt] {Phase 0};

            \draw[-{Latex[length=3mm, fill=orangered]}, thick, draw=black, double=orangered, double distance=1.5pt]
                (7.2,-1.3) -- (13.1,-1.3)
                node[midway, above=6pt, xshift=-12pt,
                    draw=orangered,
                    text=orangered,
                    fill=white,
                    rounded corners,
                    inner sep=2pt] {Phase 2};

            % Sub-captions: equal width and centered, so they cannot unbalance the box
            \node[text width=6cm, align=center] at (0,-2.7)  {(a) Triangulation: to be colored};
            \node[text width=6cm, align=center] at (6.5,-2.7) {(b) Previous advected mesh};
            \node[text width=6cm, align=center] at (13,-2.7) {(c) Triangulation: colored};
        \end{tikzpicture}%
    }
    \caption{Transfer of the indicator field (colors) onto the new mesh. The phase associated with each new element is determined by querying the previous advected mesh at the element centroid. The queried phase is then assigned to the new element. This procedure is valid because the interfaces of the two meshes are conforming.}
    \label{fig:coloring}
\end{figure}

At this stage, we have successfully reconstructed a new mesh featuring edges that conform to the advected interfaces, yielding elements of sufficient quality for the subsequent solving phase. However, this new mesh currently exists solely as a geometric entity, defined by a set of nodes and their connectivity, and physical properties must still be mapped onto the newly formed elements. Because the new mesh conforms to the previous advected mesh at the interface topology level, this reassignment process is relatively straightforward. \autoref{fig:coloring} illustrates how the indicator field can be projected onto the new mesh. Thanks to this topological conformity, the new elements can be `colored' (assigned to a phase) simply by querying the center of mass of each element, or indeed any point within its interior, with the guarantee that the entire element is completely enclosed by the correct phase in the previous advected mesh. 
\bigskip 

From an implementation standpoint, the coloring procedure is made efficient by first sorting the elements along a space-filling curve and subsequently performing the lookups on the previous mesh using a walking algorithm.

Solving the governing equations also requires knowledge of the previously computed velocity field. For material points that are retained during the remeshing phase, the velocity from the prior time step is preserved. However, mesh adaptation introduces new nodes that require velocity assignment. This is addressed by evaluating the velocity field from the old mesh, where it is continuously defined, onto the new vertex locations. This constitutes the final step in generating a high-quality mesh with fully defined physical quantities, ready for the subsequent FEM solving phase.
\medskip

In practice, as previously noted, interface protection is only maintained down to a critical threshold $w_{\min}$, below which arbitrary edge flips can occur along sub-tolerance interfaces. Under these circumstances, the proposed coloring technique introduces mass conservation errors proportional to the local mesh size. This discrepancy arises because an element may straddle two distinct phases in the previous advected mesh but is assigned a single, uniform phase color in the new triangulation.

\section{Validation}
\label{sec:validation}

\subsection{Vortex in a box}
\label{sec:vortex}

\begin{figure}[h!]
    \centering
    % First subfigure
    \begin{subfigure}[b]{0.24\textwidth}
        \centering
        \begin{tikzpicture}
            \node[anchor=south west, inner sep=0] (img) at (0,0) {\includegraphics[width=\textwidth]{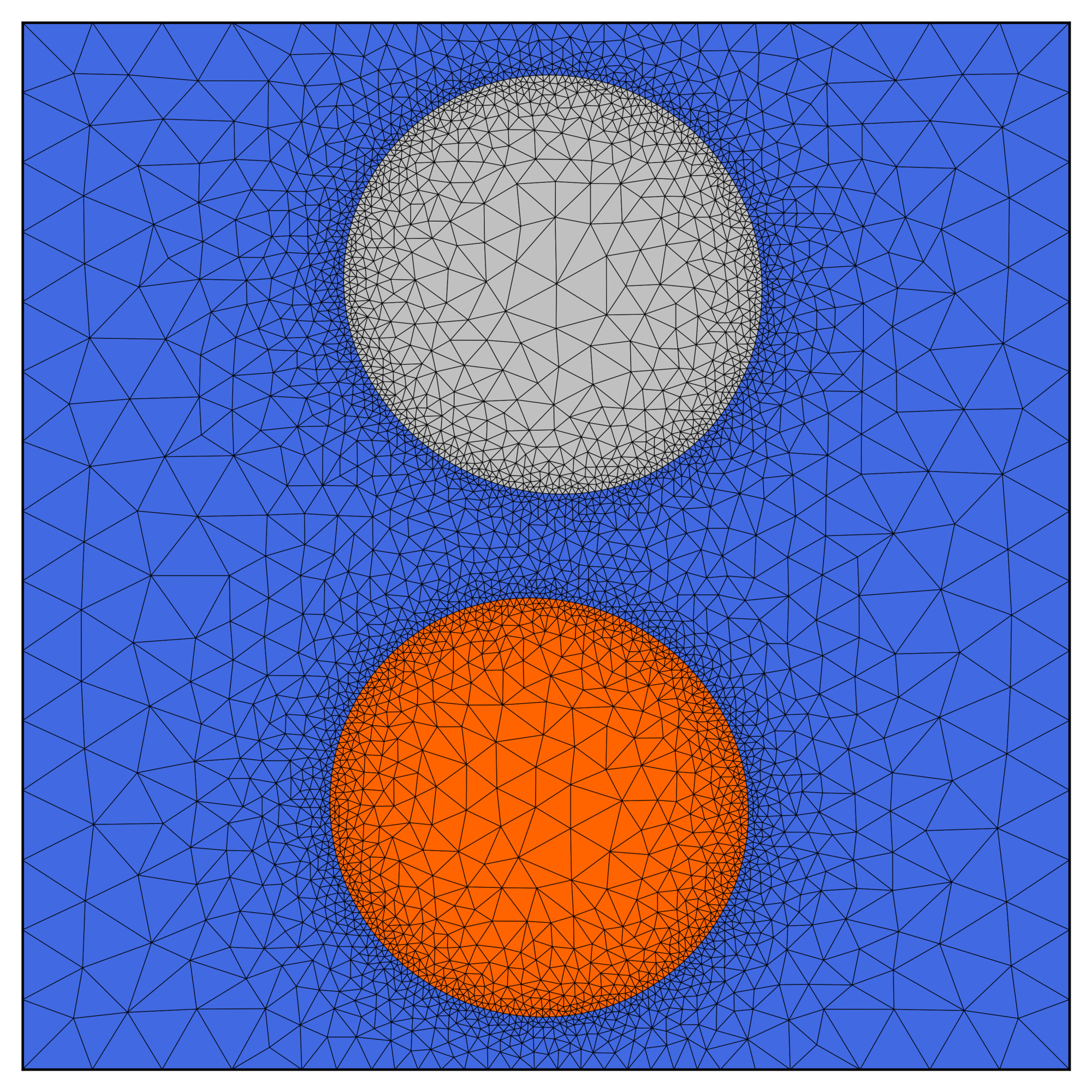}};
            \node[anchor=north west, fill=white, fill opacity=0.7, text opacity=1, inner sep=3pt] at (0.25, 3.75) {\footnotesize Nodes: $3\,617$};
        \end{tikzpicture}
        \caption{$t=0.0$}
        \label{fig:vortex_mesh1}
    \end{subfigure}
    \hfill % Pushes the first two subfigures to the outer edges
    \begin{subfigure}[b]{0.24\textwidth}
        \centering
        \begin{tikzpicture}
            \node[anchor=south west, inner sep=0] (img) at (0,0) {\includegraphics[width=\textwidth]{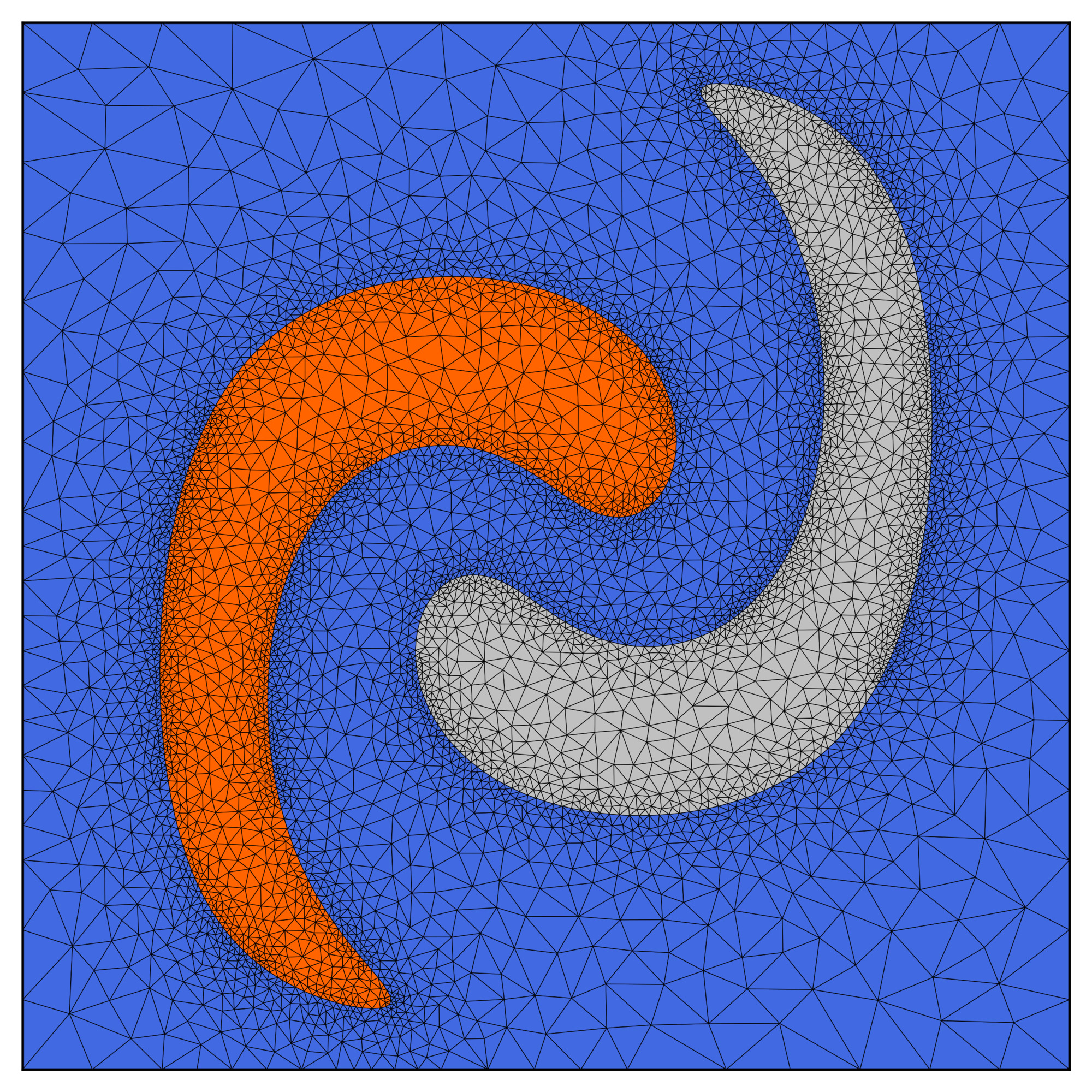}};
            \node[anchor=north west, fill=white, fill opacity=0.7, text opacity=1, inner sep=3pt] at (0.25, 3.75) {\footnotesize Nodes: $5\,629$};
        \end{tikzpicture}
        \caption{$t=0.6$}
        \label{fig:vortex_mesh2}
    \end{subfigure}
    \hfill
    % Second subfigure
    \begin{subfigure}[b]{0.24\textwidth}
        \centering
        \begin{tikzpicture}
            \node[anchor=south west, inner sep=0] (img) at (0,0) {\includegraphics[width=\textwidth]{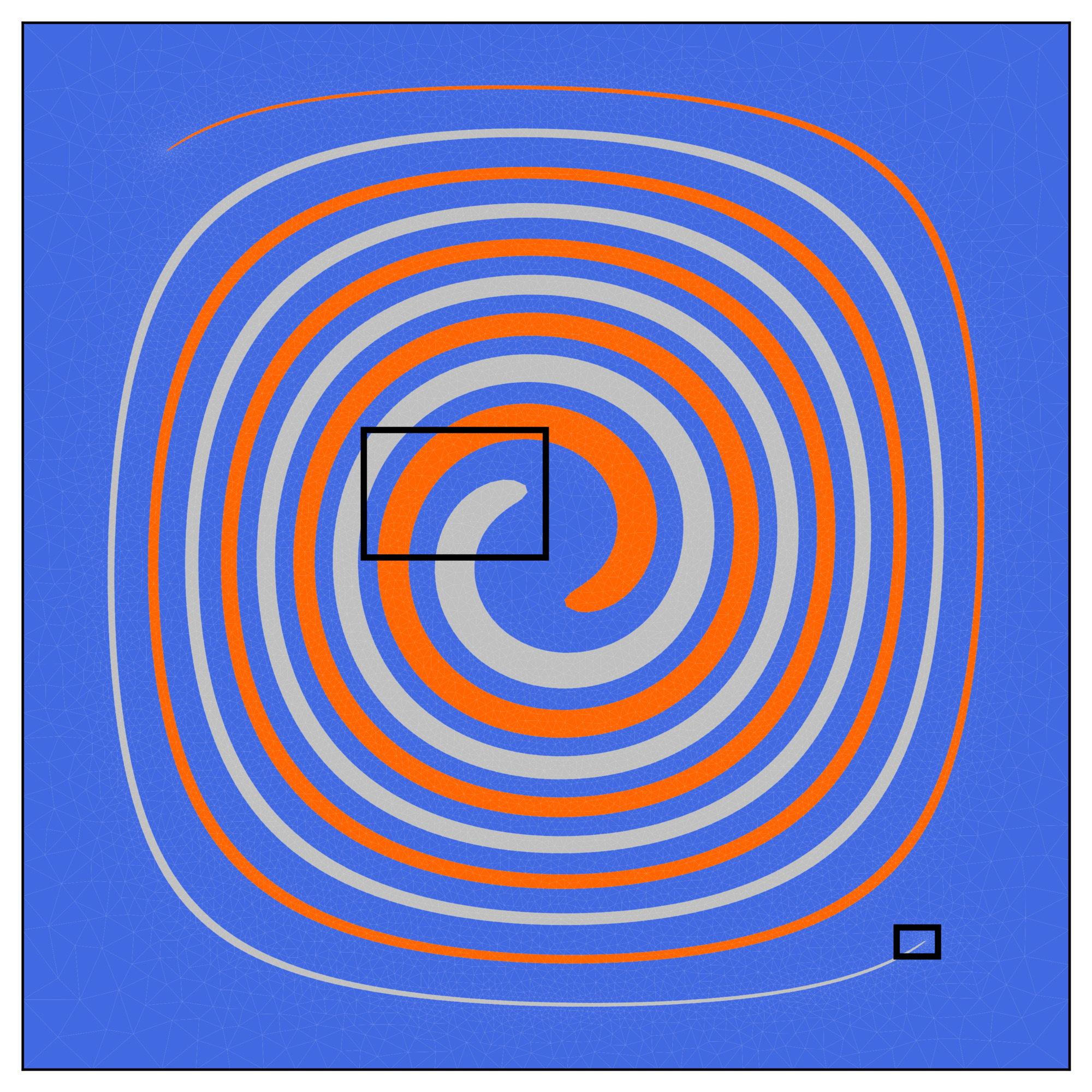}};
            \node[anchor=north west, fill=white, fill opacity=0.7, text opacity=1, inner sep=3pt] at (0.25, 3.75) {\footnotesize Nodes: $16\,300$};
            \node[anchor=north west, fill=white, fill opacity=0.7, text opacity=1, inner sep=3pt] at (0.75, 1.9) {\footnotesize \bfseries A};
            \node[anchor=north west, fill=white, fill opacity=0.7, text opacity=1, inner sep=3pt] at (2.75, 1.1) {\footnotesize \bfseries B};

        \end{tikzpicture}
        \caption{$t=5.0$}
        \label{fig:vortex_mesh3}
    \end{subfigure}
    \hfill
    \begin{subfigure}[b]{0.24\textwidth}
        \centering
        \begin{tikzpicture}
            \node[anchor=south west, inner sep=0] (img) at (0,0) {\includegraphics[width=\textwidth]{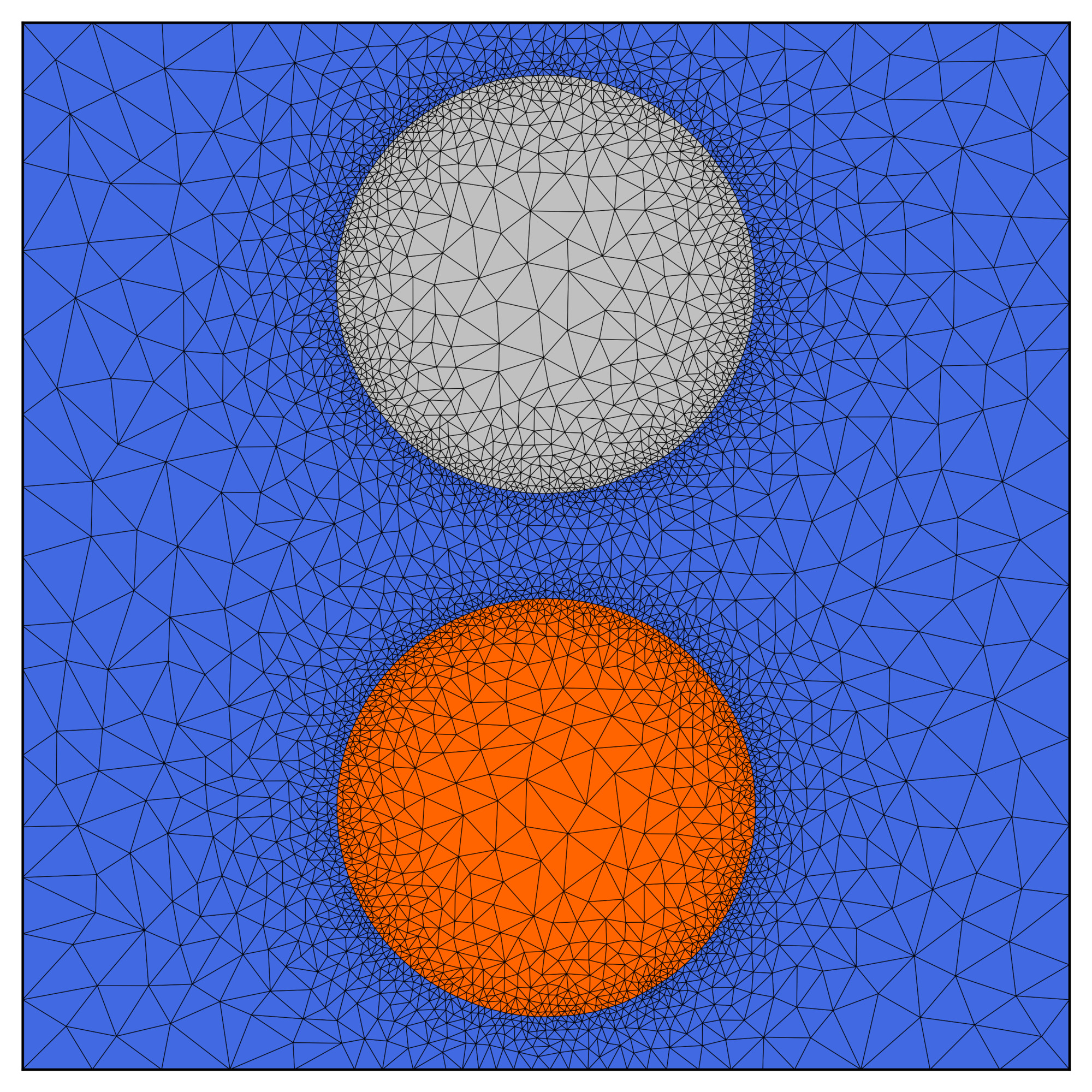}};
            \node[anchor=north west, fill=white, fill opacity=0.7, text opacity=1, inner sep=3pt] at (0.25, 3.75) {\footnotesize Nodes: $3\,768$};
        \end{tikzpicture}
        \caption{$t=10.0$}
        \label{fig:vortex_mesh4}
    \end{subfigure}

    \begin{subfigure}[b]{0.49\textwidth}
        \centering
        \begin{tikzpicture}
            \node[anchor=south west, inner sep=0] (img) at (0,0) {\includegraphics[width=\textwidth]{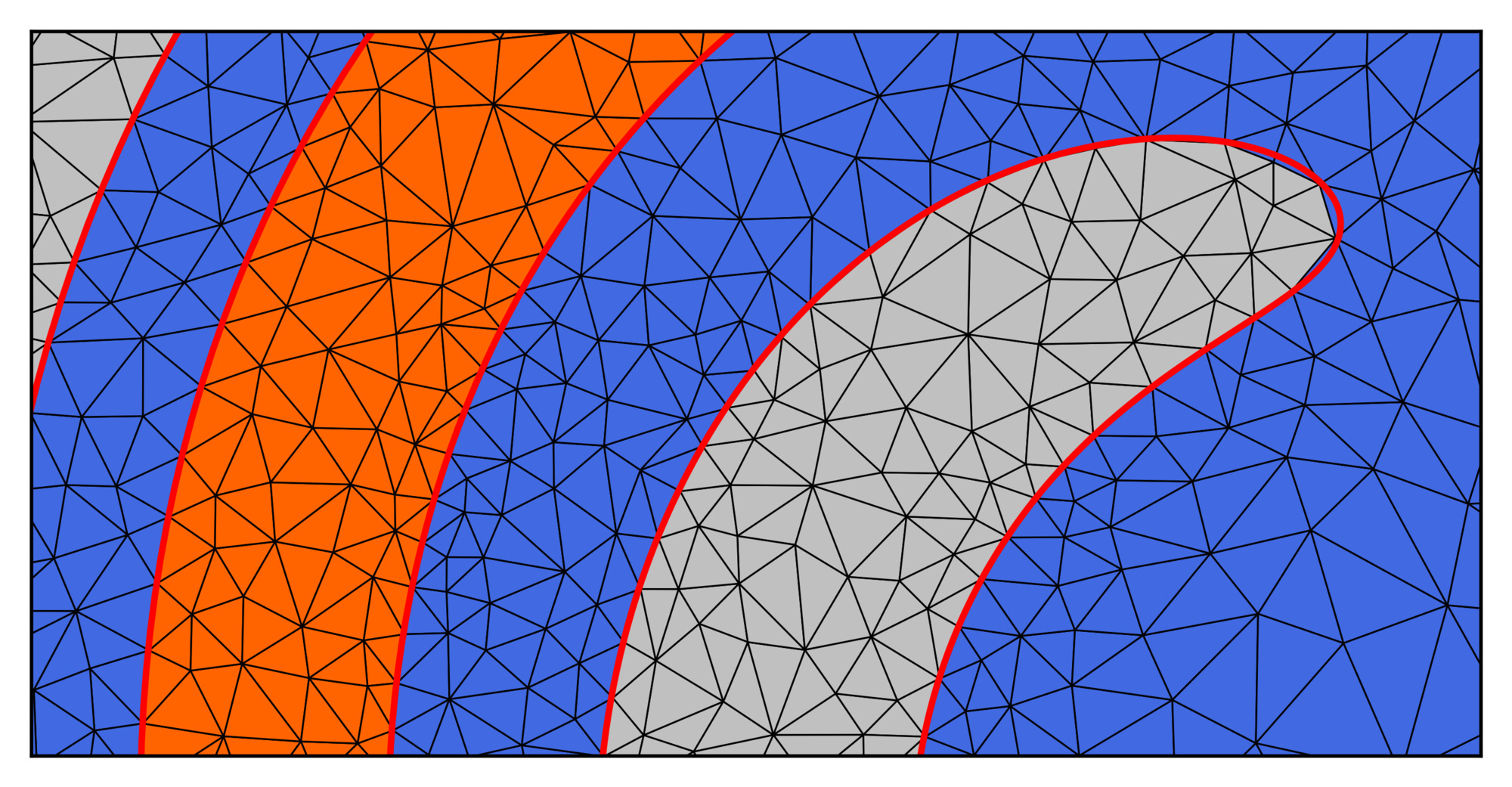}};
            \node[
                anchor=north west,
                fill=white,
                fill opacity=0.7,
                text opacity=1,
                inner sep=3pt
            ] at (7.15, 3.9) {\Large\bfseries A}; 
        \end{tikzpicture}
        \caption{$t=5.0$. Mesh at the front of the vortex.}
        \label{fig:vortex_mesh5}
    \end{subfigure}
    \hfill % Space out the bottom three horizontally (No blank line here!)
    \begin{subfigure}[b]{0.49\textwidth}
        \centering
        \begin{tikzpicture}
            \node[anchor=south west, inner sep=0] (img) at (0,0) {\includegraphics[width=\textwidth]{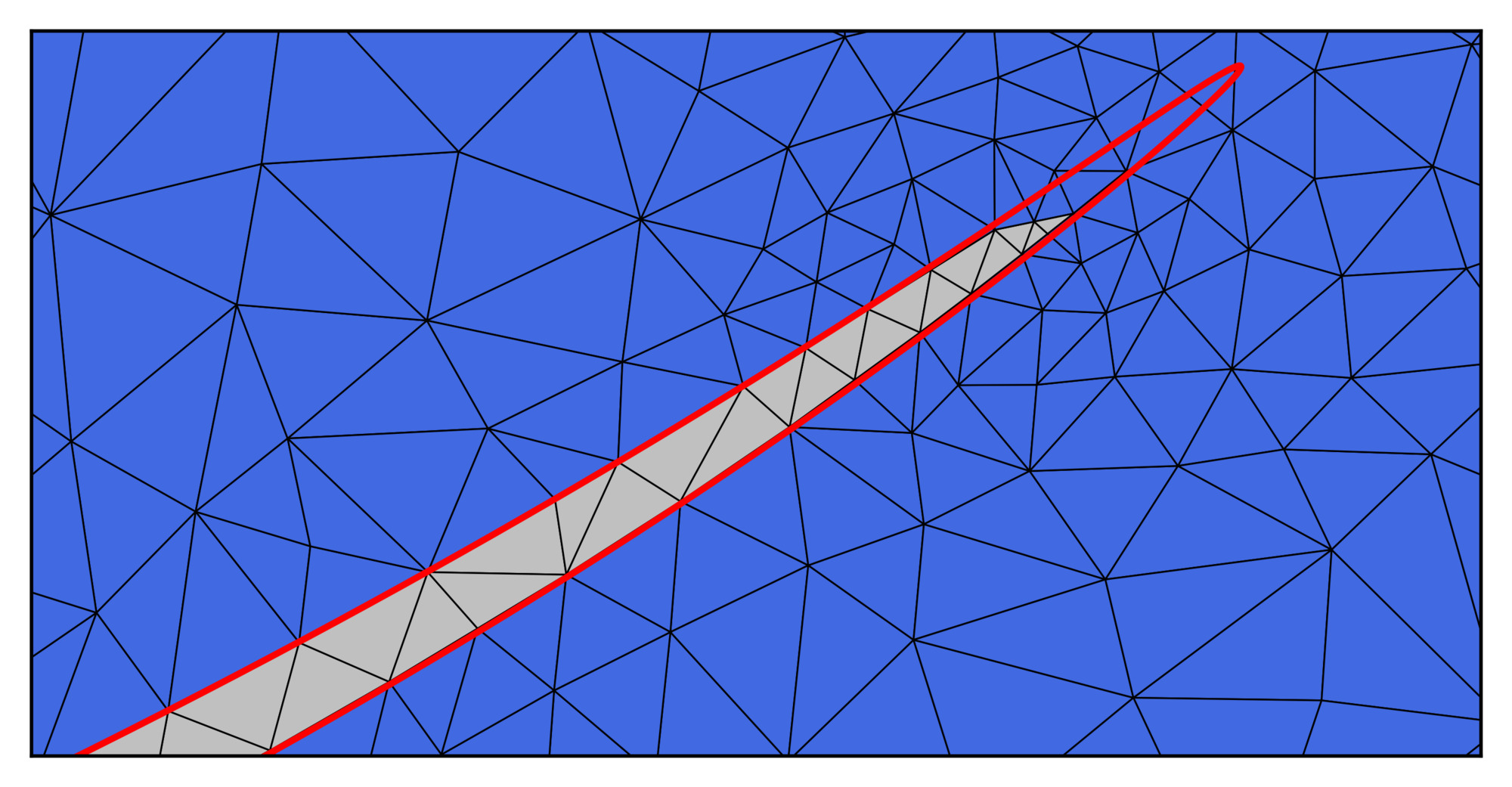}};
            \node[
                anchor=north west,
                fill=white,
                fill opacity=0.7,
                text opacity=1,
                inner sep=3pt
            ] at (0.3, 3.9) {\Large\bfseries B};   
        \end{tikzpicture}
        \caption{$t=5.0$. Mesh at the tail of the vortex.}
        \label{fig:vortex_mesh6}
    \end{subfigure}
    \caption{Mesh adaptation under an analytical advection field. The mesh is locally refined to preserve thin topological features, while the bulk elements are adapted to adhere to a prescribed size gradient surrounding the interfaces. The size field is governed by $s_{\max}=0.07$, $s_{\min}=s_{\max}/10$, $d_{\max}=3s_{\max}$, $d_{\min}=0$ and $w_{\min}=s_{\min}/5$. The subsequent coarsening of the mesh is evident as the two droplets return to their initial configurations upon reversal of the velocity field. The droplet shape at $t=5.0$ is compared with markers {\myrect[red]} which were advected with $\Delta t = 10^{-4}$ and an RK4 scheme to serve as numerical reference.}
    \label{fig:vortex_mesh}
\end{figure}

To first demonstrate the capabilities of our method regarding interface preservation and mesh adaptation, we consider two circular droplets, depicted in silver and orange in \autoref{fig:vortex_mesh}, subjected to an analytical velocity field defined by:
\begin{equation}
    \mathbf{u}(x, y) = 
        \begin{bmatrix}
            -\sin(\pi x)^2\sin(2\pi y) \\
            \sin(2\pi x) \sin(\pi y)^2  
        \end{bmatrix}.
\end{equation}

The nodes are advected for $t \in [0, 5)$ according to $\mathbf{u}$, after which the velocity field is reversed for $t \in [5, 10]$ so that the theoretical final positions of the droplets coincide with their initial configurations. Because the prescribed velocity field is strictly divergence-free, the phase topology remains invariant throughout the simulation. The Adams-Bashforth scheme presented in \autoref{sec:governing_equations} is employed for the temporal discretization of the advection step, with $\Delta t=10^{-3}$.
\medskip

The numerical results confirm that our method successfully preserves fluid filaments and robustly recovers sharp boundaries through successive remeshing steps. Furthermore, the local mesh adaptation algorithm smoothly regularizes the bulk elements in the vicinity of refined interfaces, yielding a smooth element-size transition. At $t=10$, the droplets return to their initial states, demonstrating the effectiveness of the progressive mesh coarsening algorithm during the interface retraction phase ($t \geq 5$), which executes seamlessly without compromising the topological integrity of the objects. The total number of nodes is displayed at the top of each subfigure for reference, highlighting the computational advantage of the localized mesh adaptation framework. Specifically, a significant reduction in the number of algebraic degrees of freedom is achieved as the interfaces retract and the point cloud dynamically thins out. Regarding highly elongated features, such as the trailing vortex filaments, our capacity for accurate capture depends fundamentally on the initial geometric discretizations of the droplets at $t=0$. Because the refinement algorithm purely subdivides existing linear segments, it does not introduce higher-order curvature or project newly generated nodes back onto the analytical circular profile.  
\medskip

\begin{figure}[h!]
    \centering
    \begin{tikzpicture}
        \node[anchor=south west, inner sep=0] (img) at (0,0) {\includegraphics[width=10cm]{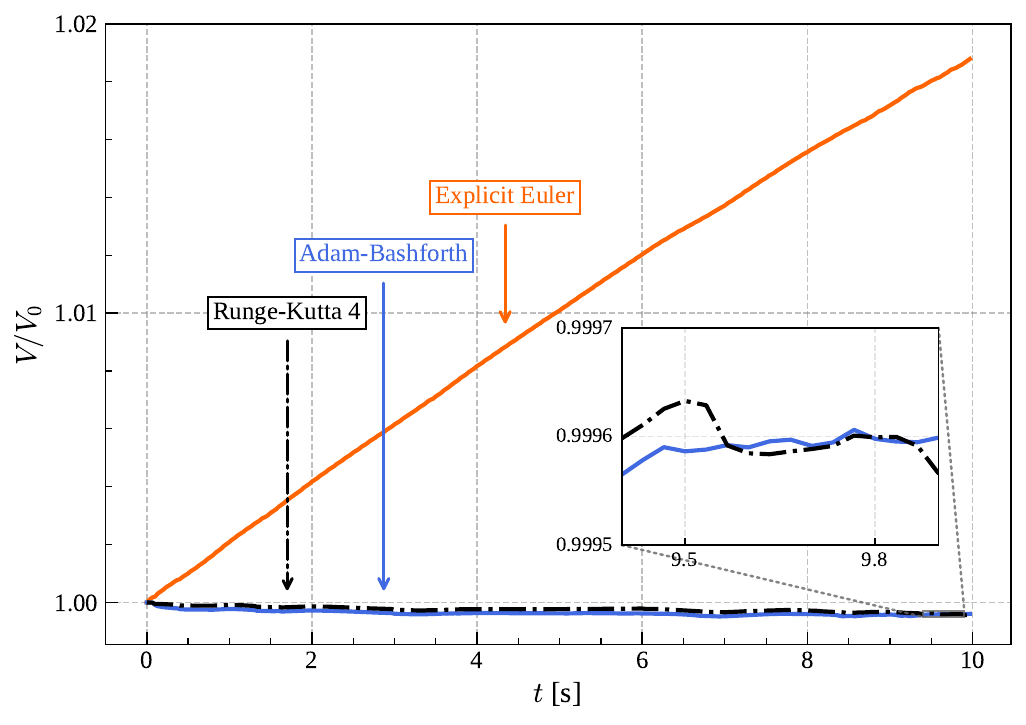}};
    \end{tikzpicture}
    \caption{Temporal evolution of the upper droplet volume for three temporal integration
        schemes: explicit Euler, Adams-Bashforth (AB2) and fourth-order Runge-Kutta (RK4), with $\Delta t=10^{-3}$.
    AB2 introduces no additional computational overhead within this framework, yet
    substantially improves mass preservation.}
    \label{fig:vortex_volume1}
\end{figure}

\begin{figure}[h!]
    \centering
    \begin{tikzpicture}
        \node[anchor=south west, inner sep=0] (img) at (0,0) {\includegraphics[width=10cm]{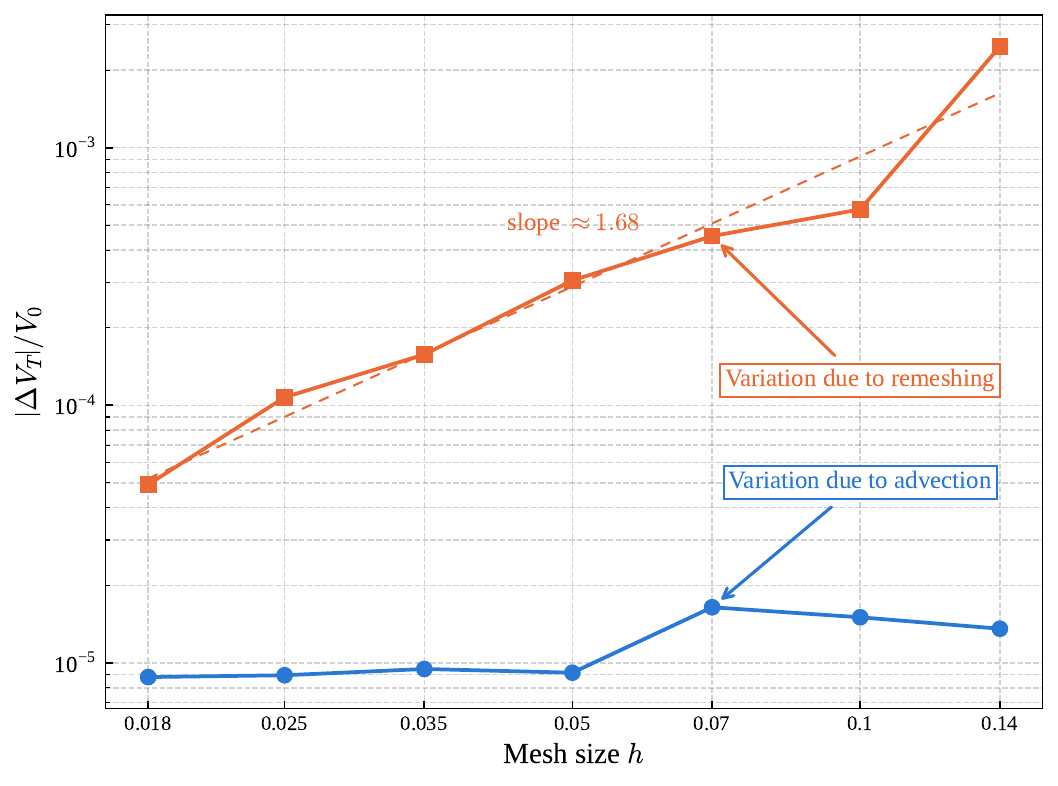}};
    \end{tikzpicture}
    \caption{Volume conservation as a function of the mesh size, using the AB2 scheme.
    The losses associated with the advection mechanism and with the surface adaptation,
    specifically the coarsening, are shown separately. Most of the volume loss originates
    from the coarsening, and this error decreases under refinement.}
    \label{fig:vortex_volume2}
\end{figure}

Using a divergence-free velocity field, we assess volume conservation after both the advection and remeshing steps. The results in \autoref{fig:vortex_volume1} show that the first-order explicit Euler scheme performs poorly, with the associated volume loss directly attributable to the temporal discretization. By contrast, the higher-order Adams-Bashforth 2 (AB2) and Runge-Kutta 4 (RK4) schemes substantially reduce volume loss and yield nearly identical profiles, suggesting that the simulation is sufficiently resolved in time. RK4 is used here as a reference, as its practical implementation is challenging when mesh topology changes occur during intermediate sub-steps, as discussed in \autoref{sec:governing_equations}. This difficulty is bypassed here by prescribing an analytical velocity field. The residual volume loss is attributed to interface coarsening: as discussed in \autoref{sec:interfaces}, removing an interface node introduces a small geometrical error controlled by the flatness tolerance. Its dependence on the interface mesh size is illustrated in \autoref{fig:vortex_volume2}, using the AB2 scheme.

\subsection{Rayleigh-Taylor with three phases}\label{sec:rt3}
The second benchmark proposed in this study is a gravity-driven three-phase flow characterized by an unstable initial configuration. Such stratified systems are well known to generate Rayleigh-Taylor instabilities. The numerical results are compared against the work of Garoosi et al.~\cite{garoosi}, who developed a Volume-of-Fluid method to simulate multiphase flows. The setup of the test case is presented in \autoref{tab:three_phase_rt_parameters}.
\medskip

\begin{table}[H]
\centering
\begin{tabular}{lccc}
\toprule
\textbf{Parameter} & & \textbf{Value} & \textbf{Units} \\
\midrule
%\textit{Physical Properties} & & &\\
Density of Phase 1 (bottom) & $\rho_1$ & $1.0$ & $\mathrm{kg/m^3}$\\
Density of Phase 2 (middle) & $\rho_2$ & $2.0$ & $\mathrm{kg/m^3}$\\
Density of Phase 3 (top) & $\rho_3$ & $4.0$ & $\mathrm{kg/m^3}$\\
Uniform kinematic viscosity & $\nu$ & $0.01$ & $\mathrm{m^2/s}$\\
Gravitational acceleration & $g$ & $17.64$ & $\mathrm{m/s^2}$\\
Interfacial surface tension & $\gamma_{1/2}, \gamma_{2/3}$ & $0.0$ & $\mathrm{N/m}$\\
\midrule
%\textit{Geometric Setup} & & & \\
Characteristic length scale & $H$ & $1.0$ & $\mathrm{m}$\\
Domain dimensions & $\Omega$ & $[0, H] \times [0, 3H]$ & $\mathrm{m}$\\
Lower interface profile & $\Gamma_{1/2}(x)$ & $1 + 0.1\cos(2\pi x)$ \\
Upper interface profile & $\Gamma_{2/3}(x)$ & $2 + 0.1\cos(2\pi x)$ \\
\bottomrule
\end{tabular}
\caption{Physical properties and geometric configurations for the three-phase Rayleigh-Taylor instability benchmark.}
\label{tab:three_phase_rt_parameters}
\end{table}

\begin{figure}[H]
    \centering
    \includegraphics[width=10cm]{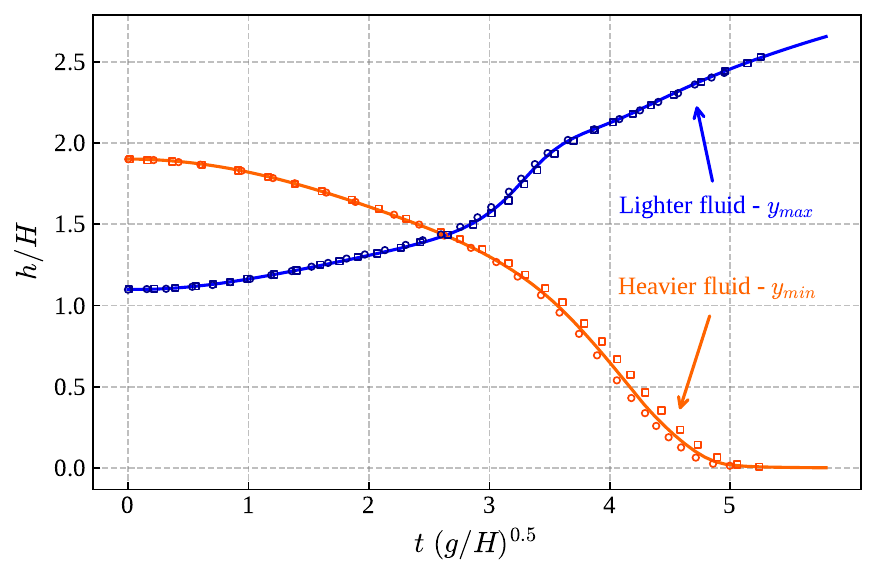}
    \caption{Temporal evolution of the minimum and maximum vertical positions for the dense and light phases, compared against the benchmark results of Garoosi et al.~\cite{garoosi}. The reference study evaluates two numerical schemes: a Lagrangian fluid discretization (MPS) \myemptydisk[darkblue]{1.8} \myemptydisk[orangered]{1.8} and a VOF
\myemptysquare[darkblue]{1.8} \myemptysquare[orangered]{1.8} approach. Our numerical results \myrect[blue] \myrect[orange] demonstrate excellent agreement with both methodologies.}
    \label{fig:3phases_positions}
\end{figure}

\begin{figure}[H]
    \centering
    % First subfigure
    \begin{subfigure}[b]{0.22\textwidth}
        \begin{tikzpicture}
            \node[anchor=south west, inner sep=0] (img) at (0,0) {\includegraphics[width=\textwidth]{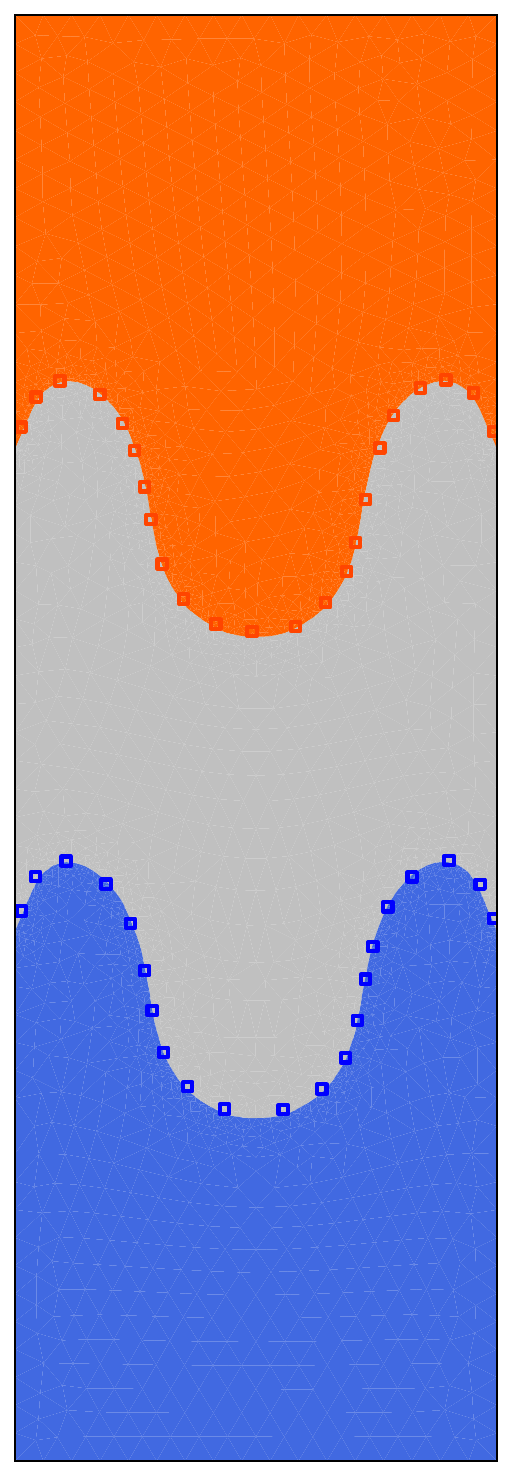}};
            \node[anchor=north west, fill=white, fill opacity=0.7, text opacity=1, inner sep=3pt] at (0.25, 10.25) {\footnotesize Nodes: $2\,941$};
        \end{tikzpicture}
        \caption{$t=0.375$}
        \label{fig:rt3fluids_sub1}
    \end{subfigure}
    \hfill % Adds horizontal space between the two top subfigures
    \begin{subfigure}[b]{0.22\textwidth}
        \begin{tikzpicture}
            \node[anchor=south west, inner sep=0] (img) at (0,0) {\includegraphics[width=\textwidth]{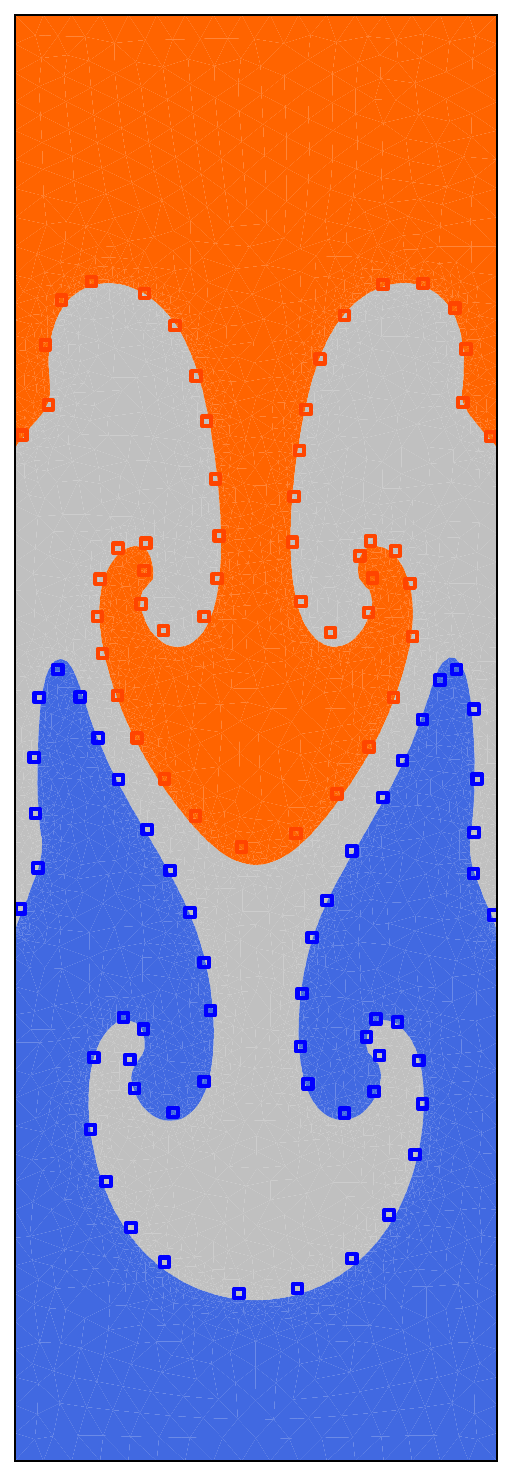}};
            \node[anchor=north west, fill=white, fill opacity=0.7, text opacity=1, inner sep=3pt] at (0.25, 10.25) {\footnotesize Nodes: $5\,930$};
        \end{tikzpicture}
        \caption{$t=0.750$}
        \label{fig:rt3fluids_sub2}
    \end{subfigure}
    \hfill % Adds horizontal space between the two top subfigures
    \begin{subfigure}[b]{0.22\textwidth}
        \begin{tikzpicture}
            \node[anchor=south west, inner sep=0] (img) at (0,0) {\includegraphics[width=\textwidth]{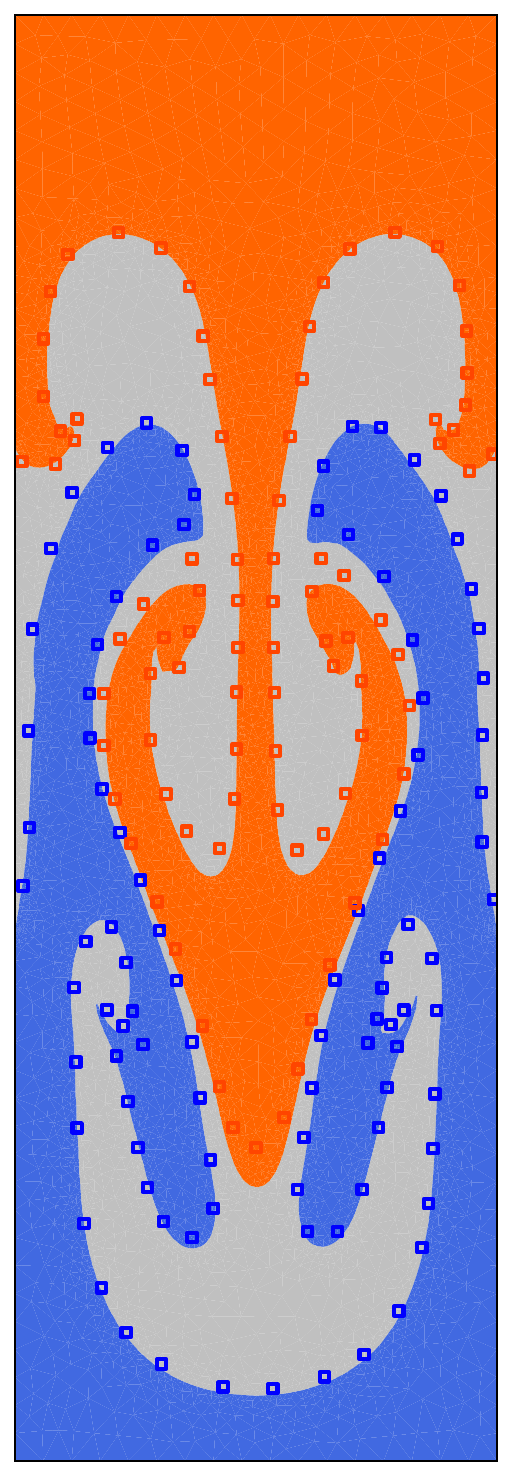}};
            \node[anchor=north west, fill=white, fill opacity=0.7, text opacity=1, inner sep=3pt] at (0.25, 10.25) {\footnotesize Nodes: $8\,698$};
        \end{tikzpicture}
        \caption{$t=0.974$}
        \label{fig:rt3fluids_sub3}
    \end{subfigure}
    \hfill % Adds horizontal space between the two top subfigures
    \begin{subfigure}[b]{0.22\textwidth}
        \begin{tikzpicture}
            \node[anchor=south west, inner sep=0] (img) at (0,0) {\includegraphics[width=\textwidth]{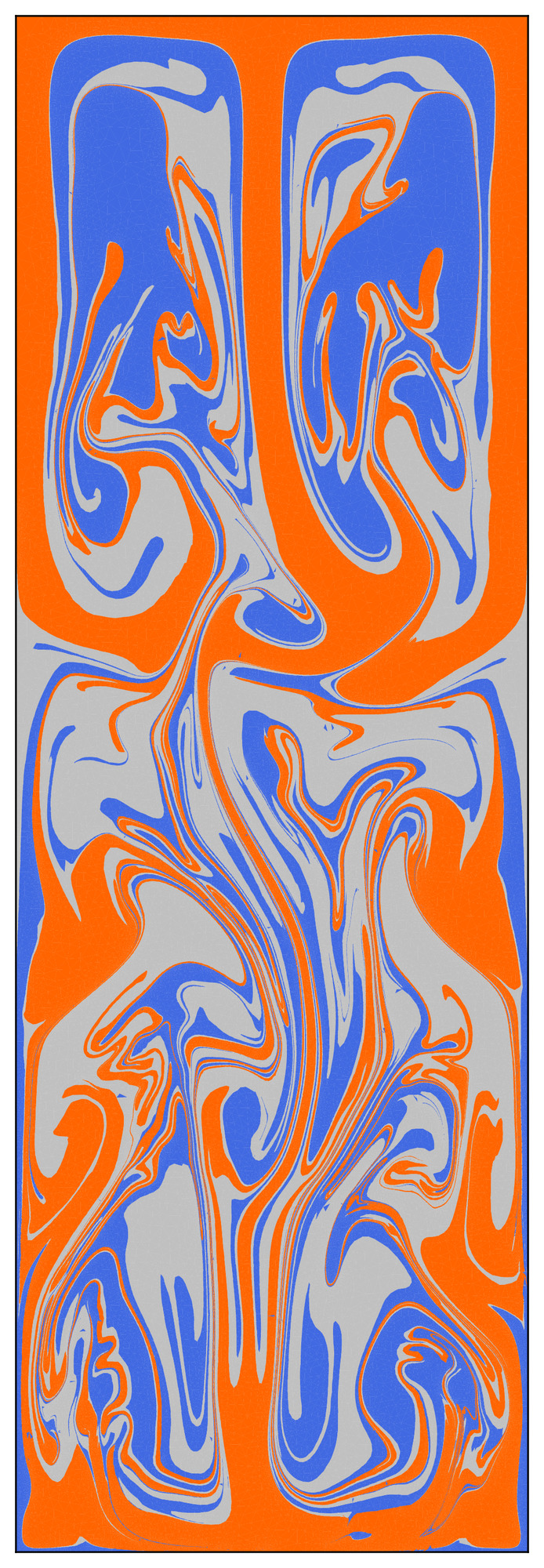}};
            \node[anchor=north west, fill=white, fill opacity=0.7, text opacity=1, inner sep=3pt] at (0.25, 10.25) {\footnotesize Nodes: $249\,075$};
        \end{tikzpicture}
        \caption{$t=2.000$}
        \label{fig:rt3fluids_sub4}
    \end{subfigure}
    \caption{A Rayleigh-Taylor instability with 3 distinct phases of densities $\rho \in \{1, 2, 4\}$. The initial configuration is perturbed to make the instabilities appear. Our results \mysquare[blue] \mysquare[orange] are compared with those of \cite{garoosi}, where Garoosi et al.\ use here a Volume-of-Fluid approach. The positions of the interfaces $\Gamma_{1/2}$ \myemptysquare[darkblue]{1.8} and $\Gamma_{2/3}$ \myemptysquare[orangered]{1.8} obtained by the authors are considered for comparison.}
    \label{fig:rt3fluids}
\end{figure}

\autoref{fig:3phases_positions} compares the minimum and maximum vertical positions of the respective phases over time. Results of the benchmark are presented in \autoref{fig:rt3fluids} alongside the data reported by Garoosi et al. Strong agreement is observed; notably, the two methods presented by those authors (VOF and MPS) exhibit a slight offset relative to one another, while our PFEM results fall between the two curves. While the VOF method operates on a uniform grid with a spatial resolution of $h=1/300$, we employ a baseline interface resolution of $s_{\min} = 1/83$. This allows us to significantly reduce the total number of nodes by clustering them exclusively in the vicinity of the interfaces. This strategy proves to be highly efficient when interfaces remain well-localized. However, during the mixing phase, thin filaments permeate the entire domain, causing the node count to increase rapidly due to the strict minimum interface resolution constraint of $w_{\min}=1/830$. \autoref{fig:rt3fluids_sub4} illustrates this surge in the number of nodes, while also demonstrating the capability of the method to capture exceptionally thin features in the fluid.
\medskip

\begin{figure}[h!]
    \centering
    \begin{tikzpicture}
        \node[anchor=south west, inner sep=0] (img) at (0,0) {\includegraphics[width=0.9\textwidth]{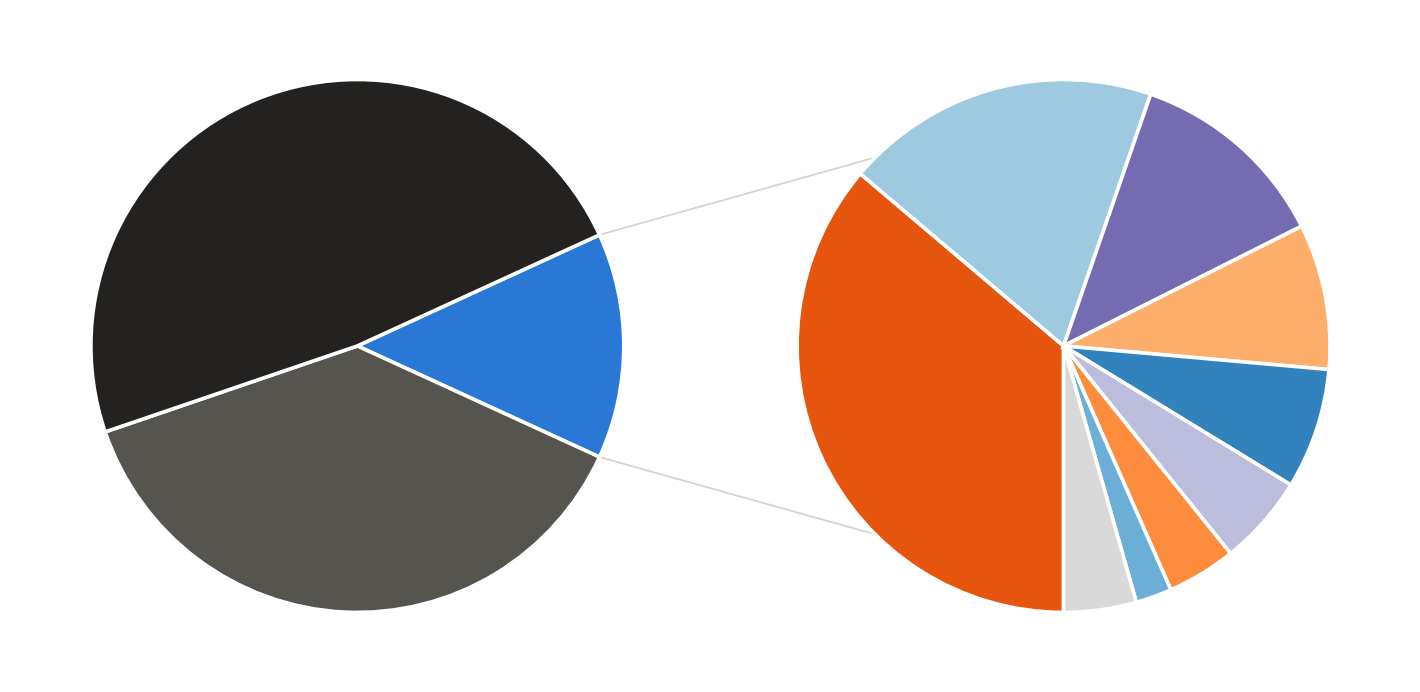}};

        % solve
        \node[white, anchor=south] at (3.4, 4.8) {\scriptsize $48.4\%$};
        \node[white, anchor=south] at (3.4, 4.4) {\scriptsize Linear Solve};
        
        % assembly
        \node[white, anchor=south] at (3.6, 2.2) {\scriptsize $37.9\%$};
        \node[white, anchor=south] at (3.6, 1.8) {\scriptsize FE Assembly};

        % Mesh Operations 
        \node[white, anchor=south] at (5.6, 3.6) {\scriptsize $13.7\%$};
        \node[white, anchor=south] at (5.6, 3.2) {\scriptsize Meshing};

        % Triangulation 
        \node[black, anchor=south] at (10.6, 5.4) {\scriptsize $19.1\%$};
        \node[black, anchor=south] at (10.6, 5) {\scriptsize Triangulation};
        
        % Mesh Refinement 
        \node[black, anchor=south] at (9.8, 3.2) {\scriptsize $36.2\%$};
        \node[black, anchor=south] at (9.8, 2.8) {\scriptsize Mesh Refinement};

        % Interface meshing 
        \draw[black, thick] (12.9, 5.6) -- (13.3, 6) -- (16, 6);
        \node[black, anchor=south] at (14.75, 5.95) {\scriptsize $12.3\%$ - Interface Adapt.};

        % coloring  
        \draw[black, thick] (13.75, 4.2) -- (14.1, 4.4) -- (16, 4.4);
        \node[black, anchor=south] at (15, 4.35) {\scriptsize $8.8\%$ - Coloring};

        % interpolated
        \draw[black, thick] (13.7, 2.9) -- (14.05, 2.75) -- (16, 2.75);
        \node[black, anchor=south] at (15.2, 2.7) {\scriptsize $7.3\%$ - Interpol.};

        % volume filtering 
        \draw[black, thick] (13.2, 1.9) -- (13.55, 1.65) -- (16, 1.65);
        \node[black, anchor=south] at (15, 1.6) {\scriptsize $5.5\%$ - Vol. Filtering};

        % Volume coarsening 
        \draw[black, thick] (12.5, 1.3) -- (12.9, 1.1) -- (16, 1.1);
        \node[black, anchor=south] at (14.5, 1.05) {\scriptsize $4.2\%$ - Vol. Coarsening};

        % Node Advection 
        \draw[black, thick] (12, 1.1) -- (12.4, 0.6) -- (16, 0.6);
        \node[black, anchor=south] at (14.2, 0.55) {\scriptsize $2.2\%$ - Mesh Advection};

        % Other
        \draw[black, thick] (11.5, 1) -- (11.3, 0.5) -- (8.5, 0.5);
        \node[black, anchor=south] at (9.5, 0.45) {\scriptsize $4.4\%$ - Other}; 

    \end{tikzpicture}
    \caption{Average computational cost (execution time) associated with the different steps of the algorithm. The profiling was performed on the \textit{three-phase Rayleigh-Taylor} test case, which runs in about $9\,$min on a standard machine equipped with an Intel Core Ultra 7 265H processor.}
    \label{fig:piechart}
\end{figure}

To provide an overview of the computational cost of a single PFEM iteration, \autoref{fig:piechart} reports the average time spent in each stage of the method for the present test case. Since this work focuses on methodological development rather than computational optimization, these timings should be interpreted as indicative of the relative cost of each stage rather than as representative of a fully optimized implementation. As an order of magnitude, the simulation presented above required approximately \(9\,\mathrm{min}\) on a standard machine equipped with an Intel Core Ultra 7 265H processor.
\medskip

The remeshing procedure accounts for approximately \(14\%\) of the total iteration time, a moderate overhead considering that localized refinement reduces the overall number of mesh nodes. Within the interface adaptation algorithm presented in \autoref{sec:interfaces}, most operations, primarily encroachment detection and subsequent edge splitting, are local and can therefore be performed in parallel. The main computational cost of the meshing stage is the volume refinement of \autoref{alg:chew_refinement}, which is currently sequential and for which a parallel implementation is under development. The cost of the Delaunay triangulation could also be further reduced using more efficient implementations such as HXT \cite{hxt}. The resulting linear systems are solved using PARDISO \cite{pardiso}. The fluid assembly is performed by MigFlow \cite{thomas_pfemdem}, an ongoing research code that is currently being optimized. Because the mesh topology may change at every time step, however, the system matrix must be assembled at each iteration. Overall, these timings indicate that the proposed remeshing strategy introduces a limited computational overhead while leaving several opportunities for further optimization of the computational pipeline.

\subsection{Isolated rising bubble}

A classic test case for validating multiphase numerical solvers is the simulation of an isolated bubble, initially at rest, rising solely under the influence of gravity. Bubbles represent challenging features to model; they can adopt a wide variety of geometric configurations depending on the prevailing flow regime, and frequently exhibit topological transitions that are difficult to resolve numerically. Furthermore, the hydrodynamic coupling between the bubble and the surrounding fluid significantly dictates its deformation, which in turn governs key physical quantities such as the terminal rise velocity and the interfacial pressure jump induced by local curvature. High-fidelity tracking of the phase interface is therefore paramount to accurately capturing the global dynamics of the flow. A robust characterization of these distinct regimes is provided in the seminal work of Grace \cite{grace}, which outlines the expected flow behavior under specified fluid properties.
\medskip

While bubbles that undergo explicit topological transformations are analyzed later in \autoref{sec:filaments}, the current benchmark focuses on the \textit{ellipsoidal} regime, where the bubble experiences severe deformation without breaking, directly affecting its rise kinematics. The numerical results are compared against the recognized benchmark of Hysing et al.~\cite{hysing}, which compiles data from three distinct multiphase solvers; two based on the Level-Set method \cite{tp2d, freelife} and one employing the Arbitrary Lagrangian-Eulerian formulation \cite{moonmd}. The numerical parameters for the experiments are provided in \autoref{tab:hysing_case1_parameters}. 
\medskip

\begin{table}[h!]
    \centering
    \begin{tabular}{lccc}
        \toprule
        \textbf{Parameter} & & \textbf{Value} & \textbf{Units} \\
        \midrule
        %\textit{Physical Properties} & & & \\
        Surrounding fluid density & $\rho_1$ & $1000$ & $\mathrm{kg/m^3}$\\
        Bubble density & $\rho_2$ & $100$ & $\mathrm{kg/m^3}$ \\
        Surrounding fluid dynamic viscosity & $\mu_1$ & $10$ & $\mathrm{kg/(m \cdot s)}$\\
        Bubble dynamic viscosity & $\mu_2$ & $1$ & $\mathrm{kg/(m \cdot s)}$ \\
        Gravitational acceleration & $g$ & $0.98$ & $\mathrm{m/s^2}$ \\
        Surface tension coefficient & $\gamma$ & $24.5$ & $\mathrm{N/m}$ \\
        \midrule
        %\textit{Geometric and Temporal Setup} & & & \\
        Domain dimensions & $\Omega$ & $[0, 1] \times [0, 2]$ & $\mathrm{m}$ \\
        Initial bubble radius & $r_0$ & $0.25$ & $\mathrm{m}$ \\
        Initial bubble center & $(x_0, y_0)$ & $(0.5, 0.5)$ & $\mathrm{m}$ \\
        Maximum simulation time & $T_{\max}$ & $3.0$ & $\mathrm{s}$ \\
        \midrule
        %\textit{Dimensionless Numbers} & & & \\
        Reynolds number & $\mathrm{Re}$ & $35$ &  \\
        Eötvös number & $\mathrm{Eo}$ & $10$ &  \\
        \bottomrule
    \end{tabular}
    \caption{Physical and numerical parameters for the first benchmark case proposed by Hysing et al.~\cite{hysing}.}
    \label{tab:hysing_case1_parameters}
\end{table}

To quantitatively evaluate our results against the benchmark data, two primary metrics are considered. The first is the bubble rise velocity, $U_y(t)$, which is intrinsically coupled to the dynamics of the surrounding fluid; its accurate prediction therefore validates the global hydrodynamic solver rather than merely the geometric interface-tracking capability in isolation. The second is the circularity, $\mathcal{C}(t)$, defined as the ratio of the perimeter of an equivalent-area disk to the actual interfacial perimeter of the bubble. This metric provides a non-dimensional quantification of the temporal deformation and stretching experienced by the phase boundary. Conforming to the benchmark, the circularity is computed on the effective numerical volume at time $t$.

\begin{figure}[H]
    \centering
    % First subfigure
    \begin{subfigure}[b]{0.47\textwidth}
        \centering
        \includegraphics[width=\textwidth]{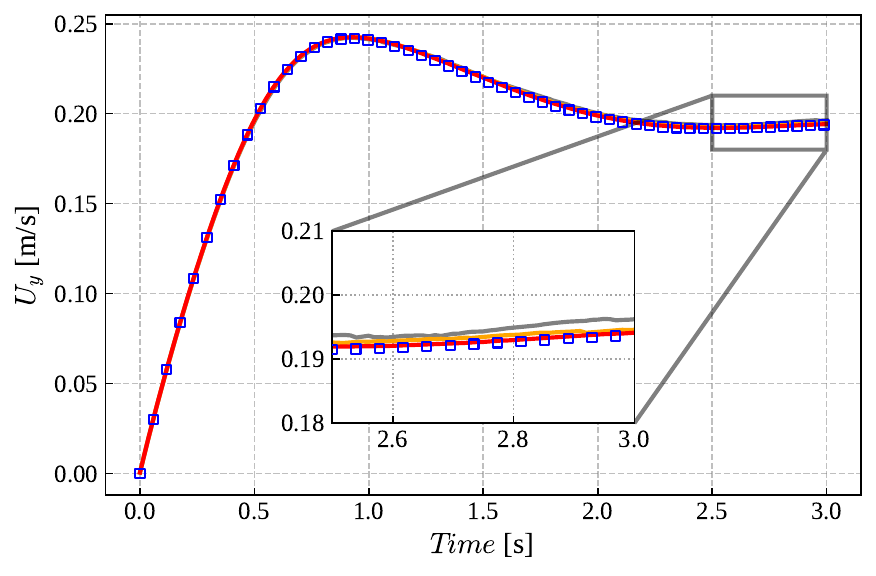}
        \caption{Vertical velocity for three mesh resolutions (c, d, e).}
        \label{fig:hysing1_velocity}
    \end{subfigure}
    \hfill % Pushes the first two subfigures to the outer edges
    % Second subfigure
    \begin{subfigure}[b]{0.47\textwidth}
        \centering
        \includegraphics[width=\textwidth]{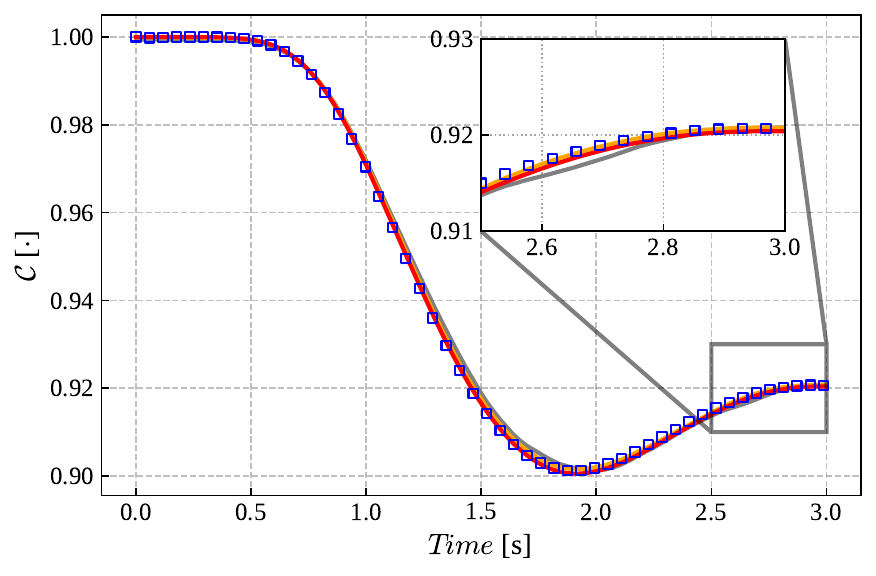}
        \caption{Circularity for three mesh resolutions (c, d, e).}
        \label{fig:hysing1_circularity}
    \end{subfigure}

    \vspace{0.3cm} % Adds a nice little vertical gap between the two rows
    \begin{subfigure}[b]{0.32\textwidth}
        \centering
        \begin{tikzpicture}
            % Place the image at (0,0)
            \node[anchor=south west, inner sep=0] (img) at (0,0) {\includegraphics[width=\textwidth]{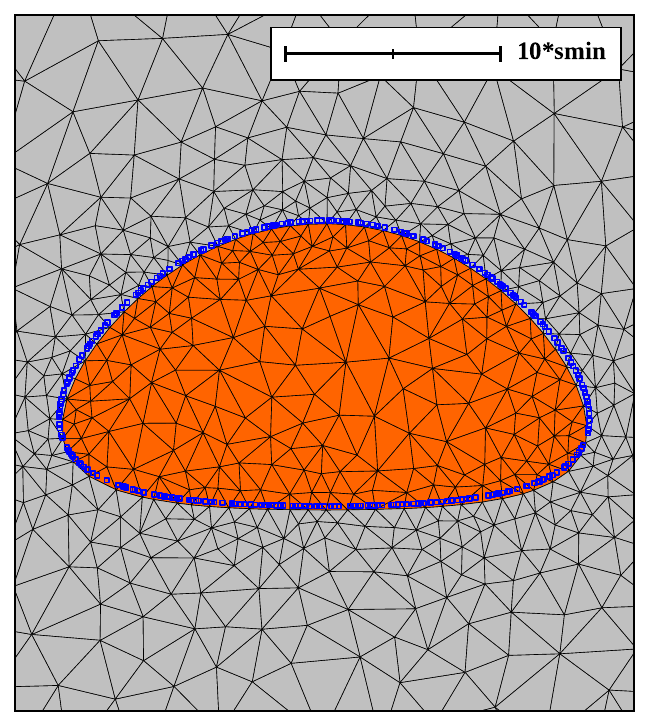}};
            % Place your label overlay
            \node[anchor=north west, fill=white, fill opacity=0.7, text opacity=1, inner sep=3pt] at (0.25, 5.72) {\footnotesize Nodes: $595$};
        \end{tikzpicture}
        \caption{$t=3.0$, $s_{\min}=1/36$ \myrect[gray].}
        \label{fig:hysing1_shape1}
    \end{subfigure}
    \hfill % Space out the bottom three horizontally (No blank line here!)
    \begin{subfigure}[b]{0.32\textwidth}
        \centering
        \begin{tikzpicture}
            \node[anchor=south west, inner sep=0] (img) at (0,0) {\includegraphics[width=\textwidth]{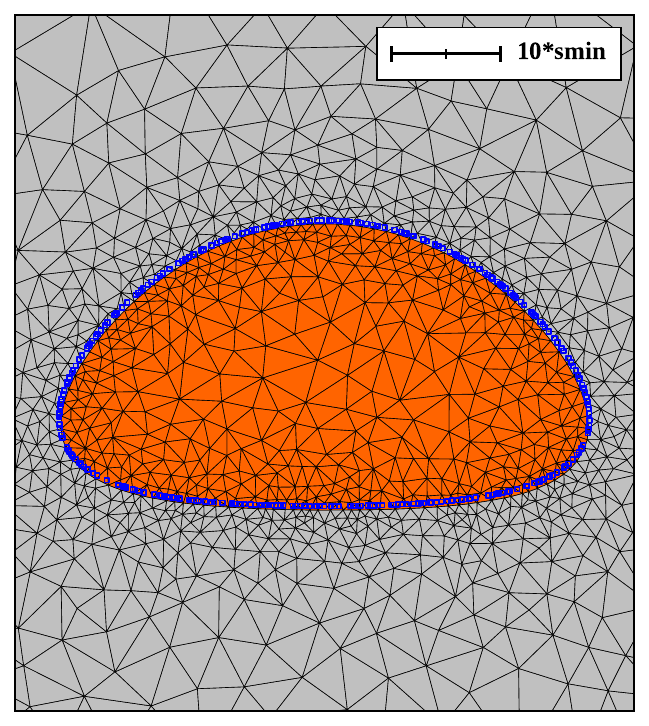}};
            \node[anchor=north west, fill=white, fill opacity=0.7, text opacity=1, inner sep=3pt] at (0.25, 5.72) {\footnotesize Nodes: $1\,168$};
        \end{tikzpicture}
        \caption{$t=3.0$, $s_{\min}=1/71$ \myrect[orange].}
        \label{fig:hysing1_shape2}
    \end{subfigure}
    \hfill % Space out the bottom three horizontally (No blank line here!)
    \begin{subfigure}[b]{0.32\textwidth}
        \centering
        \begin{tikzpicture}
            \node[anchor=south west, inner sep=0] (img) at (0,0) {\includegraphics[width=\textwidth]{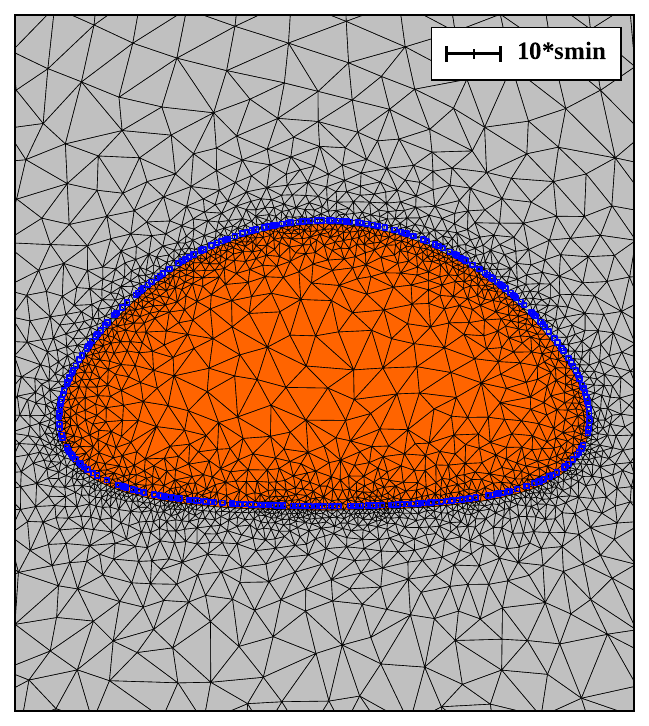}};
            \node[anchor=north west, fill=white, fill opacity=0.7, text opacity=1, inner sep=3pt] at (0.25, 5.72) {\footnotesize Nodes: $2\,697$};
        \end{tikzpicture}
        \caption{$t=3.0$, $s_{\min}=1/143$ \myrect[red].}
        \label{fig:hysing1_shape3}
    \end{subfigure}
    \caption{Comparison of the proposed PFEM model against the TP2D model results, obtained with $h=1/320$ \myemptysquare[blue]{1.8}, which closely agrees with the other two reference models of the study \cite{hysing}. Simulation is performed for three distinct interfacial mesh resolutions.}
    \label{fig:hysing1}
\end{figure}

\autoref{fig:hysing1} demonstrates the performance of our model under the conditions specified in \autoref{tab:hysing_case1_parameters}. The PFEM approach, combined with local mesh adaptation, yields results consistent with the reference models evaluated. Notably, the most refined simulation closely matches the velocity and circularity predicted by the TP2D model. As shown in \autoref{fig:hysing1_shape1} and \autoref{fig:hysing1_shape2}, our method accurately captures the complex bubble dynamics using significantly fewer nodes, offering a significant advantage over classical diffuse-interface approaches. The third simulation confirms that refining the mesh in the vicinity of the bubble drives convergence toward the benchmark solutions, while dynamic mesh adaptation effectively maintains a low overall node count.

\section{Discussion}
\label{sec:discussion}

\subsection{Mesh and flow independence}
\label{sec:filaments}

\begin{figure}[H]
    \centering
    % First subfigure
    \begin{subfigure}[b]{0.95\textwidth}
        \centering
        \begin{tikzpicture}
            \node[anchor=south west, inner sep=0] (img) at (0,0) {\includegraphics[width=\textwidth]{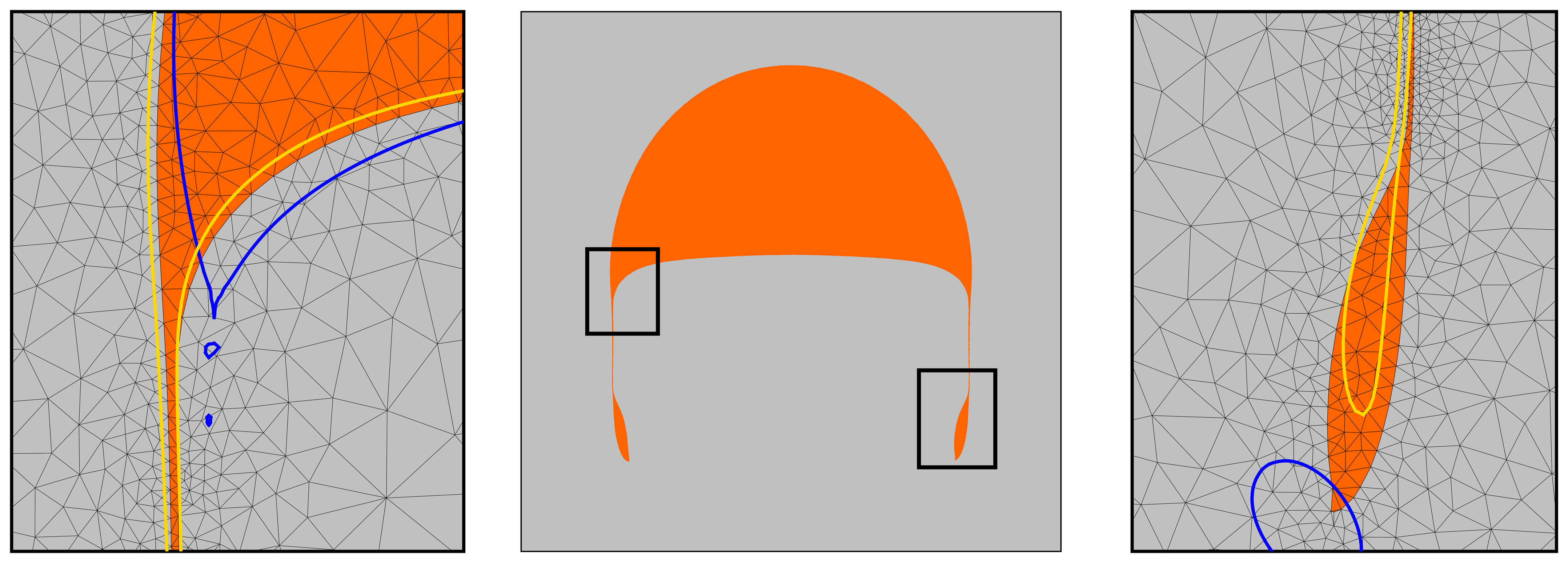}};
        \end{tikzpicture}
    \caption{No topological change. The size field at the interface is $s_{\min}=1/143$.}
        \label{fig:filaments_control1}
    \end{subfigure}
    \par\bigskip
    % Second subfigure
    \begin{subfigure}[b]{0.95\textwidth}
        \centering
        \begin{tikzpicture}
            \node[anchor=south west, inner sep=0] (img) at (0,0) {\includegraphics[width=\textwidth]{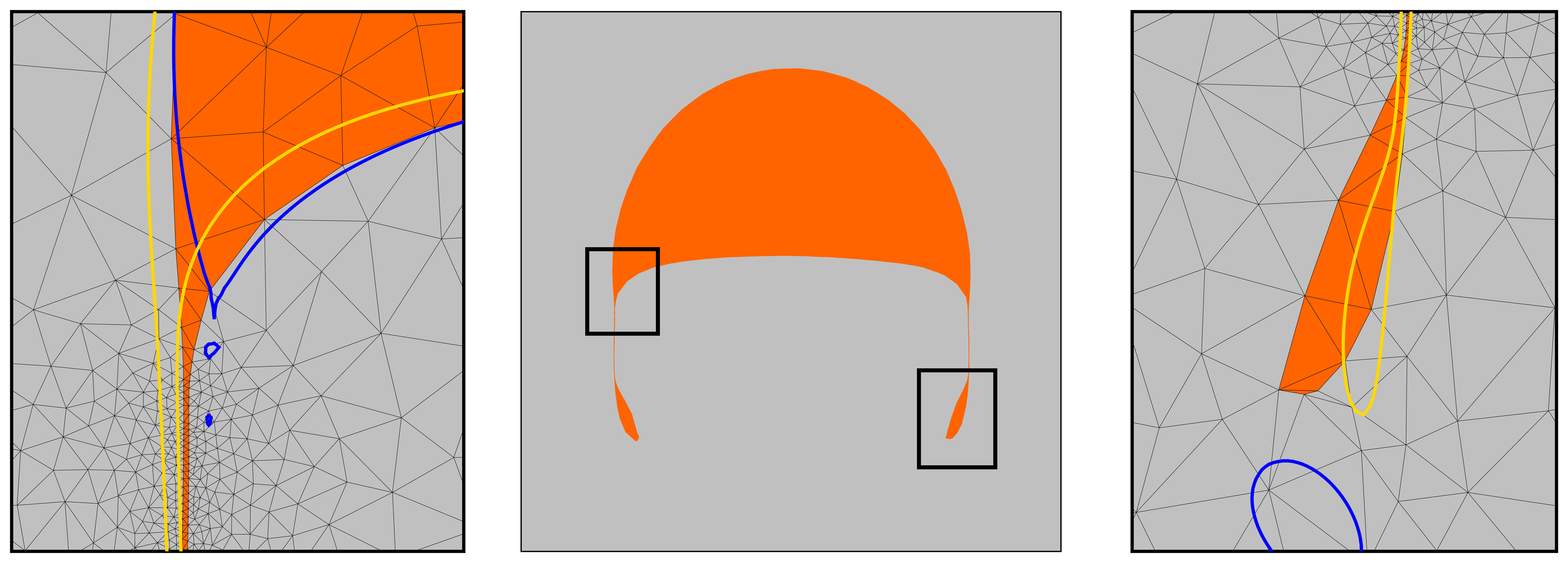}};
        \end{tikzpicture}
        \caption{No topological change. The size field at the interface is $s_{\min}=1/35$.}
        \label{fig:filaments_control2}
    \end{subfigure}
    \par\bigskip
    % third subfigure
    \begin{subfigure}[b]{0.95\textwidth}
        \centering
        \begin{tikzpicture}
            \node[anchor=south west, inner sep=0] (img) at (0,0) {\includegraphics[width=\textwidth]{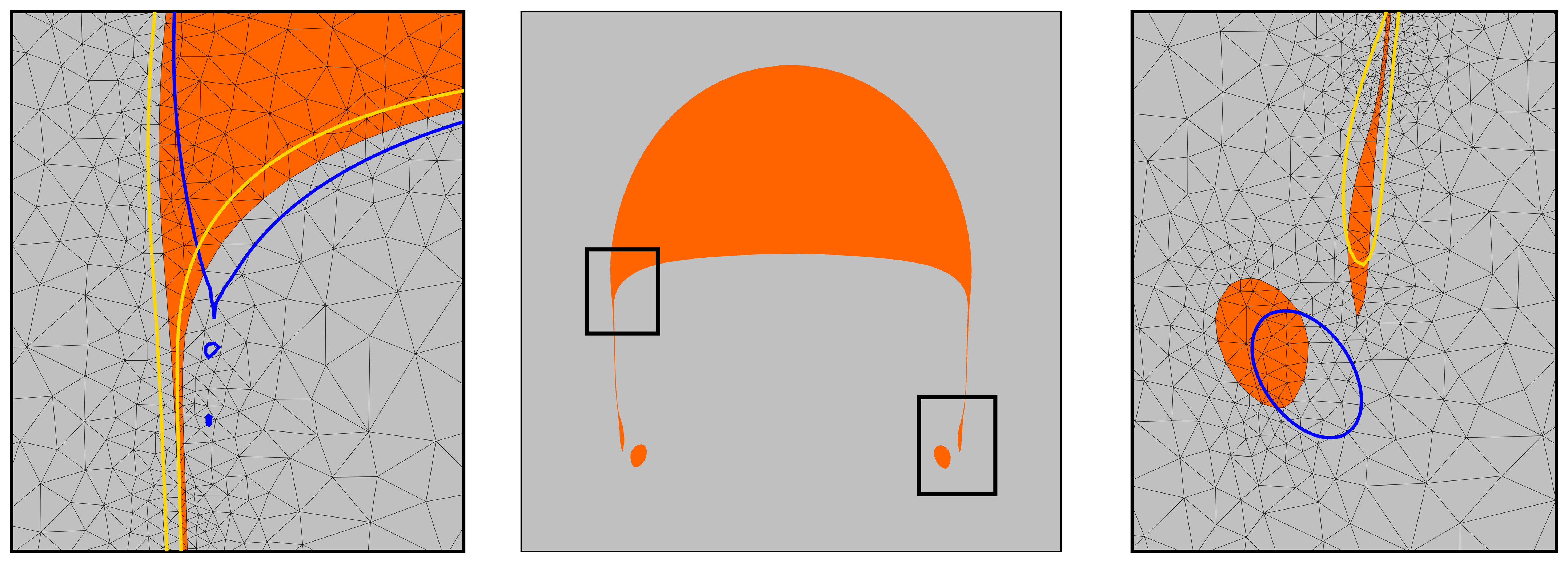}};
        \end{tikzpicture}
        \caption{Topological change enforced with an edge-flip. The size field at the interface is $s_{\min}=1/143$.}
        \label{fig:filaments_control3}
    \end{subfigure}
    \caption{Capability of the method to control fluid topology independently of the global mesh size. (a) The fluid is resolved and the topology is preserved to prevent filament breakup. (b) Although the mesh is significantly coarser, the topology remains unaffected, unlike in classical approaches, retaining the same qualitative behavior as in (a). (c) Topological changes can be explicitly enforced while remaining fully controlled and mesh-independent. The results are compared against two benchmark models from the literature, MooNMD {\myrect[gold]} \cite{moonmd} and TP2D {\myrect[blue]} \cite{tp2d}, demonstrating the ability to capture distinct topological regimes.}
    \label{fig:filaments_control}
\end{figure}

The primary objective of this work is to minimize the dependence of the computed physics on the computational mesh. From a physical perspective, the sharp-interface formulation of the incompressible Navier-Stokes equations does not intrinsically specify when a thin filament should undergo a topological transition: in physical flows, such transitions are driven by mechanisms at scales that fall below the computational mesh resolution. We contend that they should be governed by physically motivated criteria rather than triggered arbitrarily by the local grid size. To this end, a mesh adaptation strategy was developed to decouple the baseline resolution $s_{\min}$ from the length scale $w_{\min}$ at which a topological transition is permitted to occur. Preserving the interface topology down to the user-defined threshold $w_{\min}$ prevents the remeshing procedure from prematurely severing thin structures, while offering a systematic framework for incorporating sub-grid models that account for unresolved physical mechanisms and provide a rational basis for filament rupture (see, e.g., \cite{filaments_of_death}). \autoref{fig:filaments_control} demonstrates the effectiveness of this approach in controlling the topology independently of the global mesh size.
\bigskip

We revisit the isolated rising bubble case introduced in \autoref{tab:hysing_case1_parameters}, using modified fluid properties: $\rho_2=1$, $\mu_2=0.1$, and $\gamma=1.96$. These parameters increase the Eötvös number to $\mathrm{Eo}=125$, driving a transition from the \textit{ellipsoidal cap} to the \textit{skirted} bubble regime. Because satellite bubbles and thin filaments emerge in this regime, it provides an ideal benchmark for studying topological changes. We consider three distinct scenarios.
\bigskip

\autoref{fig:filaments_control1} serves as a baseline reference: the grid resolution resolves the regime, and topological changes are prohibited by our model. \autoref{fig:filaments_control2} runs with the exact same parameters as \autoref{fig:filaments_control1}, but employs a significantly coarser baseline mesh. Despite this coarsening, the topology remains consistent throughout the simulation. Whereas conventional methods suffer from numerical artifacts on coarse meshes, our approach qualitatively preserves the physical behavior with very few nodes. Thus, the qualitative bubble behavior is decoupled from the mesh size $s_{\min}$. Finally, \autoref{fig:filaments_control3} demonstrates that clean, controlled topological transitions can be enforced within the framework, in this case, by performing a single edge flip. Depending on whether this topological transition is executed, our model can reproduce distinct behaviors that align with different models from the literature.
\bigskip

\subsection{A 16-phase flow}

\begin{figure}[H]
    \centering
    % First subfigure
    \begin{subfigure}[b]{0.4\textwidth}
        \centering
        \includegraphics[width=\textwidth]{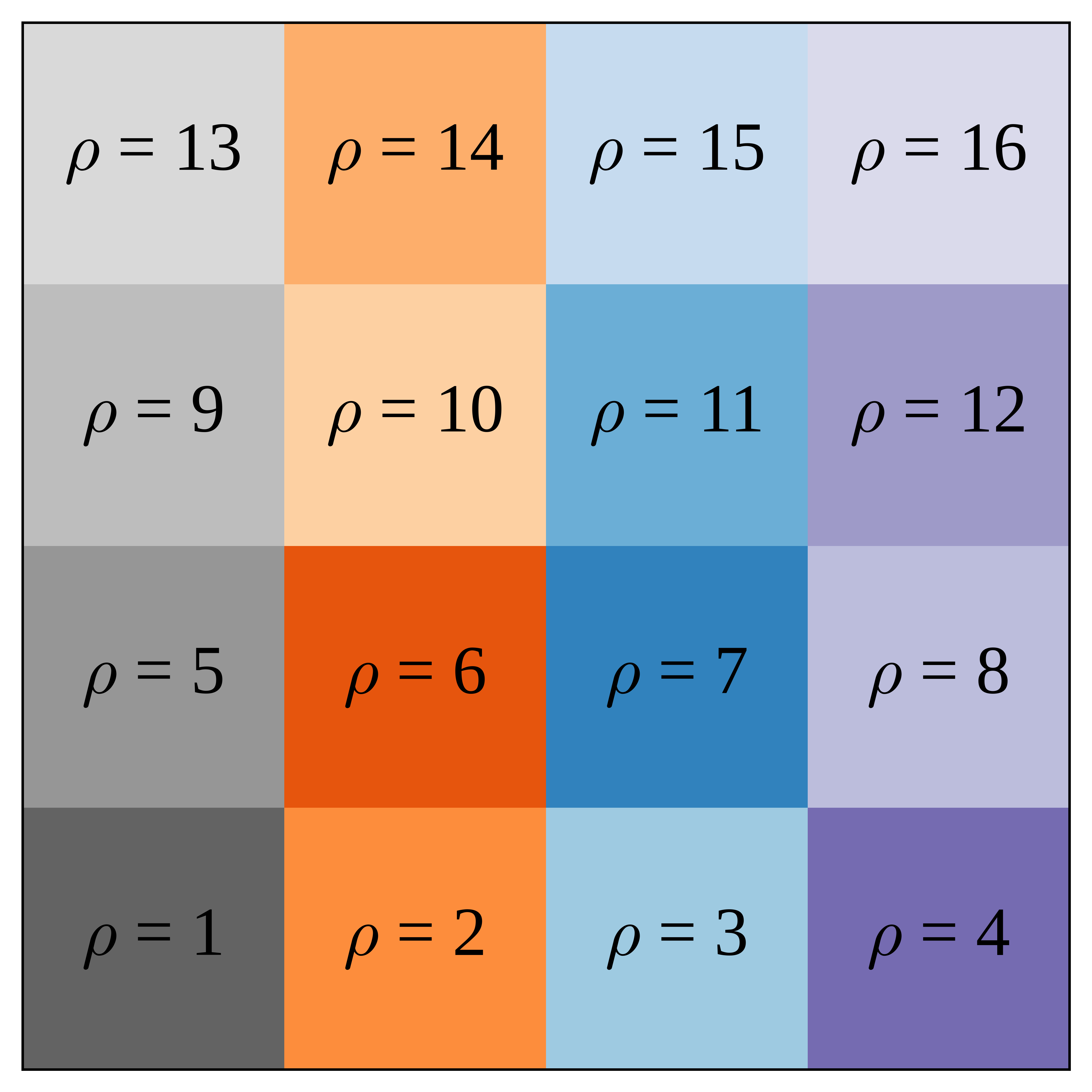}
        \caption{$t=0.0$}
        \label{fig:16fluids_sub1}
    \end{subfigure}
    \hspace{1cm} % Adds horizontal space between the two top subfigures
    % Second subfigure
    \begin{subfigure}[b]{0.4\textwidth}
        \centering
        \includegraphics[width=\textwidth]{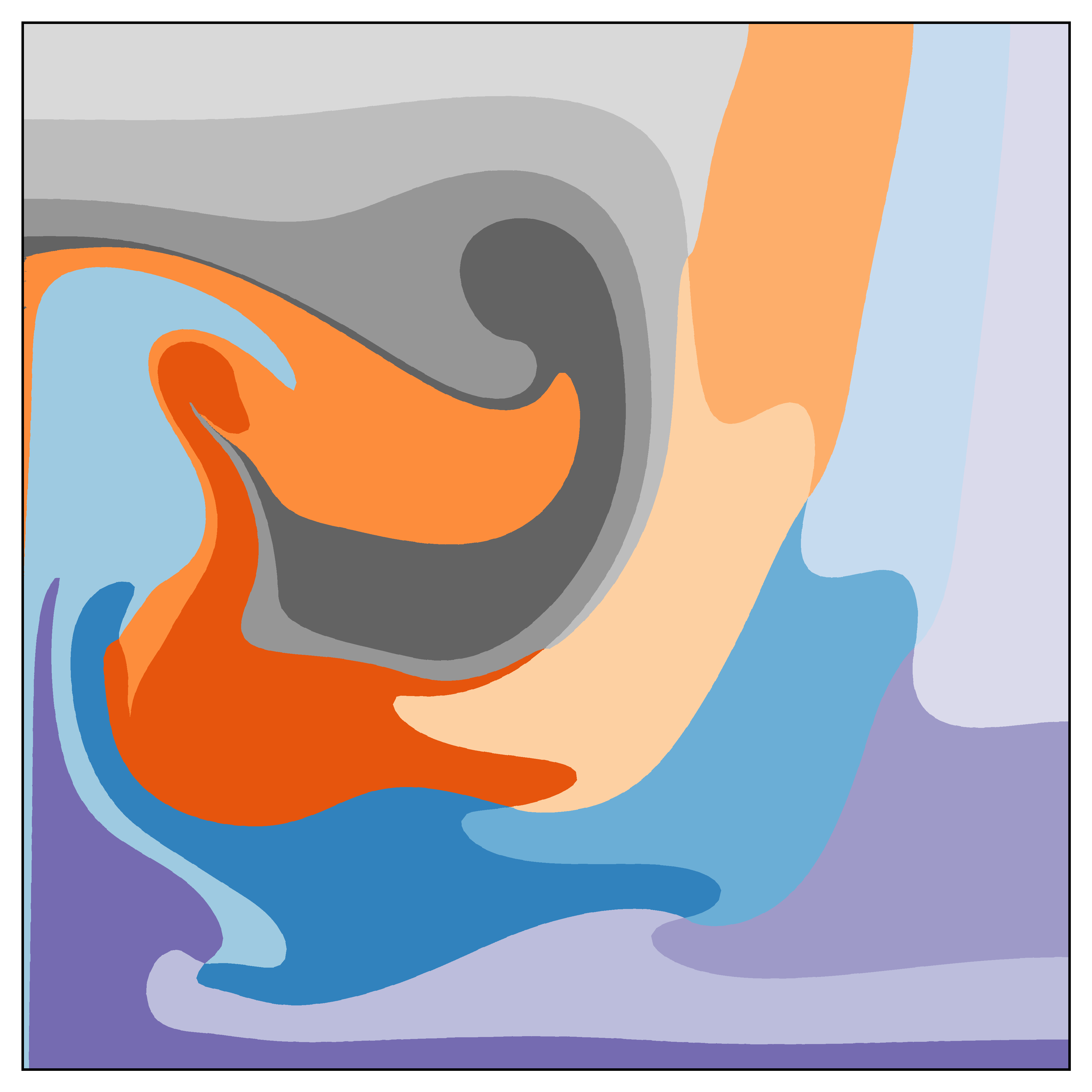}
        \caption{$t=0.8$}
        \label{fig:16fluids_sub2}
    \end{subfigure}

    \vspace{1em} % Adds vertical space between the top and bottom rows

    % Third subfigure
    \begin{subfigure}[b]{0.4\textwidth}
        \centering
        \includegraphics[width=\textwidth]{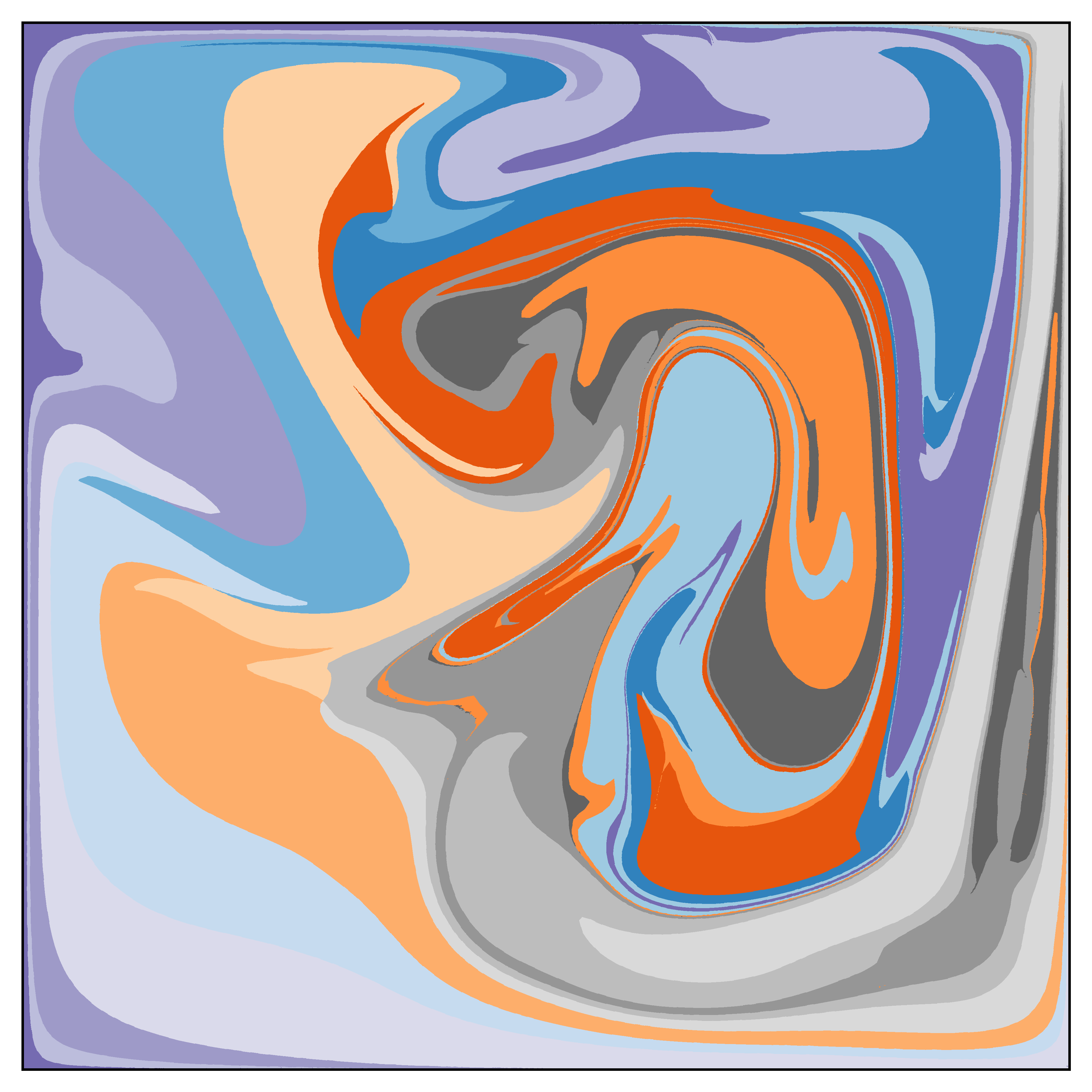}
        \caption{$t=1.6$}
        \label{fig:16fluids_sub3}
    \end{subfigure}
    \hspace{1cm} % Adds horizontal space between the two bottom subfigures
    % Fourth subfigure
    \begin{subfigure}[b]{0.4\textwidth}
        \centering
        \includegraphics[width=\textwidth]{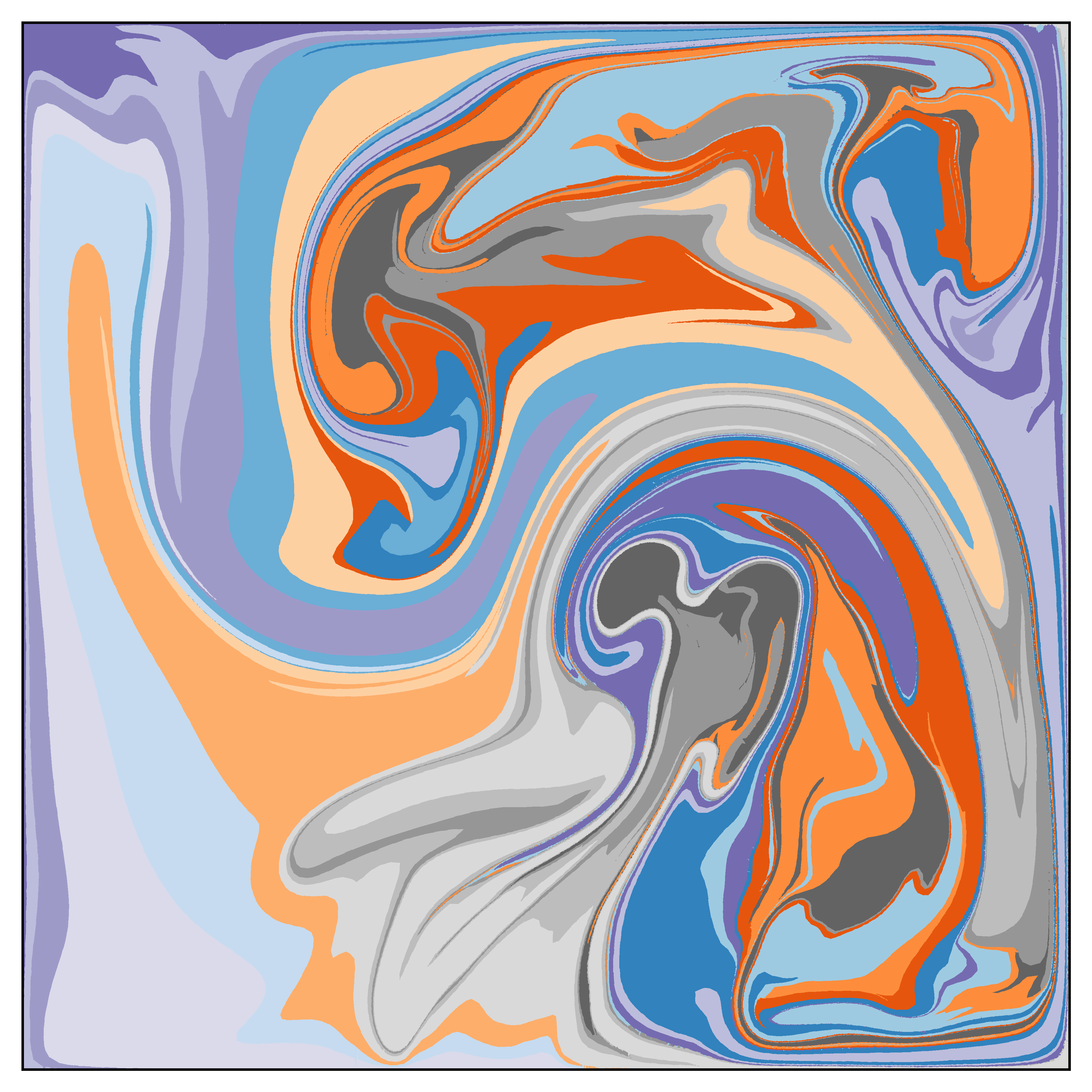}
        \caption{$t=2.4$}
        \label{fig:16fluids_sub4}
    \end{subfigure}

    \caption{A Rayleigh-Taylor simulation with 16 distinct phases of densities $\rho \in \{1, \dots, 16\}$. The flow is gravity-driven and no surface tension is considered. Thin layers of filaments are conserved by our meshing procedure.}
    \label{fig:16fluids}
\end{figure}

This test case aims to demonstrate that the proposed remeshing procedure, and more broadly the global PFEM framework, is completely agnostic to the number of phases. We simulate a system of 16 distinct fluid phases with densities ranging from $\rho=1$ to $\rho=16$. The kinematic viscosity is uniformly set to $\nu=0.01$ for all fluids, and surface tension effects are omitted. The initial configuration, shown in \autoref{fig:16fluids_sub1}, is designed such that the flow evolves naturally under the sole influence of gravity toward its equilibrium state. Temporal snapshots of the simulation are provided in \autoref{fig:16fluids}. Driven by local mesh adaptation, the method effectively captures intricate interfacial features. These structures are resolved down to a geometric threshold $w_{\min}$, beyond which topological transitions occur uncontrolled, through the edge flips of the Delaunay triangulation. Although this multiphase setup is non-physical, it successfully showcases the robustness and geometric versatility of the model.
\bigskip

\subsection{A rigid body and four phases}

The ability to explicitly conform the mesh to geometric interfaces offers a substantial advantage for multiphase flow simulations, but its application extends far beyond those cases. Regarding fluid-structure interaction (FSI), maintaining sharp interfaces is crucial for precisely capturing the coupling between the fluid and solid domains \cite{fsi_review}. Unlike diffuse or immersed boundary methods, a sharp interface allows for the exact enforcement of boundary conditions and the accurate integration of hydrodynamic forces (such as lift and drag) directly on the solid surface. While classical ALE methods track the solid by locally deforming the surrounding mesh, they are highly prone to severe mesh distortion when subjected to large-scale solid displacements. In contrast, our method tracks the fluid-solid boundary analogously to the multiphase flows discussed previously, preserving interface conformity through dynamic edge refinement. Although a comprehensive FSI solver involves complexities well beyond interface tracking, we showcase here how the geometric challenges can be elegantly handled within the PFEM framework.
\medskip

\begin{figure}[H]
    \centering
    % First subfigure
    \begin{subfigure}[b]{0.4\textwidth}
        \centering
        \includegraphics[width=\textwidth]{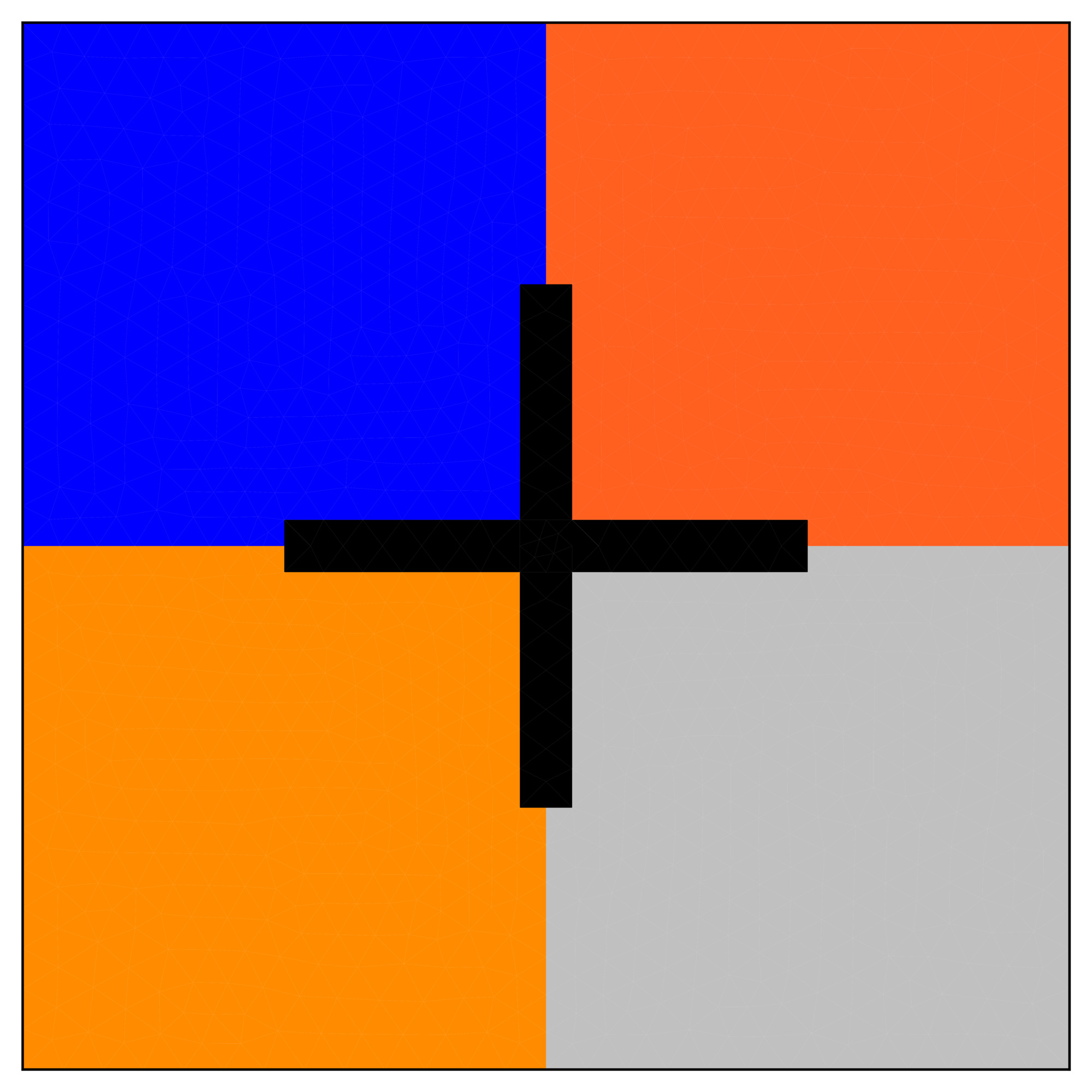}
        \caption{$t=0.0$}
        \label{fig:blender1}
    \end{subfigure}
    \hspace{1cm} % Adds horizontal space between the two top subfigures
    % Second subfigure
    \begin{subfigure}[b]{0.4\textwidth}
        \centering
        \includegraphics[width=\textwidth]{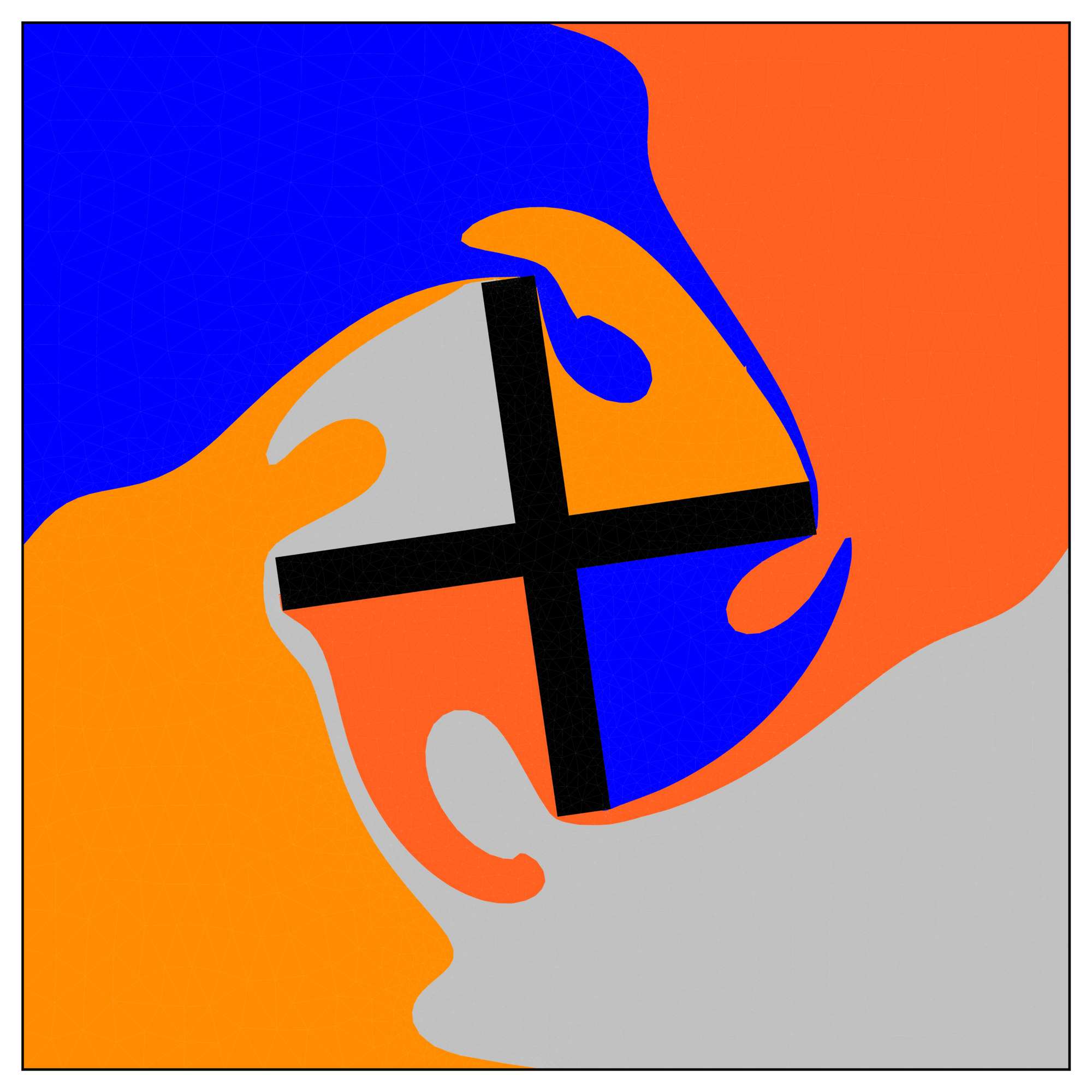}
        \caption{$t=0.3$}
        \label{fig:blender2}
    \end{subfigure}
    
    \vspace{1em} % Adds vertical space between the top and bottom rows
    
    % Third subfigure
    \begin{subfigure}[b]{0.4\textwidth}
        \centering
        \includegraphics[width=\textwidth]{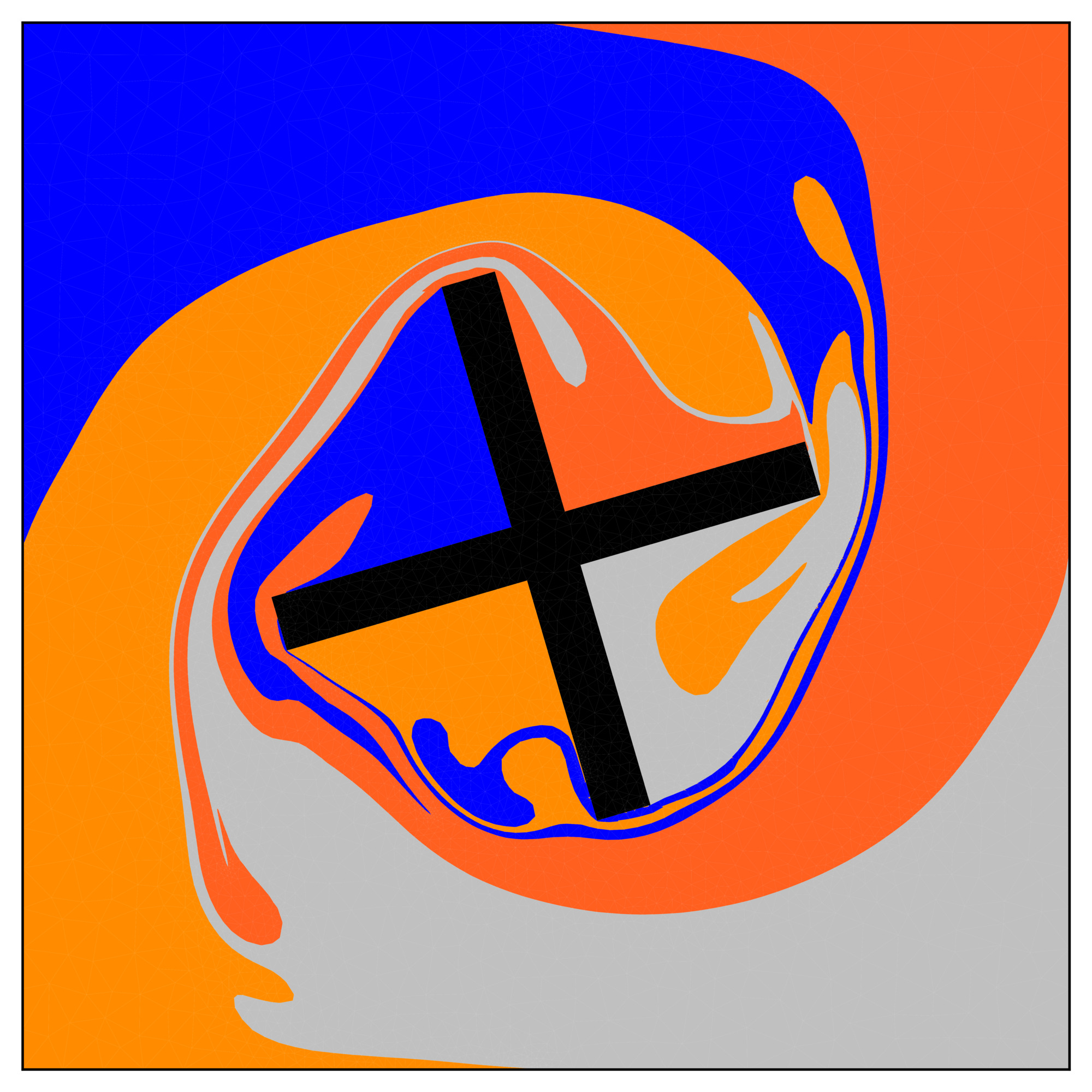}
        \caption{$t=0.6$}
        \label{fig:blender3}
    \end{subfigure}
    \hspace{1cm} % Adds horizontal space between the two bottom subfigures
    % Fourth subfigure
    \begin{subfigure}[b]{0.4\textwidth}
        \centering
        \includegraphics[width=\textwidth]{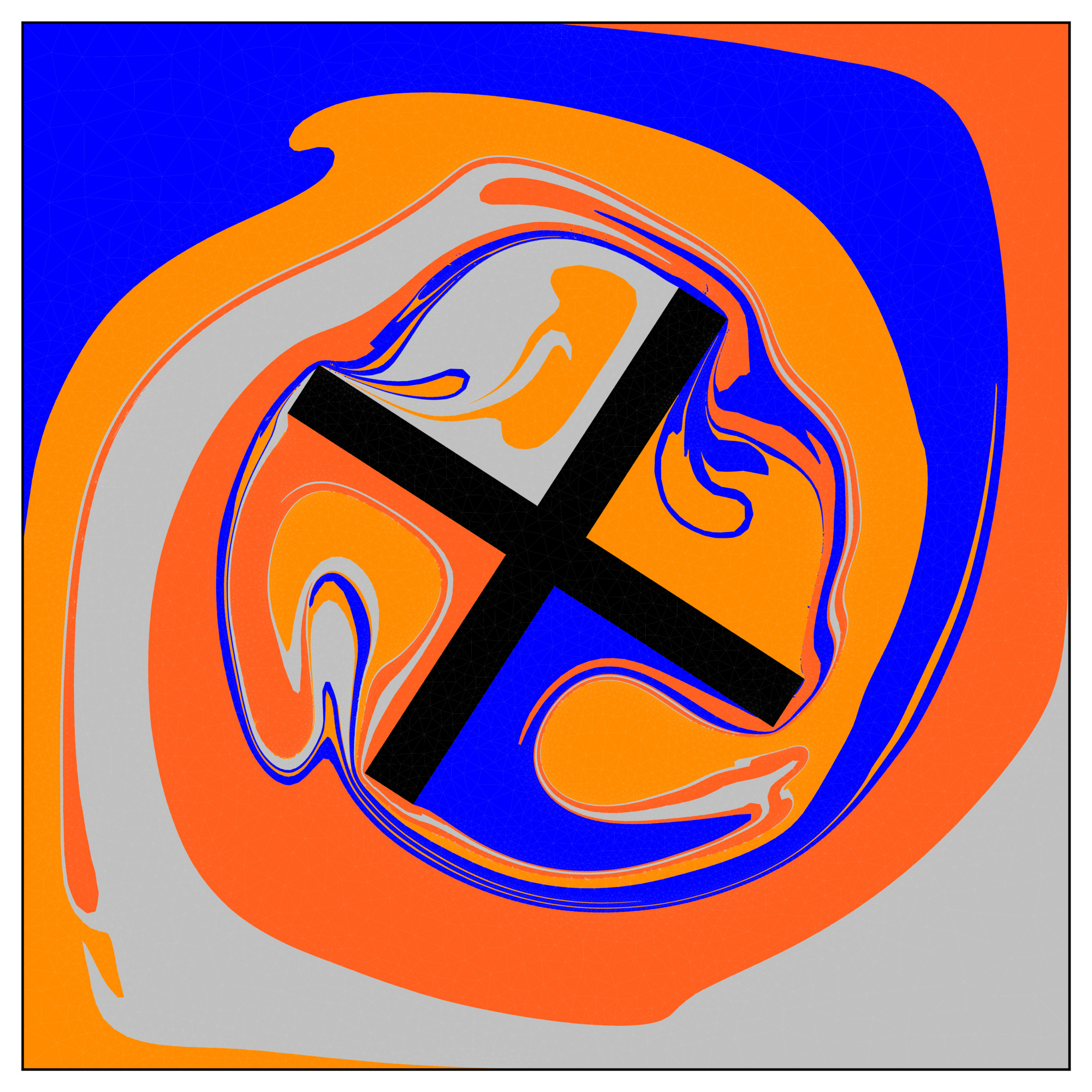}
        \caption{$t=1.0$}
        \label{fig:blender4}
    \end{subfigure}
    \caption{Demonstration of the method's capability in handling the geometric challenge behind fluid-structure interaction, within a multiphase flows configuration. A rigid body undergoes prescribed rotation, mixing four distinct fluid phases with discrete densities ranging from $\rho=1$ to $\rho=4$.}
    \label{fig:blender}
\end{figure}

We consider a test case consisting of a four-bladed rigid rotor immersed in a four-phase fluid. The rotational motion of the rigid body is prescribed analytically, and its dynamic impact on the fluid is enforced through Dirichlet boundary conditions at the interface. The simulation results are illustrated in \autoref{fig:blender}, showing that the intricate geometry of the moving solid is sharply and consistently captured by the mesh over time. Crucially, since the interior of the solid domain is also fully meshed, this framework seamlessly accommodates the integration of a structural solver, facilitating fluid-structure coupling without any grid conformity constraints.

\subsection{Two jets colliding}

\begin{figure}[H]
    \centering
    % First subfigure
    \begin{subfigure}[b]{0.6\textwidth}
        \centering
        \definecolor{jetsilver}{RGB}{192,192,192} % light phase, as rendered in Figure_48
        \begin{tikzpicture}
            \node[anchor=south west,inner sep=0] (jetimg) at (0,0)
                {\includegraphics[width=\textwidth]{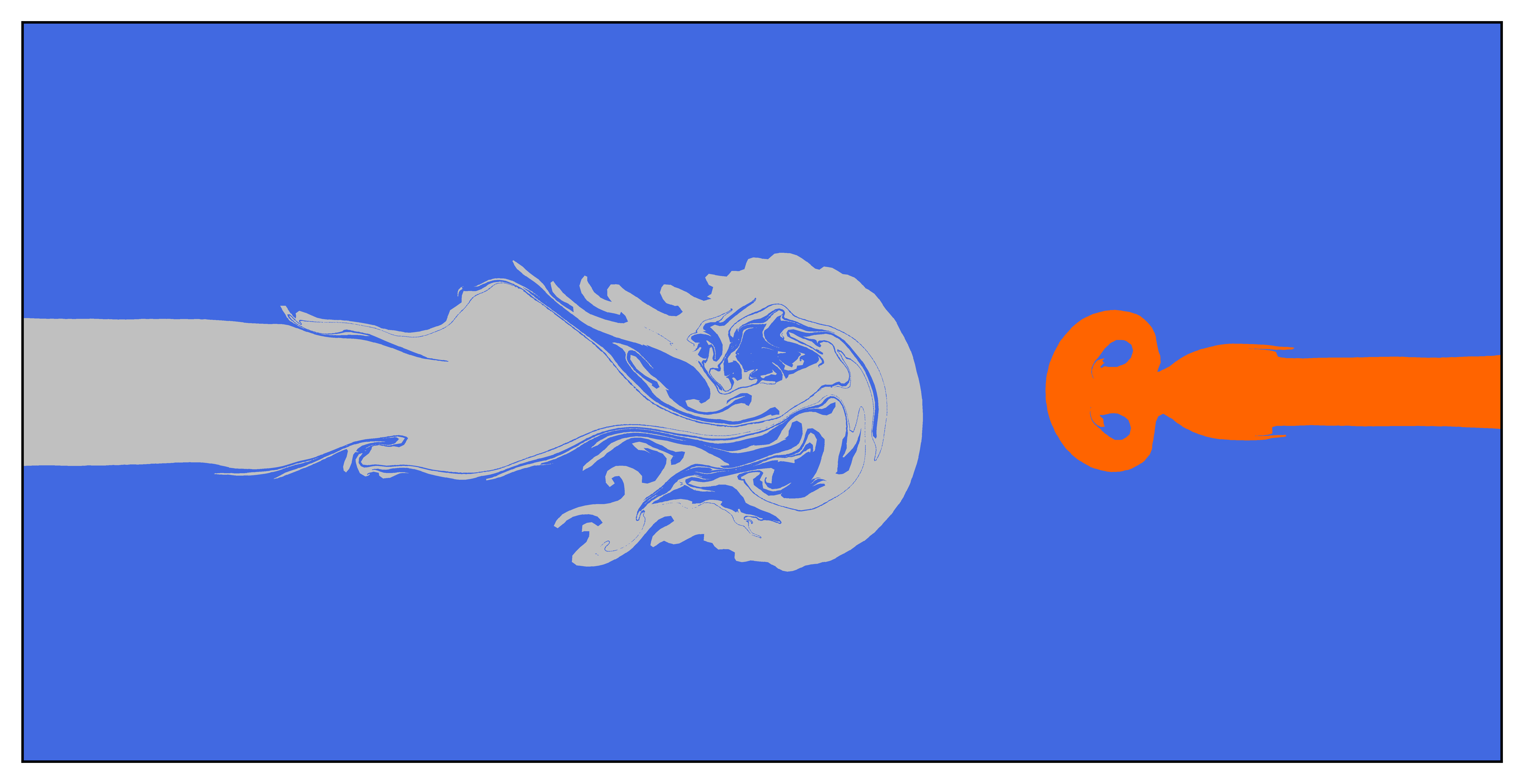}};
            % Boundary conditions drawn on top of the snapshot. All coordinates below
            % are normalised image coordinates of the plotted domain frame.
            \begin{scope}[x={(jetimg.south east)},y={(jetimg.north west)}]
                \def\xl{0.0147}\def\xr{0.9853}   % lateral walls
                \def\yb{0.03}\def\yt{0.9716}   % bottom / top boundaries
                \def\ylb{0.4062}\def\ylt{0.5943} % left inlet opening
                \def\yrb{0.4535}\def\yrt{0.5470} % right inlet opening
                % Open boundaries: hide the plotted frame and redraw it dashed
                \draw[white,line width=1.2pt]
                    (\xl,\yt) -- (\xr,\yt)  (\xl,\yb) -- (\xr,\yb);
                \draw[black,line width=1pt,dash pattern=on 2.5pt off 2pt]
                    (\xl,\yt) -- (\xr,\yt)  (\xl,\yb) -- (\xr,\yb);
                % Inlets: hide the frame where the two jets enter the domain
                \draw[jetsilver,line width=1pt] (\xl,\ylb) -- (\xl,\ylt);
                \draw[orangered,line width=1pt] (\xr,\yrb) -- (\xr,\yrt);
                % Impermeable lateral walls, interrupted at the inlets
                \draw[black,line width=1.4pt]
                    (\xl,\yb) -- (\xl,\ylb)  (\xl,\ylt) -- (\xl,\yt)
                    (\xr,\yb) -- (\xr,\yrb)  (\xr,\yrt) -- (\xr,\yt);
                % Prescribed inlet velocities
                \draw[-{Latex[length=3.5pt]},line width=1.2pt]
                    (0.022,0.5) -- (0.085,0.5)
                    node[midway,above=0.5pt,inner sep=0pt,font=\scriptsize]{$\mathbf{U}_l$};
                \draw[-{Latex[length=3.5pt]},line width=1.2pt]
                    (0.978,0.5) -- (0.930,0.5)
                    node[midway,above=0.5pt,inner sep=0pt,font=\scriptsize]{$\mathbf{U}_r$};
            \end{scope}
        \end{tikzpicture}
        \caption{$t=1.0$}
        \label{fig:jet1}
    \end{subfigure}
    
    \vspace{1em} % Adds vertical space between the top and bottom rows
    
    % Third subfigure
    \begin{subfigure}[b]{0.6\textwidth}
        \centering
        \includegraphics[width=\textwidth]{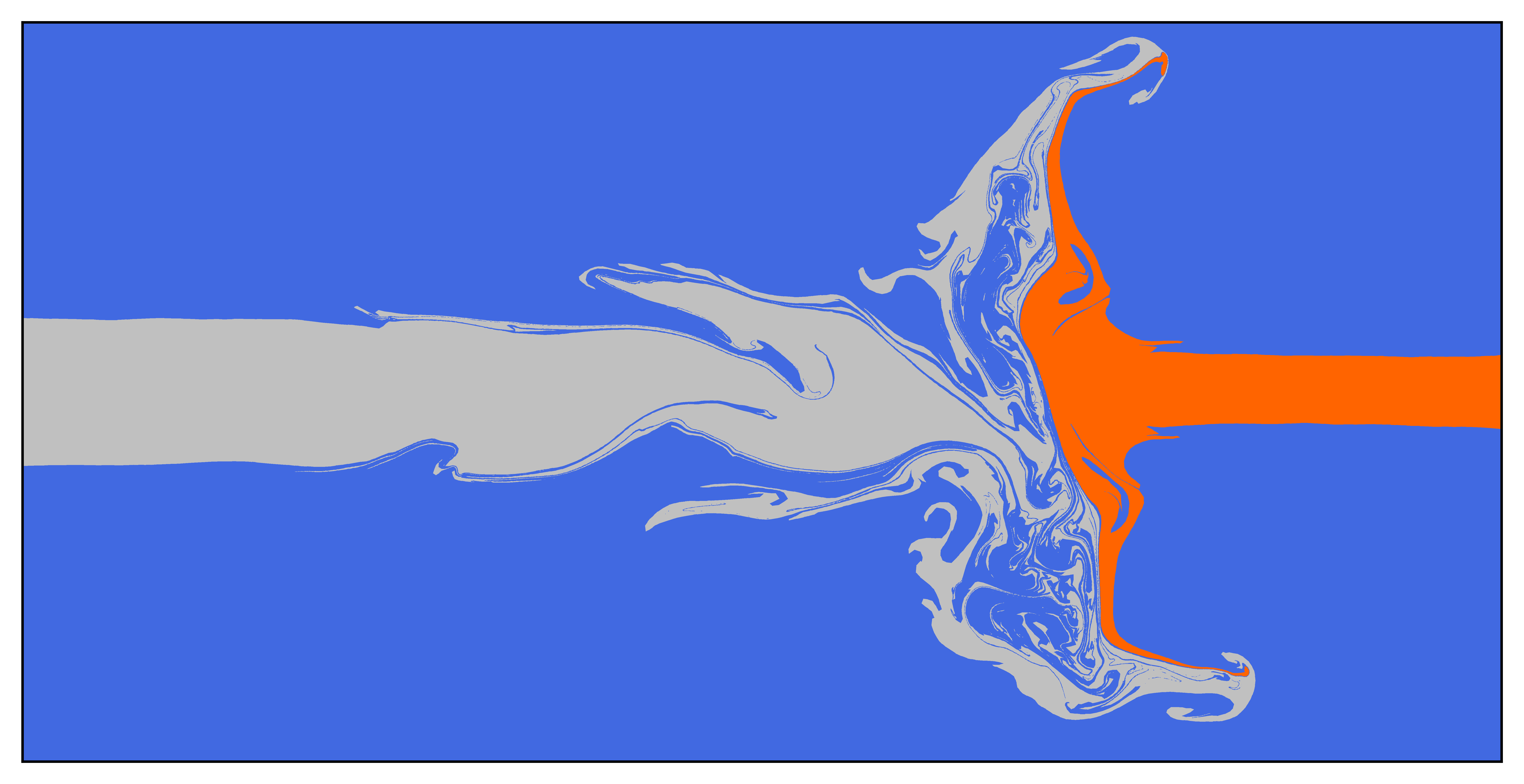}
        \caption{$t=1.3$}
        \label{fig:jet2}
    \end{subfigure}

    \vspace{1em} % Adds vertical space between the top and bottom rows
    
    % Third subfigure
    \begin{subfigure}[b]{0.6\textwidth}
        \centering
        \includegraphics[width=\textwidth]{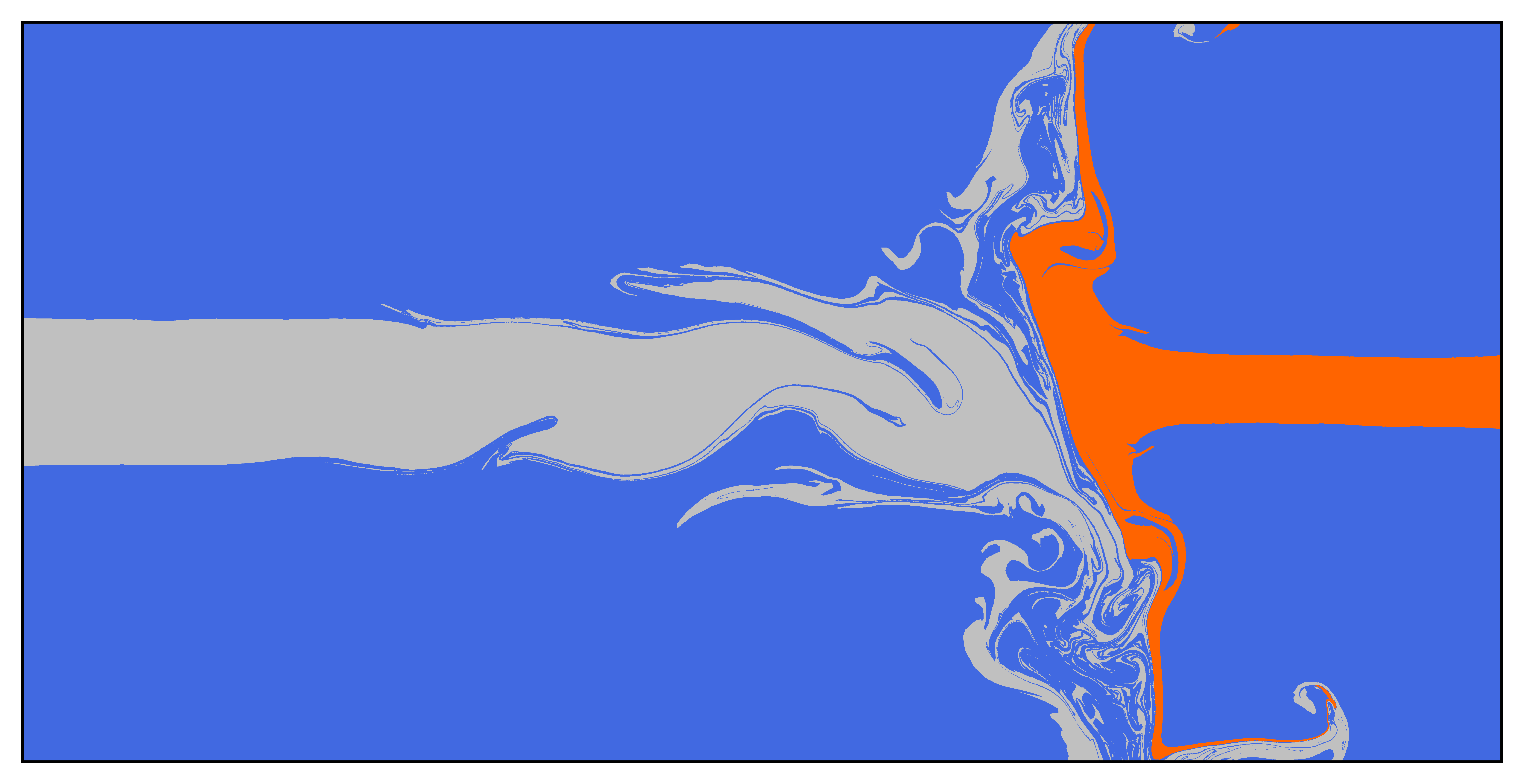}
        \caption{$t=1.4$}
        \label{fig:jet3}
    \end{subfigure}
    \caption{Two colliding jets. The domain is open at the top and bottom boundaries, and the jets enter from the left and right with velocities $\mathbf{U}_l=[2,0]^T$ and $\mathbf{U}_r=[-1,0]^T$, respectively. The boundaries are sketched in (a), where the solid lines denote the impermeable lateral walls, interrupted at the two inlets, and the dashed lines the open top and bottom boundaries. The fluid properties $(\rho,\mu)$ are given by $(0.1,5\times10^{-4})$ for \myrect[silver], $(1,10^{-4})$ for \myrect[blue], and $(2,10^{-3})$ for \myrect[orangered], respectively. Effects of surface tension are not considered, and no gravitational acceleration is at play.}
    \label{fig:jet}
\end{figure}

Flows in which mass is continuously injected into a domain at a prescribed rate arise in a wide range of engineering applications. The present test case, two immiscible jets issued from opposite lateral walls into a third, initially quiescent phase, illustrates how an inlet condition is expressed in the PFEM framework, and how the method behaves once the resulting high-vorticity structures reach an open boundary. The setup and the resulting flow are shown in \autoref{fig:jet}.
\medskip

The inlet requires no complex treatment. The nodes lying on the inlet boundary are constrained to remain on it and a velocity is prescribed there; the Lagrangian update then advects the neighboring material into the domain exactly as it advects any interior node. The incoming mass is consistent by construction: no numerical flux is evaluated, no convective term is discretized at the boundary, and the phase carried by the entering material is unambiguous. The mesh adaptation stage plays an essential role here. As the injected material moves away from the inlet, the adjacent elements stretch and the local node density drops; the insertion step restores the target size, so that a steady stream of nodes is generated at the inlet and convected downstream.
\medskip

The open boundaries at the top and bottom of the domain are a more delicate matter. At sufficiently low viscosity, high-vorticity structures are eventually convected onto these boundaries, where the normal velocity reverses locally and material re-enters the domain. In a multiphase context, the phase of that material is not determined by the data available to the solver: any answer to the question of which fluid is entering is a modeling choice rather than a computed result. This indetermination is not a consequence of the Lagrangian description; it is a property of the boundary model itself, and is encountered in identical form in interface-capturing methods. A satisfactory treatment of re-entrant flow at an open boundary crossed by a moving interface is a research question in its own right, and one largely orthogonal to the contributions of this work.

% ----------- FIGURES FOR DISCUSSION ----------------------
% ---------------------------------------------------------
\clearpage

\vspace*{\fill}
\vspace*{\fill}

\clearpage

\vspace*{\fill}
\vspace*{\fill}

\clearpage

%% -------------------------------------------------------------------------
\section{Conclusion}
\label{sec:conclusion}

In this paper, we introduced a robust, topology-preserving framework for simulating multiphase flows using the Particle Finite Element Method (PFEM). A fundamental challenge in fully Lagrangian multiphase simulations is the severe distortion of the fluid mesh over time, which necessitates frequent remeshing without destroying the discrete interfaces between phases. To address this, we developed an algorithm rooted in the mathematical duality between Delaunay triangulations and Voronoï diagrams. By strictly enforcing a diametral empty-disk condition along the piecewise-linear interfaces, achieved through strategic edge splitting and the selective filtering of encroaching volume nodes, our methodology guarantees that critical interface edges emerge natively during a standard Delaunay triangulation. This approach entirely eliminates the reliance on computationally expensive constrained triangulations while robustly preserving the physical topology of the multiphase system.
\medskip

To ensure computational efficiency and optimal element quality, this topology-preserving core was coupled with dynamic mesh adaptation techniques. We adapted Chew's refinement algorithm to operate in tandem with our interface protection rules, and we implemented a rigorous coarsening strategy governed by a geometric flatness tolerance. This allows the mesh to maintain a high-resolution clustering of nodes exclusively in the vicinity of the interfaces. Furthermore, because our algorithm guarantees topological conformity between the advected and remeshed domains, the projection of physical properties onto the new mesh becomes a trivial, mass-conserving procedure. The framework demonstrates excellent mass retention. As highlighted in our benchmark simulations, explicitly protecting the mesh topology prevents artificial edge flips that would otherwise trigger non-physical phenomena, such as premature filament pinch-off and artificial satellite bubble generation.
\medskip

Despite these successes, the proposed methodology highlights clear avenues for future research. First, while our adaptive node clustering is highly efficient for well-localized interfaces, chaotic mixing phases that generate pervasive thin filaments can still cause a rapid surge in the global node count as the algorithm enforces minimum resolution constraints. Second, the numerical validation presented here has been performed exclusively in two dimensions. The underlying theory, namely recovering a surface in the Delaunay triangulation by enforcing an empty-ball condition on each face, remains valid in three dimensions; the difficulty is practical, as splitting a face does not monotonically decrease the radius of its smallest circumsphere. A proof-of-concept 3D pipeline, in which faces are successively split through edge splitting until all of them are protected, produces correct results but proved computationally prohibitive given the number of nodes inserted. The extension therefore appears feasible, though it will likely require combining naive refinement under strict geometry preservation with additional surface remeshing techniques, and represents the next logical step for this framework.

%% -------------------------------------------------------------------------
\newpage
\section*{CRediT authorship contribution statement}
\noindent\textbf{Félix Ruyffelaere:} Conceptualization, Methodology, Software, Validation, Investigation, Visualization, Writing -- original draft. \\
\textbf{Michel Henry:} Software, Writing -- review \& editing. \\
\textbf{Jonathan Lambrechts:} Conceptualization, Supervision, Writing -- review \& editing. \\
\textbf{Jean-François Remacle:} Conceptualization, Supervision, Funding acquisition, Writing -- review \& editing.

\section*{Declaration of competing interest}
The authors declare that they have no known competing financial interests or personal relationships that could have appeared to influence the work reported in this paper.

\section*{Funding}
This work was supported by the European Research Council (ERC) under the European Union's Horizon Europe research and innovation program through the ERC Synergy Grant X-MESH (grant agreement No 101071255).

\section*{Data availability}
The code and data supporting the findings of this study are openly available at \url{https://git.immc.ucl.ac.be/fruyffelaere/nfluidspfem/-/tree/release}.

\section*{Declaration of generative AI and AI-assisted technologies in the manuscript preparation process}
During the preparation of this work the authors used generative AI tools in order to improve the readability, phrasing and language of the manuscript. After using these tools, the authors reviewed and edited the content as needed and take full responsibility for the content of the published article.

% --- PRINT BIBLIOGRAPHY ---
\clearpage
\small
\bibliographystyle{elsarticle-num} % CMAME numbered reference style (Elsevier)
\bibliography{references}    % Points to references.bib (DO NOT include the .bib extension here)
% ---------------------------

\end{document}